\documentclass[paper]{JHEP}
\usepackage{epsfig}
\usepackage{amssymb}
\def\as{\alpha_{\mbox{\tiny S}}}

\def\MSbar{\overline{\mbox{MS}}}
\def\frac#1#2{ {{#1} \over {#2} }}
\def\ME#1{\overline{{\cal M}}_{#1}}
\def\MEc#1#2{\overline{{\cal M}}_{#1}^{(#2)}}
\def\HW{{\small HERWIG}}
\def\rat#1#2{\mbox{\small $\frac{#1}{#2}$}}
\def\half{\rat{1}{2}}
\def\VEV#1{\left\langle #1\right\rangle}
\def\beq{\begin{equation}}
\def\beqn{\begin{eqnarray}}
\def\eeq{\end{equation}}
\def\eeqn{\end{eqnarray}}
\def\FWeq#1{eq.~({\bf I}.#1)}
\def\abs#1{\left|#1\right|}
\def\bb{\bar\beta}
\def\bh{\bar\theta}
\def\bk{\bar k}
\def\bl{\bar l}
\def\bp{\bar p}
\def\bs{\bar s}
\def\bt{\bar t}
\def\bu{\bar u}
\def\bx{\bar x}
\def\dsb{d\bar\sigma}
\def\tb{\tilde\beta}

\newcommand\sss{\scriptscriptstyle\rm}
\newcommand\bSigma{\overline{\Sigma}}
\newcommand\bE{{\bar E}}
\newcommand\bQ{{\bar Q}}
\newcommand\bq{{\bar q}}
\newcommand\evBB{{\Bigg|_{\rm ev}}}
\newcommand\cntBB{\Bigg|_{\rm ct}}
\newcommand\Hone{(H_1)}
\newcommand\Htwo{(H_2)}
\newcommand\Itwo{{\cal I}_2}
\newcommand\Iqtwo{{\cal I}_{\tilde {2}}}
\newcommand\xMC{|_{\sss {\rm MC}}}
\newcommand\xMCB{\Big|_{\sss {\rm MC}}}
\newcommand\xMCBB{{\Bigg|_{\sss {\rm MC}}}}
\newcommand\stepf{\Theta}
\newcommand\FW{{\bf I}}
\newcommand\Ktwo{{\bf 2}}
\newcommand\Kthree{{\bf 3}}
\newcommand\IMC{I_{\sss MC}}
\newcommand\EVprjmap{{\cal P}_{\clH\to\clS}}
\newcommand\EVprjmapi{{\cal P}_{\clH\to\clS}^{(in)}}
\newcommand\EVprjmapo{{\cal P}_{\clH\to\clS}^{(out)}}
\newcommand\clH{{\mathbb H}}
\newcommand\clS{{\mathbb S}}

\newcommand\thq{\theta_{\sss Q}}
\newcommand\vpq{\varphi}
\newcommand\xin{x_{in}}
\newcommand\yin{y_{in}}
\newcommand\thin{\theta_{in}}
\newcommand\vpin{\varphi_{in}}
\newcommand\xout{x_{out}}
\newcommand\yout{y_{out}}
\newcommand\thout{\theta_{out}}
\newcommand\vpout{\varphi_{out}}
\newcommand\pt{p_{\rm \scriptscriptstyle T}}
\newcommand\kt{k_{\rm \scriptscriptstyle T}}
\newcommand\mt{m_{\rm \scriptscriptstyle T}}
\newcommand\bkT{\bk_{\rm \scriptscriptstyle T}}
\newcommand\bkL{\bk_{\rm \scriptscriptstyle L}}
\newcommand\Gfun{{\cal G}}
\newcommand\ptt{p_{\sss {\rm T}}^{\sss (t)}}
\newcommand\pttbar{p_{\sss {\rm T}}^{\sss (\bar{t})}}
\newcommand\yt{y^{\sss (t)}}
\newcommand\pttt{p_{\sss {\rm T}}^{\sss (t\bar{t})}}
\newcommand\dphitt{\Delta\phi^{\sss (t\bar{t})}}
\newcommand\yb{y^{\sss (b)}}
\newcommand\ybbar{y^{\sss (\bar{b})}}
\newcommand\ptb{p_{\sss {\rm T}}^{\sss (b)}}
\newcommand\ptbbar{p_{\sss {\rm T}}^{\sss (\bar{b})}}
\newcommand\ptbb{p_{\sss {\rm T}}^{\sss (b\bar{b})}}
\newcommand\dphibb{\Delta\phi^{\sss (b\bar{b})}}
\newcommand\siggg{\sigma_{ab}}

\newcommand\fgg{m_{ab}}
\newcommand\thu{\thq}
\newcommand\thd{\varphi}
\newcommand\ep{\epsilon}
\newcommand\omxr{\left(\frac{1}{1-x}\right)_{\tilde\rho}}
\newcommand\omxrr{\left(\frac{1}{1-x}\right)_{\rho}}
\newcommand\omyo{\left(\frac{1}{1-y}\right)_{\tilde\omega}}
\newcommand\omyp{\left(\frac{1}{1-y}\right)_{1}}
\newcommand\opyo{\left(\frac{1}{1+y}\right)_{\tilde\omega}}
\newcommand\lomxr{\left(\frac{\log(1-x)}{1-x}\right)_{\tilde\rho}}
\newcommand\lomxrr{\left(\frac{\log(1-x)}{1-x}\right)_{\rho}}
\newcommand\pfun{{\cal P}}
\newcommand\ptmin{{\tt PTMIN}}
\newcommand\twototwo{2\to 2}
\newcommand\twotothree{2\to 3}
\preprint{
 Bicocca--FT--03--11\hfill\\
 Cavendish--HEP--03/03\hfill\\
 CERN--TH/2003--102\hfill\\
 GEF--TH--5/2003}
\title{Matching NLO QCD and Parton Showers in Heavy Flavour Production%
\footnote{Work supported in part by the UK Particle Physics and
Astronomy Research Council and by the EU Fourth Framework Programme
`Training and Mobility of Researchers', Network `Quantum Chromodynamics
and the Deep Structure of Elementary Particles',
contract FMRX-CT98-0194 (DG 12 - MIHT).}}
\author{Stefano Frixione\\
  INFN, Sezione di Genova,
  Via Dodecaneso 33, 16146 Genova, Italy\\
  E-mail: \email{Stefano.Frixione@cern.ch}}
\author{Paolo Nason\\
  INFN, Sezione di Milano,
  Piazza della Scienza 3, 20126 Milan, Italy\\
  E-mail: \email{Paolo.Nason@mib.infn.it}}
\author{Bryan R.\ Webber\\
  Theory Division, CERN, 1211 Geneva 23, Switzerland and\\
  Cavendish Laboratory, 
  Madingley Road, Cambridge CB3 0HE, U.K.\\
  E-mail: \email{webber@hep.phy.cam.ac.uk}}
\abstract{We apply the MC@NLO approach to the process of heavy flavour
  hadroproduction. MC@NLO is a method for matching next-to-leading
  order (NLO) QCD calculations and parton shower Monte Carlo (MC)
  simulations, with the following features: fully
  exclusive events are generated, with hadronisation according to the
  MC model; total rates are accurate to NLO; NLO results for
  distributions are recovered upon expansion in $\as$; hard emissions
  are treated as in NLO computations while soft/collinear emissions
  are handled by the MC simulation, with the same logarithmic accuracy
  as the MC; matching between the hard and soft regions is smooth, and
  no intermediate integration steps are necessary.  The method
  was applied previously to the hadroproduction of gauge boson pairs,
  which at NLO involves only initial-state QCD radiation and a unique
  colour structure. In heavy flavour production, it is necessary to
  include contributions from final-state QCD radiation and different
  colour flows.  We present illustrative results on top and bottom
  production at the Tevatron and LHC.}
\begin{document}
\section{Introduction}
The process of heavy flavour production in hadron collisions is a
valuable testing-ground for perturbative QCD, since the high scale
set by the quark mass should ensure that perturbative calculations are
reliable. The prediction of cross sections and final-state distributions
in heavy flavour production is also important for the design of collider
experiments and new particle searches, since this process gives rise to
irreducible backgrounds to many types of new physics.

Up to now, the theoretical emphasis has been on next-to-leading order
(NLO) calculations of total rates, single-inclusive 
distributions, and heavy quark-antiquark correlations, sometimes
with resummation of higher-order contributions that are enhanced in certain
kinematic regions. However, for many purposes, such as studies of backgrounds
to new physics, one needs a more complete characterization of the final state.
This is provided by a Monte Carlo event generator program, which combines a
calculation of the hard production process with a parton shower simulation and
a hadronization model, to yield an approximate but realistic hadron-level
event structure.

The problem with existing Monte Carlo event generators is that they
are based on a leading-order (LO) calculation of the production process
combined with a leading-logarithmic (LL) treatment of higher orders
via the parton shower approximation.  It has proved highly non-trivial
to incorporate the benefits of NLO calculations into event generators,
since the parton showers include parts of the NLO corrections, which
should not be double-counted. On the other hand, the parton showers cannot be
omitted, since they provide a reliable description of how final state hard
partons evolve into QCD jets. Furthermore, any viable hadronization model
operates on the multiparton states that are created by showering.

A recent proposal for combining NLO calculations and parton showers
is the so-called MC@NLO approach, introduced in ref.~\cite{Frixione:2002ik}
(hereafter referred to as \FW). It is based on the highly
successful subtraction method for NLO calculations. The basic idea 
is to modify the subtraction to take into account the terms that are
generated by the parton shower. This results in a set of weighted
LO and NLO parton configurations that can be fed into the parton
showering generator without fear of double counting.  Each weight
distribution is well-behaved in the sense that it has no divergences
or pathological tails that would lead to Monte Carlo inefficiency.
However, in order to reproduce the NLO corrections fully, some
of the configurations have negative weights. Event unweighting
can still be achieved efficiently, if desired, by generating a
small fraction of `counter-events' that contribute with equal but
opposite weight to events in all distributions.

The MC@NLO method was worked out in detail in \FW\ for processes
in which, at the Born level, there are no coloured partons in the
final state. An important example is gauge boson pair production, for
which a wide range of MC@NLO predictions were presented there.  To deal
with the process of heavy quark production we must take into account 
QCD radiation from final-state partons, in this case
the heavy quarks themselves.  A further new complication is the
possibility of different colour flows.

We shall see that no difficulties of principle arise from these
complications.  The only new task, albeit a laborious one, is to
calculate precisely what the shower Monte Carlo is doing at NLO,
in order to compute the modified subtractions correctly.  In our
case we use the \HW\ shower Monte Carlo~\cite{Corcella:2000bw}, 
the relevant features of which are summarized in Appendix A.

In the following section we review the main features of the MC@NLO approach.
Then, in sect.~\ref{sec:kin}, we discuss the partonic processes
that contribute to heavy flavour production in standard Monte
Carlos and in MC@NLO, and we define the kinematic variables for the 
corresponding $\twototwo$ and $\twotothree$ processes. In order to compute 
the Monte Carlo subtraction terms that form the basis of the
MC@NLO method, we have first to relate these variables to
those used in the \HW\ program.  This is done in
sect.~\ref{sec:rel}. Next we write down,
in sect.~\ref{sec:sub}, the approximate  $\twotothree$ particle
production cross sections generated by the \HW\ parton showering
algorithm, which are then used to construct the Monte
Carlo subtraction terms for heavy flavour production.
Technical details of this procedure are given in Appendix B.
Inserting the subtraction terms
in the formulae reviewed in sect.~\ref{sec:rev}
enables us to generate parton configurations that can
be fed into the shower Monte Carlo without any
double counting of NLO contributions. In order for
the \HW\ Monte Carlo to operate correctly we have to
assign a colour flow to each configuration; this
and other details of the implementation are
explained in sect.~\ref{sec:imp}.

We present the predictions of MC@NLO for top quark production
in  sect.~\ref{sec:top}. The case of bottom is
much more involved than that of top. Problems affecting 
$b$-physics simulations with standard Monte Carlos are reported 
in sect.~\ref{sec:binHerwig}; in sect.~\ref{sec:binMCatNLO}, we
discuss some of the features of MC@NLO in $b$ production, and in
particular the treatment of large logarithms of $\pt/m$, the ratio 
of the quark transverse momentum to its mass; in sects.~\ref{sec:bbcorr}
and~\ref{sec:bSingInc} we present MC@NLO predictions for $b\bar{b}$
correlations and single-inclusive distributions respectively, and
compare them to \HW\ and NLO results. Finally, conclusions and future
prospects are presented in sect.~\ref{sec:conc}.

\section{\boldmath Review of MC@NLO approach}
\label{sec:rev}

The MC@NLO formalism is defined in eq.~(4.22) of \FW, which
we denote by \FWeq{4.22}. We rewrite that
equation in the following, fully equivalent, form:
\beqn
{\cal F}_{\mbox{\tiny MC@NLO}}&=&\sum_{ab}\int dx_1 dx_2 d\phi_3\Bigg\{
{\cal F}_{\mbox{\tiny MC}}^{(3)}\Bigg(\frac{d\bSigma_{ab}^{(f)}}{d\phi_3}\evBB
-\frac{d\bSigma_{ab}}{d\phi_3}\xMCBB\Bigg)
\nonumber \\*&& \phantom{a}
+{\cal F}_{\mbox{\tiny MC}}^{(2)}\Bigg[
-\frac{d\bSigma_{ab}^{(f)}}{d\phi_3}\cntBB
+\frac{d\bSigma_{ab}}{d\phi_3}\xMCBB
+\frac{1}{\Itwo}\Bigg(\frac{d\bSigma_{ab}^{(b)}}{d\phi_2}
+\frac{d\bSigma_{ab}^{(sv)}}{d\phi_2}\Bigg)
\nonumber \\*&& \phantom{a}
+\frac{1}{\Iqtwo}\Bigg(
\frac{d\bSigma_{ab}^{(c+)}}{d\phi_2 dx}\evBB
+\frac{d\bSigma_{ab}^{(c-)}}{d\phi_2 dx}\evBB\Bigg)
-\frac{1}{\Iqtwo}\Bigg(\frac{d\bSigma_{ab}^{(c+)}}{d\phi_2 dx}\cntBB
+\frac{d\bSigma_{ab}^{(c-)}}{d\phi_2 dx}\cntBB\Bigg)
\Bigg]\Bigg\}.\phantom{aaaa}
\label{eq:rMCatNLO}
\eeqn
The quantities $d\sigma/dO$, $\IMC(O,\Kthree)$, and $\IMC(O,\Ktwo)$
appearing in \FWeq{4.22} have been replaced here by 
${\cal F}_{\mbox{\tiny MC@NLO}}$ (the MC@NLO generating functional), 
${\cal F}_{\mbox{\tiny MC}}^{(3)}$ (the MC generating functional
when starting from a $\twotothree$ hard subprocess), and 
${\cal F}_{\mbox{\tiny MC}}^{(2)}$ (the MC generating functional
when starting from a $\twototwo$ hard subprocess) respectively.
This renders more transparent the fact that the dependence upon the
observable $O$ in \FWeq{4.22} is only formal, and that the MC@NLO 
generates events without reference to any observable. We refer the 
reader to $\FW$ for the definitions of all the terms appearing in 
eq.~(\ref{eq:rMCatNLO}); however, the precise details will not be
relevant here. In what follows, we shall limit ourselves to describing the
basic features of eq.~(\ref{eq:rMCatNLO}), and their role in the 
implementation of heavy flavour production in MC@NLO.

We start by recalling that the definition of MC@NLO is derived from the
expectation value $\VEV{O}$, computed at the NLO, of a generic observable $O$,
\FWeq{4.19}. This can be read off from eq.~(\ref{eq:rMCatNLO}) simply by 
removing the MC subtraction terms $d\bSigma_{ab}\xMC$, and by replacing ${\cal
F}_{\mbox{\tiny MC}}^{(n)}$ with $O({\bf n})$, where $O({\bf n})$ is the
observable $O$ computed in an $n$-body final-state configuration
($n=2,3$). The NLO expression for $\VEV{O}$ is an integral over 
the momentum fractions $x_1$ and $x_2$ of the incoming partons,
and three-body phase-space variables $\phi_3$. Each point $(x_1,x_2,\phi_3)$
in the integration range thus corresponds to a $\twotothree$ kinematic
configuration (called $\clH$ hereafter).
Furthermore, a definite $\twototwo$ configuration is chosen (called $\clS$
hereafter), according to a mapping which we denote by $\EVprjmap$. The weight
associated with $\clH$ is given by the real-emission matrix element; the sum
of {\em all} the remaining contributions to the NLO cross section (namely, the
Born term, the virtual term, the soft and collinear counterterms, and the
finite remainders of the initial-state collinear singularity subtractions)
constitutes the weight associated with $\clS$. One can prove that it is always
possible to cast any expectation value $\VEV{O}$ in this form (see $\FW$,
sects.~4.4 and A.4), through a formal procedure that we call {\em event
projection}. It should be clear that event projection does not imply any
approximation, and that its specific form depends on the choice of
$\EVprjmap$. In the context of a pure NLO computation, this choice is
arbitrary, and its freedom has been used in the past~\cite{Mangano:jk} to
improve the convergence of the numerical integration procedure.
On the other hand, when defining an MC@NLO it is the MC itself that 
dictates the form of $\EVprjmap$.

We suppose now to have chosen a map $\EVprjmap$, and to have performed event
projection on an NLO cross section. For each point $(x_1,x_2,\phi_3)$ we get a
pair of kinematic configurations, $\clH$ and $\clS$.  Instead of using these
configurations for defining the observable $O$, as in the NLO computation of
$\VEV{O}$, we feed them into an MC, where they are treated as initial
conditions for shower evolution. This corresponds to defining the MC@NLO as in
eq.~(\ref{eq:rMCatNLO}), except for the MC subtraction terms
$d\bSigma_{ab}\xMC$ which are omitted. However, this naive attempt fails (see
$\FW$, sect.~3.3.1). Basically, when evolving $\clS$ configurations the shower
reproduces some of the $\clH$ configurations, which are therefore double
counted. The idea of {\em modified subtraction}, upon which the MC@NLO
approach is based, is to subtract these double-counted configurations at the
level of short-distance cross sections. This is the role of the MC subtraction
terms $d\bSigma_{ab}\xMC$ which appear in eq.~(\ref{eq:rMCatNLO}); they come
in pairs, since they have to account for both the emission and the
non-branching probabilities in the MC. The MC subtraction terms act as local
counterterms in eq.~(\ref{eq:rMCatNLO}), and this implies that the weight
distributions for $\clH$ and $\clS$ configurations (the terms multiplying
${\cal F}_{\mbox{\tiny MC}}^{(3)}$ and ${\cal F}_{\mbox{\tiny MC}}^{(2)}$
respectively, see also \FWeq{4.23} and \FWeq{4.24}) are separately convergent, 
thus allowing event unweighting as is customary in MC simulations.

The MC subtraction terms are obtained by formally
expanding the MC results to the first
non-trivial order in $\as$, which corresponds to an $\clH$ configuration.
Typically, $d\bSigma_{ab}\xMC$ has the form of a hard, $\twototwo$ cross 
section, times a kernel which describes (the first) parton branching. The 
shower algorithm fully specifies how to determine the $\clH$ configuration,
given the hard $\clS$ configuration and the values of the showering
variables. Thus, it implicitly defines a map between the $\clH$ and
$\clS$ configurations. We choose $\EVprjmap$, used in event projection,
to coincide with this map. Notice that this is possible only because the 
shower algorithm is independent of the hard process.

In summary, the main steps that have to be taken in order to construct
an MC@NLO are the following.
\begin{itemize}
\item[{\em i)}] Determine $\EVprjmap$.
\item[{\em ii)}] Write the NLO cross section for the relevant production
 process, and perform event projection on it, using the map $\EVprjmap$ 
 found in {\em i)}.
\item[{\em iii)}] Define the MC subtraction terms, and insert them into the 
 expression for the NLO cross section.
\end{itemize}
Of the three steps above, only {\em ii)} depends on the process in a
non-trivial way. The implementation of event projection requires
a detailed knowledge of the formalism adopted to write the cross
section to the NLO accuracy; an explicit example of such a procedure
has been given in $\FW$, but different prescriptions are clearly possible.
On the other hand, step {\em i)} is strictly process-independent,
and thus $\EVprjmap$ can be determined once and for all. However, 
$\EVprjmap$ depends on the particular shower algorithm adopted, 
and therefore different MC's define different $\EVprjmap$ maps. 
Finally, step {\em iii)} is process dependent, but only through
the hard $\twototwo$ cross sections that appear in a factorized
form in the MC subtraction terms, i.e. at the LO level, which
is fairly simple to deal with. The part of the MC subtraction
term which describes the first branching depends only on the
shower algorithm, and can therefore be studied with full
generality. 

The definition of the MC subtraction terms is in fact one of the main goals of
the present paper; final-state emissions are considered here for the first
time, since they were not relevant to the production of vector boson pairs
considered in $\FW$. Also, the formulae presented in $\FW$ for initial-state
emissions will be generalized here to account for colour structures more
complicated that those of $\FW$. These results will allow an almost
straightforward definition of the MC subtraction terms for any 
production process. A subtle point concerns the interplay between
initial- and final-state emissions, and its impact on the definition
of $\clS$-event contribution to MC@NLO. This issue is discussed in
Appendix~\ref{sec:MCsubt}.

\section{Heavy quark production}\label{sec:kin}
\subsection{Contributing processes}\label{sec:processes}
In the MC@NLO approach, the Monte Carlo is not involved in the generation
of the hard process which, apart from the modified subtraction, is treated
as in standard NLO codes. It follows that the partonic production
processes that we need consider are
\beq
q\bq\to Q\bQ\,,\;\;\;\;\;gg\to Q\bQ
\label{eq:FCRtwo}
\eeq
at ${\cal O}(\as^2)$ and ${\cal O}(\as^3)$, and
\beq
q\bq\to Q\bQ g,\;\;\;\;\;
gg\to Q\bQ g,\;\;\;\;\;
gq\to Q\bQ q,\;\;\;\;\;
g\bq\to Q\bQ\bq
\label{eq:FCRthree}
\eeq
at ${\cal O}(\as^3)$. As discussed previously, the ${\cal O}(\as^2)$
contributions in eq.~(\ref{eq:FCRtwo}) generate (some of) the 
configurations in eq.~(\ref{eq:FCRthree}) through parton showering, but
MC@NLO is defined in such a way to avoid any double counting.
The processes in eq.~(\ref{eq:FCRtwo}) are traditionally classified 
as flavour creation (FCR hereafter) to distinguish them from the
so-called flavour excitation (FEX hereafter) processes
\beq
qQ\to qQ,\;\;\;\;\;
q\bQ\to q\bQ,\;\;\;\;\;
gQ\to gQ,
\label{eq:FEXtwo}
\eeq
where charge-conjugate processes are understood to be included. In the
case of FEX processes, it is implicitly assumed that the relevant
heavy flavour is already present in the parton distribution
functions (PDFs) of the incoming hadrons.  

In standard Monte Carlo, both FCR and FEX matrix elements are used
in the generation of the hard processes.\footnote{FEX is not relevant
to top production at realistic collider energies.} This is not the case 
in NLO computations, since the partons which initiate the hard processes
are treated as massless, understanding that their actual mass is
negligible with respect to the hard scale of the process, in this
case the heavy quark mass $m$. This forbids the presence of heavy
quarks in the initial state. Furthermore, if we use NLO matrix 
elements in the generation of the hard process, as in MC@NLO, the
distinction between FCR and FEX becomes ambiguous. For example,
the NLO FCR process $gg\to Q\bQ g$ has a contribution from initial-state
quasi-collinear gluon splitting, $g\to Q\bQ$, followed by the LO
FEX process $Qg\to Qg$. Since initial-state quasi-collinear gluon
splitting forms part of the evolution of the PDFs of the incoming
hadrons implemented through parton showers, 
we are in danger of double-counting if we include both
LO FEX and NLO FCR processes in MC@NLO. On the other hand, this
example explains why FEX processes are considered in standard Monte
Carlo: they allow one to include some of the features that could not
be included by shower evolution initiated by FCR processes.
We also point out that the process $gg\to Q\bQ g$, which we
include in the MC@NLO, generates certain kinematic configurations
that can be equivalently produced starting from a hard $gg\to gg$ 
process followed by final-state gluon splitting $g\to Q\bQ$.
The gluon splitting mechanism (GSP hereafter) is known to be important in 
bottom production, giving the dominant contribution to those configurations
in which $b$ and $\bar{b}$ are close in phase space (which
is important not only for the study of correlations, but also in
$b$-jet physics), and non-negligible contributions to single-inclusive 
distributions in the intermediate-$\pt$ region.

Flavour excitation and gluon splitting pose a couple of non-trivial problems
in standard Monte Carlo event generators.\footnote{For a discussion of $b$
physics as described by Monte Carlo parton shower programs, see e.g. 
refs.~\cite{Norrbin:2000zc,Field:2002da}.} In FEX, the presence of a heavy 
quark in the initial state implies a direct dependence upon its PDF;
in the small-$\pt$ region, i.e. close to threshold,
it is unlikely that the treatment of the backward evolution performed
by the parton shower will be compatible with what was implemented in the
PDF evolution. Furthermore, the $t$-channel singularity in the
matrix elements requires the implementation of a cutoff (possibly
an effective one, see app.~\ref{sec:dyn}) which prevents the generation
of $\pt=0$ events. Both features result in predictions which have
a certain degree of implementation dependence. As far as GSP
is concerned, the corresponding events are usually obtained 
by considering pure QCD events (i.e., with no heavy quark involved in 
the hard process, the generation of which also requires a $\pt$ cutoff), 
and selecting those in which at least one $Q\bQ$ pair is obtained
through showering, a very inefficient procedure from the statistical
point of view. More details will be given in sect.~\ref{sec:binHerwig}.

Worst of all, both FEX and GSP processes are well defined only in the
case of large transverse momenta of the heavy quark. Their extrapolation
to the low transverse momentum region can at best be considered a very
rough model of higher-order heavy flavour production processes.

The implementation of heavy flavour production in MC@NLO helps to
avoid some of these problematic features of the FEX and GSP
processes. In the processes considered in NLO computations,
eqs.~(\ref{eq:FCRtwo}) and~(\ref{eq:FCRthree}), there is no such thing
as direct dependence upon the heavy quark PDF, and singularities are
cancelled at the level of short-distance cross sections, without the
need to introduce unphysical parameters acting as cutoffs.
Furthermore, one generates gluon splittings with 100\% efficiency,
since they are included in the matrix elements. Clearly, the presence
of negative weights in MC@NLO degrades the statistics, but (at least
with the fraction of negative weights we find in our implementation)
the overall efficiency is still larger than that required to obtain
a useful gluon-splitting sample with the usual Monte Carlo method.

\begin{figure}[htb]
\begin{center}
\epsfig{file=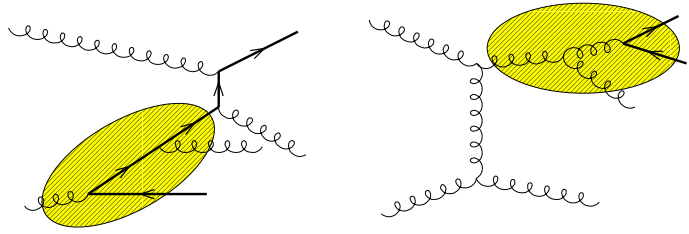,width=0.6\textwidth}
\end{center}
\caption{\label{fig:as4logs}
Graphs that give rise to enhanced higher-order terms,
and are not included in the MC@NLO implementation. Shaded
areas represent almost collinear branchings due to parton
showers.}
\end{figure}

However, an MC@NLO based on the hard processes of eqs.~(\ref{eq:FCRtwo}) 
and~(\ref{eq:FCRthree}) cannot account for {\em all} the contributions 
generated by FEX and GSP. For example, MC@NLO does not include the 
diagrams in which a gluon, rather than a heavy quark, is emitted in the 
first backward branching of the heavy quark entering a FEX hard process.
Similarly, MC@NLO does not include those GSP diagrams
in which the gluon emits another gluon before splitting.
Examples of omitted contributions are shown in fig.~\ref{fig:as4logs}.

The absence of such contributions is not surprising: the
terms missed by the MC@NLO are part of those relevant to the
region of large heavy-quark transverse momenta (large $\pt$),
where the quark mass $m$ no longer sets the hard scale. In this region,
the NLO computation is not expected to give sensible results,
since large logarithms $\log(\pt/m)$ need
to be resummed to all orders.  For single-inclusive observables, 
techniques are known for resumming these large logs (to NLL accuracy), 
and for matching the resummed result to the NLO one (for example,
FONLL~\cite{Cacciari:1998it}). However, there is at present no known
way of extending such a procedure to the fully exclusive case that
would be needed in order to include it in MC@NLO. 
This problem is relevant, for example, to bottom production at the Tevatron.
In sect.~\ref{sec:binMCatNLO} we shall show that MC@NLO gives in practice a
sensible answer also in this case, since these enhanced high-$\pt$ effects
are not so large in the kinematic region of practical interest.

\subsection{$\twototwo$ processes}
In this and in the following section, we introduce the notation
that we shall use in order to describe the hard processes in 
eqs.~(\ref{eq:FCRtwo}) and~(\ref{eq:FCRthree}).

We denote variables relating to the $\twototwo$ FCR 
processes of eq.~(\ref{eq:FCRtwo}) by barred symbols:
$\bp_1,\bp_2$ represent the incoming
(massless) momenta and $\bk_1,\bk_2$ the outgoing heavy quark
and antiquark (mass $m$) momenta.  The momentum fractions of the 
incoming partons are denoted by $\bx_{1,2}$, i.e.
\beq
\bp_1 = \bx_1P_1\;,\;\;\;
\bp_2 = \bx_2P_2\;,
\eeq
where $P_{1,2}$ are the beam momenta, and the invariants by
\beq
\bs = 2\bp_1\cdot\bp_2\;,\;\;\;
\bt =-2\bp_1\cdot\bk_1\;,\;\;\;
\bu =-2\bp_1\cdot\bk_2\;.
\eeq
Then $\bs +\bt +\bu =0$ and $\bs=\bx_1\bx_2S$, where $S$ is the overall
c.m.\ energy squared.  In the c.m.\ frame of the $\twototwo$ process,
we can write
\beq\label{eq:2to2cm}
\bp_{1,2} = \bE(1,0,0,\pm 1)\;,\;\;\;
\bk_{1,2} = (\bE,\pm\bkT,0,\pm\bkL)\;,
\eeq
where
\beq
\bs = 4\bE^2\;,\;\;\;\bt = -2\bE(\bE-\bkL)\;,\;\;\;
\bu = -2\bE(\bE+\bkL)\;.
\label{eq:stutwo}
\eeq
We also introduce the c.m.\ scattering angle $\bh$ and heavy
quark velocity
\beq
\bb=\sqrt{1-4m^2/\bs}\;,
\label{eq:bbdef}
\eeq
so that
\beq
\bt = -\half\bs (1-\bb\cos\bh)\;,\;\;\;
\bu = -\half\bs (1+\bb\cos\bh)\;.
\label{eq:tuinv}
\eeq
The hard scattering Born cross section is then
\beq
\dsb =  \sum_{ab}d\bx_1\,d\bx_2\,
f_a^{\Hone}(\bx_1)f_b^{\Htwo}(\bx_2)\,\dsb_{ab}\,,
\eeq
where
\beq
\dsb_{ab} \equiv \ME{ab}\,d\phi_2 = 
\frac{\bb}{16\pi}\,\ME{ab}\,d\cos\bh\,,
\label{eq:Bornxsec}
\eeq
$\ME{ab}$ being the spin- and colour-averaged Born matrix element squared
for the process $ab\to Q\bQ$, times the flux factor:
\beqn
\ME{q\bq} &=& g^4
\frac{N^2-1}{N^2}\,\frac{1}{2\bs}\left(\frac 12 -\frac{\bt\bu}{\bs^2}+
\frac{m^2}{\bs}\right),
\label{eq:MEqq}
\\
\ME{gg} &=& g^4
\frac{N}{N^2-1}\,\frac{1}{2\bs}
\left(\frac{\bu}{\bt} +\frac{\bt}{\bu} -\frac 1{N^2}
\frac{\bs^2}{\bt\bu}\right)\left(\frac 12 -\frac{\bt\bu}{\bs^2}+
\frac{2m^2}{\bs}-\frac{2m^4}{\bt\bu}\right).
\label{eq:MEgg}
\eeqn

\subsection{$\twotothree$ processes}\label{sec:2to3}
For the $\twotothree$ FCR processes of eq.~(\ref{eq:FCRthree})
we use unbarred symbols: $p_1, p_2$ for the incoming (massless) momenta,
$k_1, k_2$ for the outgoing heavy quark and antiquark momenta,
respectively, and $k$ for the momentum of the emitted (massless) parton.
We denote the momentum fractions of the incoming partons by $x_{1,2}$, i.e.
\beq
p_1 = x_1P_1\;,\;\;\;
p_2 = x_2P_2\;,
\eeq
and the invariants by
\beq
s   = 2 p_1\cdot p_2\;,\;\;\;
t_1 =-2 p_1\cdot k_1\;,\;\;\;
t_2 =-2 p_2\cdot k_2\;,\;\;\;
u_1 =-2 p_1\cdot k_2\;,\;\;\;
u_2 =-2 p_2\cdot k_1\;,
\eeq
so that $s=x_1x_2S$. We also introduce
\beq
v_1 =-2 p_1\cdot k\,,\;\;\;
v_2 =-2 p_2\cdot k\,,
\eeq
and 
\beq
w_1 = 2 k_1\cdot k\,,\;\;\;
w_2 = 2 k_2\cdot k\,,
\eeq
but these are not independent variables since
\beqn
&&s+t_1+u_1+v_1=s+t_2+u_2+v_2=0\,,\nonumber\\
&&s+t_1+u_2-w_2=s+t_2+u_1-w_1=0\,.
\eeqn
This notation is summarized and compared with that of ref.~\cite{Mangano:jk}
in table~\ref{tab:kin}.
\begin{table}[htb]
\renewcommand{\arraystretch}{1.2}
\begin{center}
\begin{tabular}{|c|c|c|c|}
\hline
Label & Invariant & Ref.~\cite{Mangano:jk} & Relation\\
\hline\hline
$s$   &  $2 p_1\cdot p_2$ & $s$   & \\
$t_1$ & $-2 p_1\cdot k_1$ & $q_1$ & \\
$t_2$ & $-2 p_2\cdot k_2$ & $q_2$ & \\
$u_1$ & $-2 p_1\cdot k_2$ & $\hat q_1$ & \\
$u_2$ & $-2 p_2\cdot k_1$ & $\hat q_2$ & \\
$v_1$ & $-2 p_1\cdot k$   & $t_k$ & $-s-t_1-u_1$\\
$v_2$ & $-2 p_2\cdot k$   & $u_k$ & $-s-t_2-u_2$\\
$w_1$ &  $2 k_1\cdot k$   & $w_1$ &  $s+t_2+u_1$\\
$w_2$ &  $2 k_2\cdot k$   & $w_2$ &  $s+t_1+u_2$\\
$M^2_{Q\bQ}$ & $(k_1 +k_2)^2 $  & $s_2$ &  $s+v_1+v_2$\\
\hline
\end{tabular}
\caption{\label{tab:kin}Notation for $\twotothree$ kinematics.}
\end{center}
\end{table}

\section{Relating NLO and MC kinematics}\label{sec:rel}
As discussed in sect.~\ref{sec:rev}, the implementation of MC@NLO
requires a detailed knowledge of the relation between $\twotothree$ ($\clH$)
and $\twototwo$ ($\clS$) kinematic configurations. Also, for the integration
of the MC subtraction
terms over the three-body phase space, as prescribed in 
eq.~(\ref{eq:rMCatNLO}), we need to express the MC showering variables
in terms of the phase-space variables. We solve both
problems in the same way, presenting the $\twototwo$ invariants and showering
variables as functions of the three-body invariants. The latter will 
eventually be computed by means of the three-body phase-space variables;
however, in this section we shall not introduce any explicit phase-space
parametrization, and the results are therefore fully general.

A shower Monte Carlo program generates arbitrarily complicated multiparton
configurations by an iterative process of parton emission, starting from the
coloured external lines of a hard subprocess.  In the hard subprocess these
external lines are treated as being on mass-shell, but the showering process
may drive them off the mass shell, and the kinematics have to be adjusted to
restore energy-momentum conservation.  This process is called momentum
reshuffling; the way it is implemented in the \HW\ program is described in
app.~\ref{app:reshuf}. The relationship between $\twototwo$ and $\twotothree$
kinematics discussed in the following subsections depends upon the particular
implementation of momentum reshuffling.

\subsection{Relating $\twototwo$ and $\twotothree$ kinematics}
The relationship between the variables of the $\twototwo$ hard 
process, and the $\twotothree$ variables which result after one parton 
emission is not simple, owing to momentum reshuffling. In particular,
since initial- and final-state showers are treated differently in
the reshuffling, it depends on whether the parton emission is from
an incoming or outgoing parton.

We consider the case of FCR with $\twototwo$ momenta
$\bp_1\bp_2\to \bk_1\bk_2$ and $\twotothree$ momenta 
$p_1p_2\to k_1k_2 k$. The way in which we relate the
kinematics of the $\twototwo$ and $\twotothree$ processes is by
solving for the $\twototwo$ invariants $\bs,\bt,\bu$ and momentum
fractions $\bx_{1,2}$ in terms of the $\twotothree$ invariants
$s, t_{1,2}, u_{1,2}$ and momentum fractions $x_{1,2}$.
In terms of the c.m.\ frame variables defined in eq.~(\ref{eq:2to2cm}),
this amounts to finding $\bx_{1,2}$, $\bE$, and $\bkL$.

It should be noted that the kinematics used in parton shower Monte
Carlos generally involve cutoffs that operate like effective light quark
and gluon masses. We can ignore these cutoffs in computing the
Monte Carlo subtraction terms since they only give rise to
power-suppressed corrections, comparable to hadronization effects,
in physical distributions. This point was discussed in
\FW, sect.~B.3. 

\subsubsection{Final-state emission}\label{sec:FSR}
Suppose a gluon is emitted from the heavy quark.  
The final state is then formed by the heavy quark jet (i.e. the
heavy quark plus the emitted gluon) and the heavy antiquark.
The three-momenta of the antiquark and of the heavy quark jet
are rescaled by a common factor in order to restore
energy conservation, according to the prescription described in
appendix \ref{app:reshuf}. The heavy antiquark has momentum 
(in the  hard process c.m.\ frame)
\beq
k_2 = ( \sqrt{m^2+\alpha^2\bk^2}, -\alpha\bkT, 0, -\alpha\bkL)
\eeq
where $\bk^2 = \bkT^2 +\bkL^2 =\bE^2 - m^2$ and $\alpha$
is the rescaling factor, given by the energy conservation constraint,
\beq
2\bE = \sqrt{m^2+(\alpha\bk)^2} + \sqrt{(k+k_1)^2+(\alpha\bk)^2}\;. 
\eeq
Hence for final-state emission from the heavy quark $Q$ we find
\beq
\bE = \half\sqrt{s}\;,\;\;\;
\bkL = \bE\left(\frac{t_2-u_1}{s-w_1}\right)\frac{\bb}{\beta_2}\,,
\eeq
where $\beta_2$ is the velocity of the heavy antiquark in the heavy
quark-antiquark c.m.\ frame,
\beq\label{eq:beta2}
\beta_2 = \sqrt{1-4sm^2/(s-w_1)^2}\;.
\eeq

For emission from the heavy antiquark, the labels 1 and 2 are
interchanged, so the relation between the $\twototwo$ and $\twotothree$ 
invariants depends on which outgoing parton emits the gluon. We label 
the $\twototwo$ invariants as $\bt_Q$, $\bt_\bQ$, etc., according to 
which parton is the emitter. Then we have explicitly
\beqn\label{eq:fin_inv}
\bs_Q &=& \bs_\bQ = s\nonumber\\
\bt_Q &=& -\half s\left[1-\left(\frac{t_2-u_1}{s-w_1}\right)
\frac{\bb}{\beta_2}\right]\nonumber\\
\bt_\bQ &=& -\half s\left[1-\left(\frac{t_1-u_2}{s-w_2}\right)
\frac{\bb}{\beta_1}\right]\nonumber\\
\bu_Q &=& -s-\bt_Q\,,\;\;\;\;\;\;
\bu_\bQ = -s-\bt_\bQ\,,
\eeqn
where
\beq
\beta_1 = \sqrt{1-4sm^2/(s-w_2)^2}\;.
\eeq
Note that in the soft limit ($w_{1,2}\to 0$, $t_{1,2}\to\bt$, $u_{1,2}\to\bu$)
we have $\bt_{Q,\bQ}\to\bt$ and $\bu_{Q,\bQ}\to\bu$ as expected.

The incoming parton momenta are not affected, so
$\bp_1 = p_1$, $\bp_2 = p_2$ and $\bs_{Q,\bQ}=s$.
Thus the incoming momentum fractions are unchanged
by final-state emission:
\beq\label{eq:fin_x12}
\bx_{1f}=x_1\;,\;\;\;\bx_{2f}=x_2\;.
\eeq

The formulae presented here can also be used for the branchings of massless 
partons; one simply lets $\bb\to 1$, $\beta_1\to 1$, $\beta_2\to 1$. In 
such cases, collinear limits must also be considered. When a gluon
is emitted collinearly by the heavy quark ($t_2\to\bt$, $u_1\to\bu$,
$t_1+v_1\to\bt$, $u_2+v_2\to\bu$) we have $\bt_Q\to\bt$, $\bu_Q\to\bu$.
For a collinear emission by the heavy antiquark ($t_1\to\bt$, $u_2\to\bu$,
$t_2+v_2\to\bt$, $u_1+v_1\to\bu$) we have $\bt_\bQ\to\bt$, $\bu_\bQ\to\bu$.

\subsubsection{Initial-state emission}\label{sec:ISR}

In the case of emission from the incoming partons, the invariant mass of
the heavy quark pair is not changed by momentum reshuffling, so
\beq\label{eq:ini_bs}
\bs = (k_1+k_2)^2 = s+v_1+v_2\;.
\eeq
However, the pair receives longitudinal and transverse boosts, as
described in app.~\ref{app:reshuf}. To relate the other invariants,
we consider the heavy quark momentum difference.
In the $\twototwo$ c.m.\ frame, it is
\beq
\bk_1-\bk_2 = 2(0,\bkT,0,\bkL)\;.
\eeq
Since this has no energy component, the effect of the longitudinal boost
is simply to rescale the longitudinal component by the boost factor
\beq
\gamma_L = \sqrt{1+\frac{(k_1+k_2)_L^2}{(k_1+k_2)^2}}\;.
\eeq
The transverse boost does not change the longitudinal component, so
\beq
(k_1-k_2)_L =  \gamma_L (\bk_1-\bk_2)_L =2\gamma_L \bkL\;.
\eeq
We can extract the longitudinal components using the combination
\beq
\frac{x_2p_1-x_1p_2}{\sqrt{x_1x_2s}} = (0,0,0,1)\;.
\eeq
Hence
\beq
\bkL = -\frac{(x_2p_1-x_1p_2)\cdot(k_1-k_2)}{2\gamma_L\sqrt{x_1x_2s}}\,,
\eeq
where
\beq
\gamma_L = \sqrt{1+\frac{[(x_2p_1-x_1p_2)\cdot(k_1+k_2)]^2}
{x_1x_2s(k_1+k_2)^2}}\;.
\eeq
Expressing everything in terms of invariants, we therefore have for
initial-state emission
\beq
\bE = \half\sqrt{s+v_1+v_2}\;,\;\;\;
\bkL = \bE\,\frac{x_2(t_1-u_1)+x_1(t_2-u_2)}
{2s\sqrt{x_+^2 -x_1x_2v_1v_2/s^2}}\,,
\eeq
where for future reference we have defined
\beq\label{eq:xpm}
x_{\pm} = \frac 1 2 \left(\frac{s+v_2}{s}x_1 \pm\frac{s+v_1}{s}x_2\right)\;.
\eeq
Note that in this case the result is independent of which incoming
parton emits.  However, for consistency we label $\bt=\bt_+$ or $\bt_-$
according to whether parton 1 or 2 is emitting.  Thus we have explicitly
\beqn\label{eq:ini_inv}
\bs_\pm &=& s+v_1+v_2\nonumber\\
\bt_\pm &=& -\half (s+v_1+v_2)\left[1-\frac{x_2(t_1-u_1)+x_1(t_2-u_2)}
{2s\sqrt{x_+^2 -x_1x_2v_1v_2/s^2}}\right]\nonumber\\
\bu_\pm &=& -\bs_\pm-\bt_\pm\;.
\eeqn
Again in the soft limit ($v_{1,2}\to 0$, $t_{1,2}\to\bt$, $u_{1,2}\to\bu$)
we have $\bt_\pm\to\bt$ and $\bu_\pm\to\bu$ as expected. In the case of
collinear emission from leg 1 ($v_1\to 0$, $t_2\to\bt$, $u_2\to\bu$,
$t_1+\omega_1\to\bt$, $u_1+\omega_2\to\bu$) we have 
$\bt_\pm\to\bt$, $\bu_\pm\to\bu$.
For a collinear emission from leg 2 ($v_1\to 0$, $t_1\to\bt$, $u_1\to\bu$,
$t_2+\omega_2\to\bt$, $u_2+\omega_1\to\bu$) we also have 
$\bt_\pm\to\bt$, $\bu_\pm\to\bu$.

To relate the incoming momentum fractions, we note that in this case
eq.~(\ref{eq:ini_bs}) applies, and therefore instead of eq.~(\ref{eq:fin_x12})
we have
\beq\label{eq:ini_x1x2}
\bx_{1i}\bx_{2i}=\frac{s+v_1+v_2}{s}x_1x_2\;.
\eeq
The values of $\bx_{1i}$ and $\bx_{2i}$ separately depend on the momentum
reshuffling scheme, as explained in app.~\ref{app:reshuf}:
\begin{itemize}
\item In the $p$-scheme the longitudinal momentum
of the heavy-quark pair is preserved, which means that
\beq
\bx_{1i} -\bx_{2i} = \frac{s+v_2}{s}x_1 - \frac{s+v_1}{s}x_2\,,
\eeq
and hence
\beq\label{eq:ini_x12_p}
\bx_{1i} = x_- +\sqrt{x_+^2 -x_1x_2v_1v_2/s^2}\,,\;\;\;
\bx_{2i} = \bx_{1i} - 2x_- \,,
\eeq
where $x_\pm$ are given by eq.~(\ref{eq:xpm}). Note that 
eq.~(\ref{eq:ini_x12_p}) coincides with \FWeq{A.47}; this is to be
expected, since the details of the hard process are irrelevant in 
the determination of the $x$'s.

\item In the $y$-scheme the rapidity of the heavy-quark pair
is preserved, which implies
\beq
\frac{\bx_{1i}}{\bx_{2i}} = \frac{x_1(s+v_2)}{x_2(s+v_1)}\,,
\eeq
and hence
\beq\label{eq:ini_x12_y}
\bx_{1i}=x_1\sqrt{\frac{(s+v_1+v_2)(s+v_2)}{s(s+v_1)}}\;,\;\;\;
\bx_{2i}=x_2\sqrt{\frac{(s+v_1+v_2)(s+v_1)}{s(s+v_2)}}\;,
\eeq
which corresponds to \FWeq{A.42}.
\end{itemize}

\subsection{Relating HERWIG variables to invariants}
When relating the \HW\ showering variables to the kinematics of the
$\twotothree$ hard process, we must again take careful account of 
which quantities are preserved under momentum reshuffling.

\subsubsection{Final-state emission}\label{sec:HW_fin}
For emission of a gluon from the outgoing heavy quark in the
FCR processes, the \HW\ variables
are the angular variable $\xi = k\cdot k_1/k^0 k_1^0$ and the energy
fraction $z=k_1^0/E_0=1-k^0/E_0$, all energies being evaluated in the
showering frame where $E_0^2=-\bt/2$ or $-\bu/2$ as discussed in
app.~\ref{app:init}.  The invariant quantities are the jet virtuality
\beq
(k_1+k)^2-m^2= w_1 =2z(1-z)\xi E_0^2\;,
\label{timebrinvm}
\eeq
and the ``+'' momentum fraction of the gluon with respect to the jet axis,
\beq
\zeta_1\equiv\frac{k\cdot n_2}{(k_1+k)\cdot n_2}
= (1-z)\frac{1+(1-z\xi)/\tb_1}{1+\tb_1}\,,
\label{timebrplus}
\eeq
where $\tb_1$ is the heavy-quark jet velocity in the showering frame
\beq
\tb_1 =\sqrt{1-(w_1+m^2)/E_0^2}\,,
\eeq
and $n_2$ is a lightlike vector along the direction of the heavy antiquark
in the heavy quark-antiquark c.m.\ frame:
\beq\label{eq:ndef}
n_2 = k_2 - \frac{s-w_1}{2s}(1-\beta_2)(p_1+p_2)\,,
\eeq
where $\beta_2$ is given by eq.~(\ref{eq:beta2}).
Inserting eq.~(\ref{eq:ndef}) into eq.~(\ref{timebrplus}), we find
\beq
\zeta_1 = \frac{(s+w_1)w_2+(s-w_1)[(w_1+w_2)\beta_2-w_1]}
{(s-w_1)\beta_2[(s+w_1)+(s-w_1)\beta_2]}\;.
\label{timeplinv}
\eeq
Solving eqs.~(\ref{timebrinvm}), (\ref{timebrplus}) and (\ref{timeplinv})
for $z$ and $\xi$ with $E_0^2=-\bt_Q/2$ as given in eq.~(\ref{eq:fin_inv}),
we obtain $z^{(t)}_Q$ and $\xi^{(t)}_Q$, corresponding to emission from
the heavy quark with the $t$-flow colour structure as defined in
app.~\ref{app:col}.
Similarly, solving with $E_0^2=-\bu_Q/2$, as appropriate to the $u$-flow,
gives  $z^{(u)}_Q$ and $\xi^{(u)}_Q$.  We find that
\beqn\label{eq:z_fin}
z^{(l)}_Q &=& 1-\tb_1\zeta_1 -\frac{w_1}{(1+\tb_1)|\bl_Q|}\,,
\\
\label{eq:xi_fin}
\xi^{(l)}_Q &=& \frac{w_1}{z^{(l)}_Q(1-z^{(l)}_Q)|\bl_Q|}\,,
\eeqn
where $\bl_Q = \bt_Q, \bu_Q$ for $l=t,u$. Interchanging the
labels 1 and 2 and using $\bl_\bQ$ instead of $\bl_Q$ gives  
$z^{(t,u)}_\bQ$ and $\xi^{(t,u)}_\bQ$, corresponding to emission 
from the heavy antiquark.

Note that the transverse momentum of the emitted gluon relative to the
jet axis is $\kt$ where
\beq
\kt^2 =\zeta_1 \left[(1-\zeta_1)\,w_1 -\zeta_1\, m^2\right]\,,
\label{eq:ktsq}
\eeq
and thus $\kt\simeq \sqrt 2z(1-z)Q$ at small values of $\xi$ and $m^2/E_0^2$,
where $Q=E_0\sqrt\xi$ is the \HW\ evolution variable.

\subsubsection{Initial-state emission}\label{sec:HW_ini}
In the case of an initial-state jet, the jet axis coincides with the
beam axis and so the jet invariants are simpler than in the final-state
case.  For emission from parton 1, the jet virtuality is
\beq
(p_1-k)^2 = v_1 = -2\frac{1-z}{z^2}\xi\,E_0^2\,,
\label{spacebrinvm}
\eeq
and the ``+'' component of the gluon momentum is\footnote{Note that
eqs.~(\ref{spacebrinvm}) and (\ref{spaceplinv}) are equivalent
to \FWeq{A.67} and \FWeq{A.68}, if we set $E_0^2=M_{WW}^2/2$.}
\beq
\frac{k\cdot p_2}{p_1\cdot p_2} = -\frac{v_2}{s} =\half (1-z)(2-\xi)\;.
\label{spaceplinv}
\eeq
Solving eqs.~(\ref{spacebrinvm}) and (\ref{spaceplinv}) for $z$ and $\xi$
with $E_0^2=-\bt_+/2$ or $-\bu_+/2$ as given in eq.~(\ref{eq:ini_inv}),
we obtain $z^{(t,u)}_+$ and $\xi^{(t,u)}_+$, corresponding to emission
from incoming parton 1 with these initial conditions.  In the case of
emission from an initial-state gluon, we also have the possibility
that $E_0^2=\bs_+/2$ (see app.~\ref{app:col}), and we denote the
corresponding variables by $z^{(s)}_+$ and $\xi^{(s)}_+$. Then we can 
write the solutions explicitly as
\beqn\label{eq:z_ini}
z^{(l)}_+ &=& \frac{|\bl_+|}{v_1}\left[1-\sqrt{1-2\frac{v_1}{|\bl_+|}
\left(1+\frac{v_2}{s}\right)}\right] \\ \label{eq:xi_ini}
\xi^{(l)}_+ &=& 2\left[1+\frac{v_2}{s(1-z^{(l)}_+)}\right]
\eeqn
where $\bl_+ = \bs_+, \bt_+, \bu_+$ for $l=s,t,u$.

For emission from incoming parton 2, the variables $v_1$ and $v_2$ are
interchanged. We denote the corresponding solutions by $z^{(l)}_-$ and
$\xi^{(l)}_-$.

\subsection{Dead regions}
The parton shower initial conditions imply that gluon radiation is
confined to angular regions (cones) specified by the colour flow.
This represents an approximate treatment of the destructive interference
due to colour coherence.
The dead regions outside the cones can be found from the above mapping
of the \HW\ shower variables onto invariants.

\subsubsection{Final-state emission}
For final-state jets the kinematic region available for gluon emission is
\beq
1-\sqrt{1-m^2/E_0^2} < \xi < 1\;,\;\;\;
\frac{m}{E_0\sqrt{\xi(2-\xi)}} < z < 1\;.
\eeq
However, the radiation in the forward direction is suppressed dynamically
for $\xi<m^2/E_0^2$ and therefore in \HW\ the region used (neglecting the
gluon effective mass cutoff) is
\beq\label{eq:region_fin}
m^2/E_0^2 < \xi < 1\;,\;\;\;
\frac{m}{E_0\sqrt\xi} < z < 1\;.
\eeq
The scale $E_0$ is given by
$E_0^2=-\bt/2 =\bs(1-\bb\cos\bh)/4$ or $-\bu/2$ or
$\bs/2$, depending on the colour flow (see app.~\ref{app:init}).
Thus all cases can be represented by the situation for $E_0^2=-\bt/2$,
with $\cos\bh\to-\cos\bh$ when $E_0^2=-\bu/2$ and
$\cos\bh=-1/\bb\simeq -1$ when $E_0^2=\bs/2$.

It is convenient to express the jet regions in terms of the
Dalitz plot variables of the $Q\bQ g$ final state,
\beqn
x_Q &=& 2k_1\cdot(p_1+p_2)/s = -(t_1+u_2)/s = 1-w_2/s\nonumber\\
x_{\bQ} &=& 2k_2\cdot(p_1+p_2)/s = -(u_1+t_2)/s = 1-w_1/s\\
x_g &=& 2k\cdot(p_1+p_2)/s=-(v_1+v_2)/s = 2-x_Q-x_{\bQ}\;.\nonumber
\eeqn
Then for emission from the heavy quark we have
\beqn
w_1 &=& s(1-x_{\bQ})\,,
\\
\zeta_1 &=&
\frac{x_g+(x_g^2+x_{\bQ}^2-x_Q^2)/(2x_{\bQ}\beta_2)}
{2-x_{\bQ}(1-\beta_2)}\,,
\eeqn
where $w_1$ and $\zeta_1$ are related to the \HW\ variables
$\xi$ and $z$ by eqs.~(\ref{timebrinvm}) and (\ref{timebrplus}).
For emission from the heavy antiquark, the variables $x_Q$
and $x_{\bQ}$ are interchanged.

\begin{figure}[htb]
  \begin{center}
  \epsfig{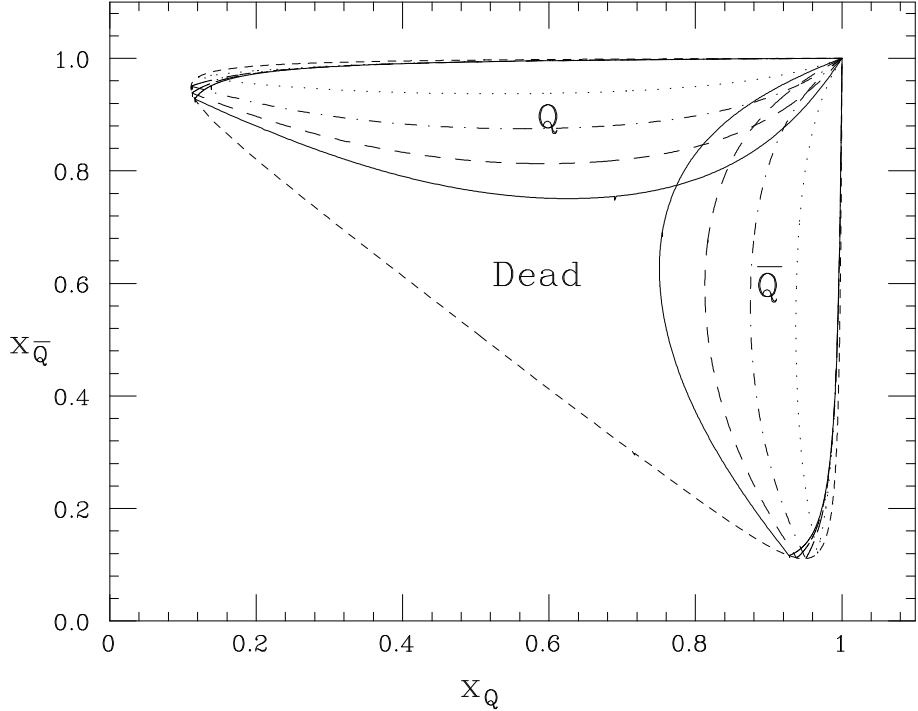}
  \caption{Dalitz plot and jet regions for final-state emission
  when $s=M_Z^2$, $m=5$ GeV,
  $\cos\bh=$0.5 (dotted), 0 (dot-dash), --0.5 (dashed)
  and --1 (solid).}\label{fig:fin}
  \end{center}
\end{figure}

The boundaries of the quark and antiquark jet regions of the
$Q\bQ g$ Dalitz plot are shown in fig.~\ref{fig:fin}.
Note that there is an overlap in the soft
region for $\cos\bh<0$.  The boundary of the physical
region is also shown (short-dashed). We see that there is a
dead region in which hard, non-collinear gluon
emission is missed by the \HW\ shower algorithm, and also narrow
near-collinear dead regions, where emission is forbidden
in order to simulate the dynamical suppression of collinear
emission from the heavy quarks, as discussed above.

\subsubsection{Initial-state emission}
In the case of an initial-state jet,
the allowed region is $\xi<z^2$ since now
$z=E_0/p_1^0$ \cite{Corcella:1999gs}. The conventional variables
are $x,y$ which give the heavy diquark mass $M_{Q\bQ}$
and the emission angle $\theta^*_g$ of the gluon in the partonic
c.m.\ frame.  For emission from incoming parton 1 we find:
\beq\label{eq:xydef}
x \equiv M^2_{Q\bQ}/s = 1+\frac{v_1+v_2}{s}\;,\;\;\;
y \equiv \cos\theta^*_g = \frac{v_2-v_1}{v_1+v_2}\;.
\eeq
For emission from parton 2, $v_1$ and $v_2$ are interchanged,
i.e.\ $y\to -y$.  Inserting $\xi=z^2$ in
eqs.~(\ref{spacebrinvm})--(\ref{spaceplinv}) then gives the  
boundaries of the jet regions in the $x,y$ plane.

\begin{figure}[htb]
  \begin{center}
  \epsfig{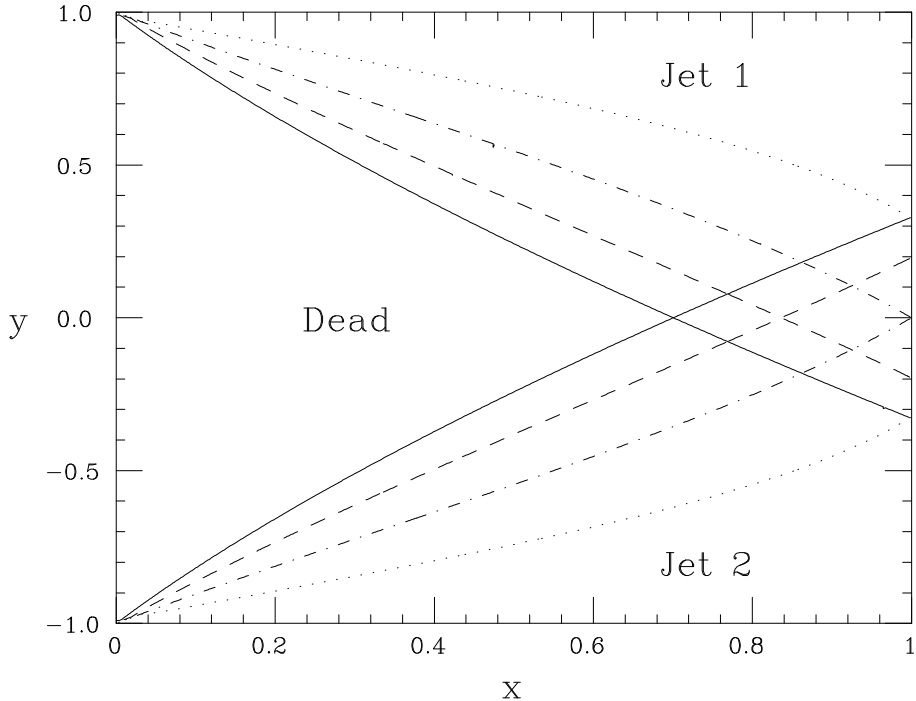}
  \caption{Phase space and jet regions for initial-state emission
  when $s=M_Z^2$, $m=5$ GeV,
  $\cos\bh=$0.5 (dotted), 0 (dot-dash), --0.5 (dashed)
  and --1 (solid).}\label{fig:ini}
  \end{center}
\end{figure}

The jet boundaries are shown in fig.~\ref{fig:ini}.
Note that there is again an overlap in the soft
region for $\cos\bh<0$.  There are no collinear dead regions,
because the incoming partons are treated as massless.

\section{MC cross sections expanded to NLO}\label{sec:sub}
In this section, we present the cross section that we would obtain by keeping
the first non-trivial order in the $\as$ expansion of the \HW\ Monte Carlo
result. This quantity, which we denote by $d\sigma\xMC$, is directly related
to the MC subtraction terms that enter the definition of MC@NLO.  We shall not
give here the rather technical details of the construction of the MC
subtraction terms, which we report in app.~\ref{sec:MCsubt}. We shall limit
ourselves to highlighting the main differences with respect to $\FW$, which
result from the more complicated singularity and colour structure of the heavy
flavour cross section.

The result we are seeking can be written as follows:
\beq
d\sigma\xMCB=\sum_{ab}\sum_{L}\sum_{l}d\sigma_{ab}^{(L,l)}\xMCB\,,
\label{MCatas}
\eeq
where the first sum in eq.~(\ref{MCatas}) runs over parton processes.
The index $L$ runs over the emitting legs and, consistently with
sect.~\ref{sec:FSR} and~\ref{sec:ISR}, it assumes the values
$+$, $-$, $Q$, and $\bQ$. The index $l$ runs over the colour structures,
and it can take the values $s$, $t$, and $u$ (see sect.~\ref{sec:HW_fin} 
and~\ref{sec:HW_ini}). Using the same formal expansion as in \FWeq{A.58},
we obtain
\beqn
d\sigma_{ab}^{(+,l)}\xMCB &=&
\frac{1}{z_+^{(l)}}
f_a^{\Hone}(\bx_{1i}/z_+^{(l)})f_b^{\Htwo}(\bx_{2i})\,
d\hat{\sigma}_{ab}^{(+,l)}\xMCB d\bx_{1i}\,d\bx_{2i}\,,
\label{eq:spl}
\\
d\sigma_{ab}^{(-,l)}\xMCB &=&
\frac{1}{z_-^{(l)}}
f_a^{\Hone}(\bx_{1i})f_b^{\Htwo}(\bx_{2i}/z_-^{(l)})\,
d\hat{\sigma}_{ab}^{(-,l)}\xMCB d\bx_{1i}\,d\bx_{2i}\,,
\label{eq:smn}
\\
d\sigma_{ab}^{(Q,l)}\xMCB &=&
f_a^{\Hone}(\bx_{1f})f_b^{\Htwo}(\bx_{2f})\,
d\hat{\sigma}_{ab}^{(Q,l)}\xMCB d\bx_{1f}\,d\bx_{2f}\,,
\label{eq:sQ}
\\
d\sigma_{ab}^{(\bQ,l)}\xMCB &=&
f_a^{\Hone}(\bx_{1f})f_b^{\Htwo}(\bx_{2f})\,
d\hat{\sigma}_{ab}^{(\bQ,l)}\xMCB d\bx_{1f}\,d\bx_{2f}\,.
\label{eq:sQb}
\eeqn
In the case of $l=s$, the sum of eqs.~(\ref{eq:spl}) and~(\ref{eq:smn})
coincides with \FWeq{A.58}. The cases $l=t,u$, and eqs.~(\ref{eq:sQ}) 
and~(\ref{eq:sQb}) were not considered in $\FW$, since they correspond
to colour flows not relevant to gauge boson pair production, and emissions
from strongly-interacting final-state partons. 

The short-distance cross sections $d\hat{\sigma}$ in 
eqs.~(\ref{eq:spl})--(\ref{eq:sQb}) 
have a form similar to \FWeq{A.63} and \FWeq{A.64}, namely a
factor depending on the \HW\ showering variables $\xi$ and $z$, times a 
Born cross section. As discussed above, $\xi$ and $z$, and the 
$\twototwo$ invariants $\bs,\bt$ and $\bu$ entering the Born cross sections, 
must be expressed in terms of the $\twotothree$ integration variables used in 
the NLO code. In the case of emissions from both the initial- and final-state
partons, we have four pairs of functional relations between $\xi$ and $z$ and
the NLO integration variables, corresponding to $z_+^{(l)}$, $z_-^{(l)}$,
$z_Q^{(l)}$ and $z_{\bQ}^{(l)}$ (and analogously for $\xi$), given by
eqs.~(\ref{eq:z_ini},\ref{eq:xi_ini}) and (\ref{eq:z_fin},\ref{eq:xi_fin}).
We have the following non-vanishing contributions:

\noindent
$\bullet$~$q\bar{q}$~initial state
\beqn
d\hat\sigma_{q\bar{q}}^{(+,t)}\xMCB &=& \frac{\as}{2\pi}\,
\frac{d\xi_+^{(t)}}{\xi_+^{(t)}}dz_+^{(t)}
P^{(0)}_{qq}(z_+^{(t)})\,\dsb_{q\bar{q}}
\stepf\left((z_+^{(t)})^2-\xi_+^{(t)}\right)
\label{sqqbarpone}
\\
d\hat\sigma_{q\bar{q}}^{(-,t)}\xMCB &=& d\hat\sigma_{q\bar{q}}^{(+,t)}\xMCB
\left(z_+^{(t)}\to z_-^{(t)},\xi_+^{(t)}\to \xi_-^{(t)}\right)
\\
d\hat\sigma_{q\bar{q}}^{(Q,t)}\xMCB &=& \frac{\as}{2\pi}\,
\frac{d\xi_Q^{(t)}}{\xi_Q^{(t)}}dz_Q^{(t)}
P^{(0)}_{qq}(z_Q^{(t)})\,\dsb_{q\bar{q}}
\stepf\left(1-\xi_Q^{(t)}\right)
\stepf\left((z_Q^{(t)})^2-\frac{2m^2}{|\bt_Q|\xi_Q^{(t)}}\right)
\\
d\hat\sigma_{q\bar{q}}^{(\bQ,t)}\xMCB &=& d\hat\sigma_{q\bar{q}}^{(Q,t)}\xMCB
\left(\bt_Q\to\bt_\bQ, z_Q^{(t)}\to z_{\bQ}^{(t)},
\xi_Q^{(t)}\to \xi_{\bQ}^{(t)}\right)
\eeqn

\noindent
$\bullet$~$qg$~initial state
\beqn
d\hat\sigma_{qg}^{(+,t)}\xMCB &=& \frac{\as}{4\pi}\,
\frac{d\xi_+^{(t)}}{\xi_+^{(t)}}dz_+^{(t)}
P^{(0)}_{gq}(z_+^{(t)})\,\dsb_{gg}^{(t)}
\stepf\left((z_+^{(t)})^2-\xi_+^{(t)}\right)
\\
d\hat\sigma_{qg}^{(+,u)}\xMCB &=& d\hat\sigma_{qg}^{(+,t)}\xMCB
\left(\dsb_{gg}^{(t)}\to\dsb_{gg}^{(u)},
z_+^{(t)}\to z_+^{(u)},\xi_+^{(t)}\to \xi_+^{(u)}\right)
\\
d\hat\sigma_{qg}^{(+,s)}\xMCB &=& \frac{\as}{4\pi}\,
\frac{d\xi_+^{(s)}}{\xi_+^{(s)}}dz_+^{(s)}
P^{(0)}_{gq}(z_+^{(s)})\,\dsb_{gg}
\stepf\left((z_+^{(s)})^2-\xi_+^{(s)}\right)
\\
d\hat\sigma_{qg}^{(-,t)}\xMCB &=& d\hat\sigma_{q\bar{q}}^{(-,t)}\xMCB
\left(P^{(0)}_{qq}\to P^{(0)}_{qg}\right)
\eeqn
where $\dsb_{gg}^{(t,u)}$ are the colour $t$- and $u$-flow
contributions defined in eq.~(\ref{eq:dsbtu});

\noindent
$\bullet$~$gg$~initial state
\beqn
d\hat\sigma_{gg}^{(+,l)}\xMCB &=& d\hat\sigma_{qg}^{(+,l)}\xMCB
\left(P^{(0)}_{gq}\to P^{(0)}_{gg}\right)
\\
d\hat\sigma_{gg}^{(-,l)}\xMCB &=& d\hat\sigma_{gg}^{(+,l)}\xMCB
\left(z_+^{(l)}\to z_-^{(l)},\xi_+^{(l)}\to \xi_-^{(l)}\right)
\\
d\hat\sigma_{gg}^{(Q,t)}\xMCB &=& \frac{\as}{2\pi}\,
\frac{d\xi_Q^{(t)}}{\xi_Q^{(t)}}dz_Q^{(t)}
P^{(0)}_{qq}(z_Q^{(t)})\,\dsb_{gg}^{(t)}\stepf\left(1-\xi_Q^{(t)}\right)
\stepf\left((z_Q^{(t)})^2-\frac{2m^2}{|\bt_Q|\xi_Q^{(t)}}\right)
\\
d\hat\sigma_{gg}^{(Q,u)}\xMCB &=& d\hat\sigma_{gg}^{(Q,t)}\xMCB
\left(\dsb_{gg}^{(t)}\to\dsb_{gg}^{(u)},
z_Q^{(t)}\to z_Q^{(u)},\xi_Q^{(t)}\to \xi_Q^{(u)}\right)
\label{sggQtwo}
\\
d\hat\sigma_{gg}^{(\bQ,l)}\xMCB &=& d\hat\sigma_{gg}^{(Q,l)}\xMCB
\left(\bt_Q\to\bt_\bQ, \bu_Q\to\bu_\bQ, z_Q^{(l)}\to z_{\bQ}^{(l)},
\xi_Q^{(l)}\to \xi_{\bQ}^{(l)}\right)
\label{sggQbl}
\eeqn
Note that in each equation the Born cross section $\dsb_{ab}$ or
$\dsb_{ab}^{(l)}$ has to be computed using the relevant definitions of 
$\bs$, $\bt$ and $\bu$; more details are given in app.~\ref{sec:MCsubt}.
The argument of $\as$ used in the computation of these cross sections is 
the same one as that used in the computation of the NLO short-distance 
cross sections; as discussed in sect.~\ref{sec:rev}, the generation of the 
hard process in MC@NLO has nothing to do with the analogous generation 
occurring in the MC. On the other hand, the factor of $\as$ which
explicitly appears in eqs.~(\ref{sqqbarpone})--(\ref{sggQbl}) is due 
to parton showering, and its argument should in principle be chosen
according to eq.~(\ref{eq:emsca}). However, the choice of this
argument is strictly speaking a matter beyond NLO. This is discussed 
more fully in \FW, sect.~B.3. As done there, for simplicity we choose 
the same scale as that used for NLO terms.

The analogous results for $\bar{q}q$, $\bar{q}g$, $gq$, and $g\bar{q}$
can easily be obtained from the equations above. The situation is
summarized in table~\ref{tab:proc}.
\begin{table}[htb]
\renewcommand{\arraystretch}{1.2}
\begin{center}
\begin{tabular}{|c||c|c|c|}
\hline
$ab$ & $q\bar{q}\to Q\bQ$ & $\bar{q}q\to Q\bQ$ & $gg\to Q\bQ$ \\\hline\hline
$q\bar{q}$ & $\pm(t)$, $Q(t)$, $\bQ(t)$ & &\\\hline
$\bar{q}q$ & & $\pm(u)$, $Q(u)$, $\bQ(u)$ & \\\hline
$qg$       & $-(t)$ & & $+(s,t,u)$ \\\hline
$\bar{q}g$ & & $-(u)$ & $+(s,t,u)$ \\\hline
$gq$       & & $+(u)$ & $-(s,t,u)$ \\\hline
$g\bar{q}$ & $+(t)$ & & $-(s,t,u)$ \\\hline
$gg$       & & & $\pm(s,t,u)$, $Q(t,u)$, $\bQ(t,u)$ \\\hline
\end{tabular}
\end{center}
\caption{\label{tab:proc}
Short-distance contributions to MC subtraction terms, from Born processes
$q\bar{q}\to Q\bQ$, $\bar{q}q\to Q\bQ$, and $gg\to Q\bQ$. Each entry
lists the emitting legs (+, --, $Q$, $\bQ$); for each emitting leg,
we report in parentheses the different contributions $l$,
according to the possible colour flows (corresponding to $E_0^2=|\bl|/2$).
}
\end{table}

These results allow one to obtain the MC subtraction terms
$d\bSigma_{ab}\xMC$ entering the MC@NLO definition, eq.~(\ref{eq:rMCatNLO}),
as explained in app.~\ref{sec:MCsubt}.

\section{Implementation of MC@NLO}\label{sec:imp}

The practical implementation of MC@NLO for heavy flavour production
proceeds in a similar way to that for gauge boson pair production,
described in sect.~4.5 of \FW. The integrals necessary to determine the 
required numbers of $\clH$ ($\twotothree$ parton) and $\clS$ ($\twototwo$ 
parton) configurations are computed according to \FWeq{4.23} and \FWeq{4.24}, 
together with the corresponding equations in which the integrands are taken
in absolute value. Configurations are generated according to the
MC-subtracted weight distributions in eq.~(\ref{eq:rMCatNLO}), and
then unweighted with weight $\pm 1$ according to the sign of the
weight, using the {\small SPRING-BASES} package~\cite{Kawabata:1995th}.
Weighted events could also be generated, but this option is not
implemented at the moment.

The selected configurations and their weights are written on a
file which is input to the \HW\ event generator. \HW\ then performs
the parton showering and hadronization as explained in app.~\ref{sec:showers}
and \ref{sec:hadrize} respectively.  The generated events with weight +1 are
to be treated as real events for the purposes of histogramming
and/or detector simulation, whereas those with weight --1 (a fraction
of the order of 10\% for top and 20\% for bottom) are ``counter-events'', 
which must be treated as real events in detector simulation but have to be
subtracted from histogram bins to which they contribute.

The only significant differences from the implementation of MC@NLO
described in \FW\ are that a colour flow must be assigned to
each parton configuration and that this flow, together with 
the event weight and the parton identities and momenta, are
handled by the ``Les Houches'' generic user process 
interface~\cite{Boos:2001cv}. Details of these two new features 
are given in the following subsections.

\subsection{Colour flow assignment}\label{sec:flow}
The assignment of colour flows for $\clS$ 
configurations is done according to the \HW\ prescription explained in
app.~\ref{app:col}: where the colour flow is ambiguous,
it is assigned according to the $N\to\infty$ limit.
The same prescription is extended to $\clH$ 
configurations as follows.

\begin{figure}
\begin{center}
\epsfig{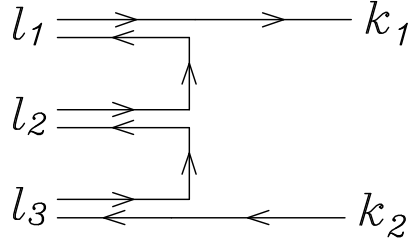}
\end{center}
\caption{\label{fig:ggcolflow}
Basic colour flow configuration involving
three gluons and a heavy quark-antiquark pair.}
\end{figure}

Consider first the process $gg\to Q\bar{Q} g$.
In the large-$N$ limit, the basic planar colour flow amplitude
involving three gluons and a heavy quark-antiquark pair is depicted in
fig.~\ref{fig:ggcolflow}.
Double-directed lines are gluons, and single-directed lines are quarks.
The momenta $l_1$, $l_2$ and $l_3$ are all defined to be outgoing.
In the notation of sect.~\ref{sec:2to3}, the six independent colour-ordered
amplitudes for the $gg\to Q\bar{Q} g$ process are obtained by
replacing $l_1,\,l_2,\,l_3 \longrightarrow -p_1,\,-p_2,\,k$ in
all possible ways.

\begin{figure}
\begin{center}
\epsfig{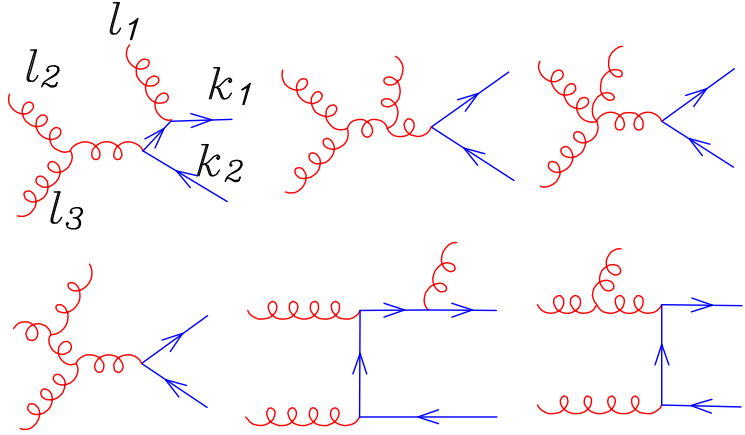}\hskip 0.05\textwidth
\end{center}
\caption{\label{fig:ggplanar}
Planar graphs contributing to the colour flow configuration
of fig.~\ref{fig:ggcolflow}.}
\end{figure}

The Feynman graphs
that allow for the colour pattern of fig.~\ref{fig:ggcolflow} are
shown in fig.~\ref{fig:ggplanar}.
The amplitudes are easily computed from the large-$N$ limit Feynman rules,
which amount to the following graphical prescriptions for the vertices:
\beqn
\mbox{$q\bar q$ gluon:} &&
\frac{1}{\sqrt{2}}\;\;
\raisebox{-0.2\height}{\includegraphics[scale=0.45,clip=false]{ffg.eps}}
\;\;
i\gamma^\mu \\
\mbox{three gluons:} &&
\frac{-i}{\sqrt{2}}\;\;
\raisebox{-0.4\height}{\includegraphics[scale=0.35,clip=false]{gluon3.eps}}
\;\;
\left[g^{\mu_1 \mu_2}(p^{\mu_3}_1-p^{\mu_3}_2)+g^{\mu_2 \mu_3}
(p^{\mu_1}_2-p^{\mu_1}_3)
+g^{\mu_3 \mu_1}(p^{\mu_2}_3-p^{\mu_2}_1)\right]
\nonumber \\
\mbox{four gluons:} &&
\raisebox{-0.4\height}{\includegraphics[scale=0.35,clip=false]{gluon4.eps}}
\;\;
\frac{-i}{2}
\left(g^{\mu_1 \mu_2}g^{\mu_3 \mu_4}+g^{\mu_2 \mu_3}g^{\mu_4 \mu_1}
-2g^{\mu_1 \mu_3}g^{\mu_2 \mu_4}\right)\;,
\eeqn
where all momenta and indices have to be assigned counterclockwise,
and momenta are all incoming.

We define
\beq
f^{(g)}(k_1 ,l_1, l_2, l_3, k_2)
=\sum_{\rm spin, col.} \left|\sum_i A_i\right|^2\,,
\eeq
the square of the sum of the
amplitudes in fig.~\ref{fig:ggplanar} in the large-$N$ limit,
summed over spin and
colours. The ordering of the momenta in the arguments of $f^{(g)}$
corresponds to the ordering of the colour connection, starting
from the outgoing heavy quark.

\begin{figure}
\begin{center}
\epsfig{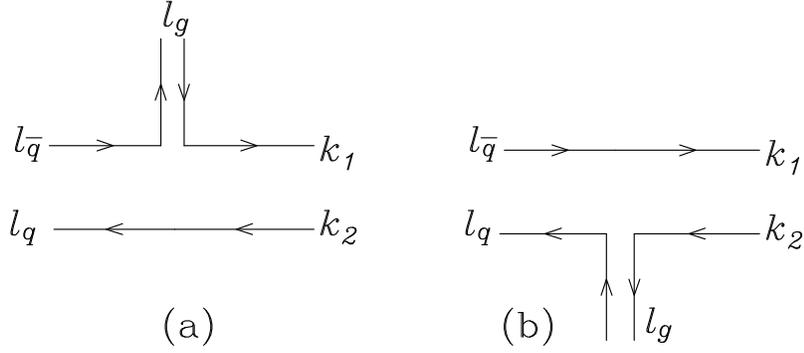}
\end{center}
\caption{\label{fig:qqcolflow}
Basic colour flow configuration involving
a gluon, a heavy quark-antiquark pair and a light quark-antiquark pair.}
\end{figure}

There are two basic colour orderings for processes involving a
gluon, a light quark-antiquark pair, and a heavy quark-antiquark pair.
They are depicted in fig.~\ref{fig:qqcolflow}.
Here we define
two squared amplitudes
\beq\label{qgplanar}
f^{Qg\bar{q}q\bar{Q}}(k_1,k_2,l_g,l_q,l_{\bar{q}}),\quad
f^{Q\bar{q}qg\bar{Q}}(k_1,k_2,l_g,l_q,l_{\bar{q}}),
\eeq
corresponding to colour flows (a) and (b) of fig.~\ref{fig:qqcolflow},
respectively.
The two squared amplitudes are related by charge conjugation
\beq
f^{Q\bar{q}qg\bar{Q}}(k_1,k_2,l_g,l_q,l_{\bar{q}})=
f^{Qg\bar{q}q\bar{Q}}(k_2,k_1,l_g,l_{\bar{q}},l_q)\;.
\eeq
Notice that in our notation we call $q$ and $\bar{q}$ the light flavour
lines corresponding to an {\em outgoing} quark or antiquark respectively.
The corresponding colour flow squared amplitudes for $q\bar{q}\to Q\bar{Q}g$
are obtained from the amplitudes of eq.~(\ref{qgplanar}) by the replacement
$l_{\bar{q}}=-p_1$, $l_q=-p_2$ and $l_g=k$.
The squared amplitudes for the process $q g \to Q\bar{Q} q$
are instead obtained with the replacement
$l_{\bar{q}}=-p_1$, $l_g=-p_2$, $l_q=k$.

The correctness of our calculation of the colour flow amplitudes was
checked by comparing the sum of all colour squared amplitudes with the
known full squared amplitude, for very large values of $N$.

As an example, consider the assignment of colour flow for the
process  $q g \to Q\bar{Q} q$.  The configuration of momenta
$p_1,p_2,k_1,k_2,k$ is chosen according to the full ($N=3$)
expressions with modified subtraction as explained in sect.~\ref{sec:rev}.
Then the squared amplitudes for the large-$N$ colour flows in
fig.~\ref{fig:qqcolflow} are computed:
\beq
f_a = f^{Qg\bar{q}q\bar{Q}}(k_1,k_2,-p_2,k,-p_1)\,,\quad
f_b = f^{Qg\bar{q}q\bar{Q}}(k_2,k_1,-p_2,-p_1,k)\;.
\eeq
The colour flows (a) and (b) are then assigned with probabilities
\beq
P_a = \frac{f_a}{f_a+f_b} \,,\quad P_b = \frac{f_b}{f_a+f_b}\;.
\eeq

\subsection{Interface to HERWIG MC}\label{sec:interface}

The colour flow selected as described above is encoded in a single
integer {\tt IC} which is written on a file together with the event
weight and the parton momenta and identities.  The possible
values of the colour flow code {\tt IC} and their meanings are
given in app.~\ref{sec:codes}.

The file of parton configurations is read by \HW\ as input to
the Les Houches generic user process interface~\cite{Boos:2001cv}.
A negative value of the process code {\tt IPROC} signals to \HW\ that,
instead of generating a partonic hard subprocess as outlined in
app.~\ref{sec:hardsub}, it should load and use subprocess information
in the Les Houches common blocks {\tt HEPRUP} and {\tt HEPEUP}.

First a file header is read by the interface subroutine {\tt UPINIT},
which checks that the file has been generated with parameter
values consistent with those set in this \HW\ run and initialises
the run common block {\tt HEPRUP}. The parton configurations are
then read sequentially by the interface subroutine {\tt UPEVNT}, which loads
the event common block {\tt HEPEUP} with all information necessary
to generate an event, including the interpretation of the colour flow code
{\tt IC} as outlined above.  

\section{Results on top quark production}\label{sec:top}
In this section we present results for top quark production,
obtained with the MC@NLO implementation\footnote{A public 
version of the program is available on the MC@NLO web page 
{\tt http://www.hep.phy.cam.ac.uk/theory/webber/MCatNLO/}.} described above. 
We do not attempt here to give a phenomenological description that corresponds 
to a specific experimental configuration; rather, we wish to show the
differences between MC@NLO, standard \HW\ MC, and NLO
results. For this reason, we present distributions obtained
either by integrating over the whole phase space, or else
with cuts on the heavy quark variables, rather than on
those of their decay products. We first show results for the LHC,
i.e.\ for $pp$ collisions at $\sqrt{S}=14$~TeV.
We set the top mass to $m=173$~GeV, and the renormalization and factorization
scales equal to the top transverse mass, $\sqrt{m^2+\pt^2}$.
We adopt the low-$\as$ set of MRST99 parton densities~\cite{Martin:1999ww},
since this set has a $\Lambda_{\sss QCD}$ value which is rather close
to that used as \HW\ default. This value of $\Lambda_{\sss QCD}$
($\Lambda_5^{\sss \overline{MS}}=164$~MeV) is used for all our
MC@NLO, MC, and NLO runs. In the case of standard \HW\ MC 
runs, we give each event (we generate unweighted events) 
a weight equal to $\sigma_{tot}/N_{tot}$, with $N_{tot}$ the total number of
events generated. For fully exclusive distributions with no cuts
applied, this is equivalent to normalizing \HW\ results to the total
NLO rate $\sigma_{tot}$ (with our choice of parameters, we obtain
$\sigma_{tot}=6.668$~pb at the Tevatron, and $\sigma_{tot}=736.6$~pb at 
the LHC). All the MC@NLO and MC results (but not, of course, the NLO ones) 
include the hadronization of the partons in the final state.

\begin{figure}[htb]
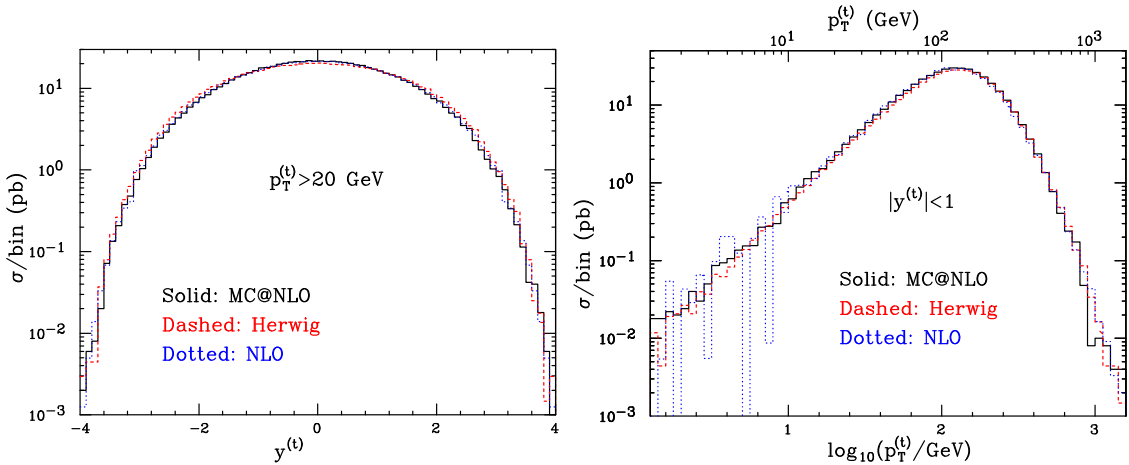

  \begin{center}
    \epsfig{figure=top_lhc_yt_cut.eps,width=0.49\textwidth}
    \epsfig{figure=top_lhc_ptt_cut.eps,width=0.49\textwidth}
\caption{\label{fig:top_lhc_yandpt} 
  MC@NLO (solid), \HW\ (dashed) and NLO (dotted) results
  for the rapidity (left panel) and the transverse momentum (right
  panel) of the top quark at the LHC, with acceptance cuts. 
  \HW\ results have been normalized as explained in the text.
}
  \end{center}
\end{figure}
We present in fig.~\ref{fig:top_lhc_yandpt} the rapidity (left panel) and the 
transverse momentum (right panel) distributions of the top quark. The rapidity
(transverse momentum) result has been obtained after applying the cut 
$\ptt>20$~GeV ($\abs{\yt}<1$). The solid, dashed and dotted histograms show 
the MC@NLO, MC and NLO results, respectively. These distributions
are fairly inclusive, and we expect them to be reliably predicted by NLO
QCD in a wide range. From the figure, we see that the NLO and MC@NLO
results are extremely close to each other in the whole ranges considered.
\HW\ results are also fairly close to the NLO and MC@NLO ones,
giving only a slightly broader rapidity distribution. The same pattern can 
be found for the invariant mass of the $t\bar t$ pair. We conclude that, for
these kinds of observables, NNLO effects are very small, and NLO, MC 
and MC@NLO are almost equivalent. This also implies that any possible 
reshuffling of the momenta, due to the hadronization phase in MC@NLO, 
has negligible impact on the $t$ and $\bar t$. The right panel of
fig.~\ref{fig:top_lhc_yandpt} also clearly shows a characteristic feature
of the comparisons between MC@NLO and NLO results, namely that the former
are numerically more stable than the latter. This feature, which is even
more evident in the case of $b$ production, is due to the fact that, 
as eq.~(\ref{eq:rMCatNLO}) documents, in MC@NLO all cancellations between
large numbers occur at the level of short-distance cross sections,
rather than in histograms as in the case of NLO computations.

\begin{figure}[htb]
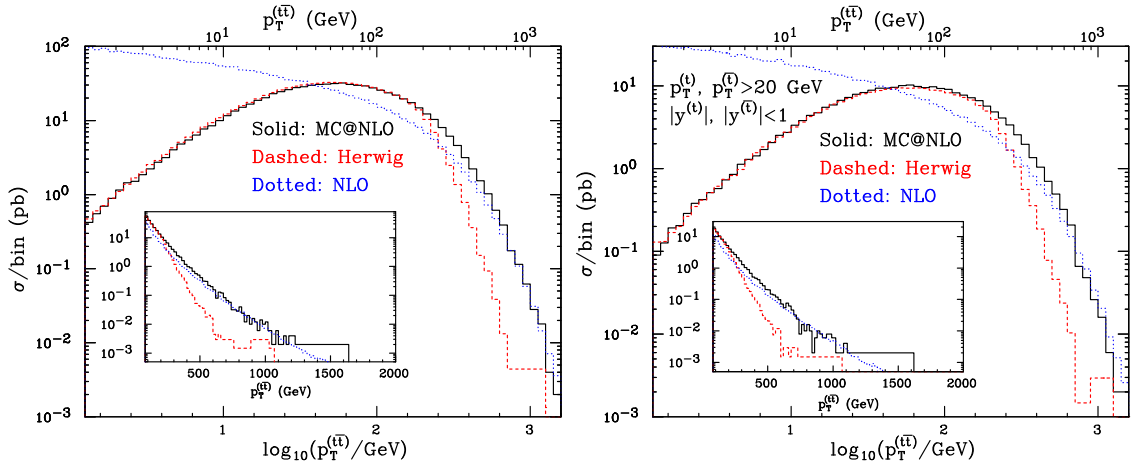

  \begin{center}
    \epsfig{figure=top_lhc_ptpair.eps,width=0.49\textwidth}
    \epsfig{figure=top_lhc_ptpair_cut.eps,width=0.49\textwidth}
\caption{\label{fig:top_lhc_ptpair} 
  As in fig.~\ref{fig:top_lhc_yandpt}, for the transverse momentum of the
  $t\bar t$ pair, without (left panel) and with (right panel) acceptance
  cuts.
}
  \end{center}
\end{figure}

We now turn to the case of more exclusive quantities, such as
correlations between $t$ and $\bar t$ variables.
In fig.~\ref{fig:top_lhc_ptpair} we present the modulus of the 
vector sum of the transverse momenta of the $t$ and $\bar t$,
which we denote by $\pttt$. NLO computations
cannot predict this observable in the region $\pttt\simeq 0$, because
of a logarithmic divergence for $\pttt\to 0$; on the other hand, NLO
is expected to give reliable predictions at large $\pttt$. The MC behaves
in the opposite way; thanks to the cascade emission of soft and collinear
partons, it can effectively resum the distribution around $\pttt=0$.
However, its results are not reliable in the large-$\pttt$ region, which 
is mainly populated by events in which a very hard parton recoils against 
the $t\bar t$ pair.

The complementary behaviour of the NLO and MC approaches can be seen clearly 
in fig.~\ref{fig:top_lhc_ptpair}, regardless of the cuts on the rapidities
and transverse momenta of the heavy quarks. In the tail
of the $\pttt$ distribution, the NLO cross section
is much larger than the MC one, simply because hard emissions are correctly
treated only in the former. The presence of the dead zones shown in
figs.~\ref{fig:fin} and \ref{fig:ini} makes it very difficult
to generate a very hard parton recoiling against the $t\bar t$ pair 
in the MC, which therefore gives rates much below the NLO
result in this region. For $\pttt\to 0$, the difference between
the two histograms shows the effect of all-order resummation; clearly,
no meaningful comparison between NLO and data can be attempted in this
region. It is therefore reassuring that the MC@NLO result
interpolates the MC and NLO results smoothly. In the small-$\pttt$ region,
the shape of the MC@NLO curve is identical to that of the
MC result. This is evidence of the fact that MC and MC@NLO
resum large logarithms at the same level of accuracy, as argued
in \FW.  When $\pttt$ grows large, the MC@NLO
tends to the NLO result, as expected. Again, hadronization has
no significant impact on the $t\bar t$ system.

\begin{figure}[htb]
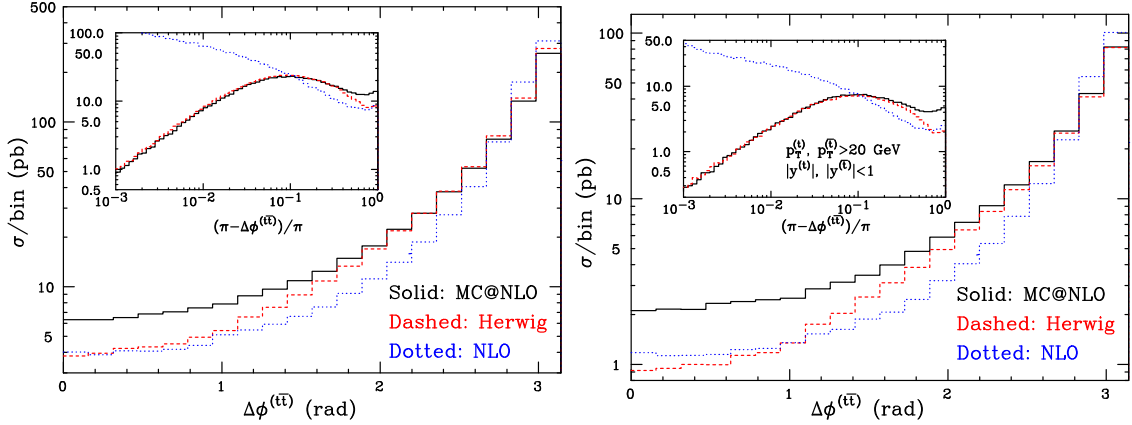

  \begin{center}
    \epsfig{figure=top_lhc_dphi.eps,width=0.49\textwidth}
    \epsfig{figure=top_lhc_dphi_cut.eps,width=0.49\textwidth}
\caption{\label{fig:top_lhc_dphi} 
  As in fig.~\ref{fig:top_lhc_ptpair}, for the difference in the azimuthal 
  angles of the $t$ and $\bar t$.
}
  \end{center}
\end{figure}

In fig.~\ref{fig:top_lhc_dphi} we present the distribution of the difference
between the azimuthal scattering angles (i.e., those in the plane transverse 
to the beam direction) of the $t$ and $\bar t$, which we denote
by $\dphitt$. This distribution cannot be reliably 
predicted by fixed-order QCD computations in the region
$\dphitt\simeq\pi$; in fact, the NLO prediction diverges logarithmically
for $\dphitt\to\pi$. This can be seen in the insets of the plots,
where the cross section has been plotted versus
\mbox{$(\pi-\dphitt)/\pi$} on a logarithmic scale, in order to visually
enhance the region $\dphitt\simeq\pi$. We can see that in this region the
MC@NLO and MC results have identical shapes, as in the case of 
the observable $\pttt$ near zero discussed above. The other end of the
spectrum, i.e. the tail $\dphitt\simeq 0$, is populated
by configurations in which a hard jet recoils against the
$t\bar{t}$ pair, but also by configurations in which the $t$ and $\bar{t}$ 
have small transverse momenta. The NLO calculation gives a good
description of the former region, but not the latter,
while the MC can treat reliably the latter region, but not the former.
MC@NLO, on the other hand, is expected to handle both regions reliably, while
still avoiding any double counting.  We also notice that, when cuts on the
transverse momenta of the heavy quarks are applied, the contribution from
multiple soft or collinear emissions becomes less important, and so the MC
does less well at $\dphitt\simeq 0$, as may be seen in the right-hand panel of
fig.~\ref{fig:top_lhc_dphi}.

\begin{figure}[htb]
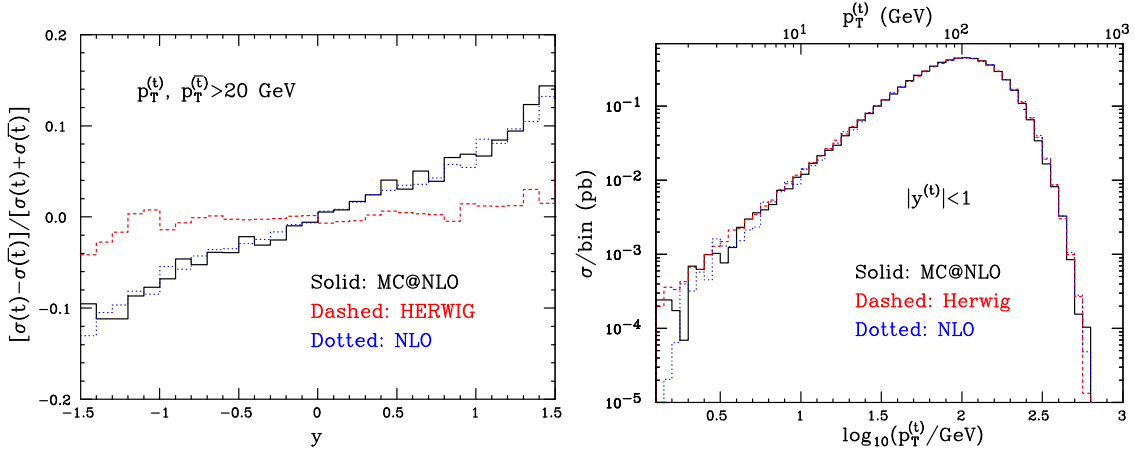

  \begin{center}
    \epsfig{figure=top_tev_yasy_cut.eps,width=0.49\textwidth}
    \epsfig{figure=top_tev_ptt_cut.eps,width=0.49\textwidth}
\caption{\label{fig:top_tev_yandpt} 
  MC@NLO (solid), \HW\ (dashed) and NLO (dotted) results
  for the rapidity asymmetry (left panel) and the transverse momentum 
  (right panel) of the top quark at the Tevatron. \HW\ results have been 
  normalized as explained in the text.
}
  \end{center}
\end{figure}

\begin{figure}[htb]
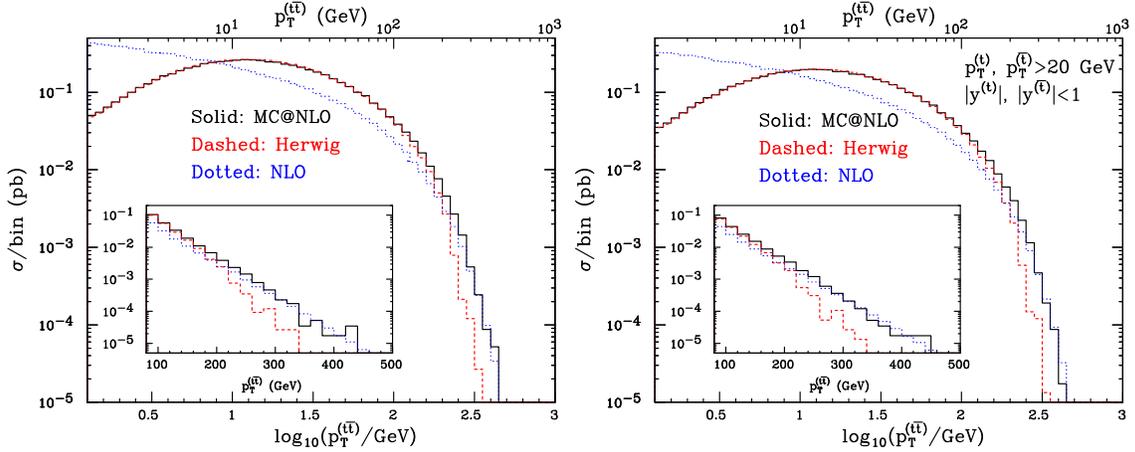

  \begin{center}
    \epsfig{figure=top_tev_ptpair.eps,width=0.49\textwidth}
    \epsfig{figure=top_tev_ptpair_cut.eps,width=0.49\textwidth}
\caption{\label{fig:top_tev_ptpair} 
  As in fig.~\ref{fig:top_tev_yandpt}, for the transverse momentum of 
  the $t\bar t$ pair, without (left panel) and with (right panel) acceptance
  cuts.
}
  \end{center}
\end{figure}

\begin{figure}[htb]
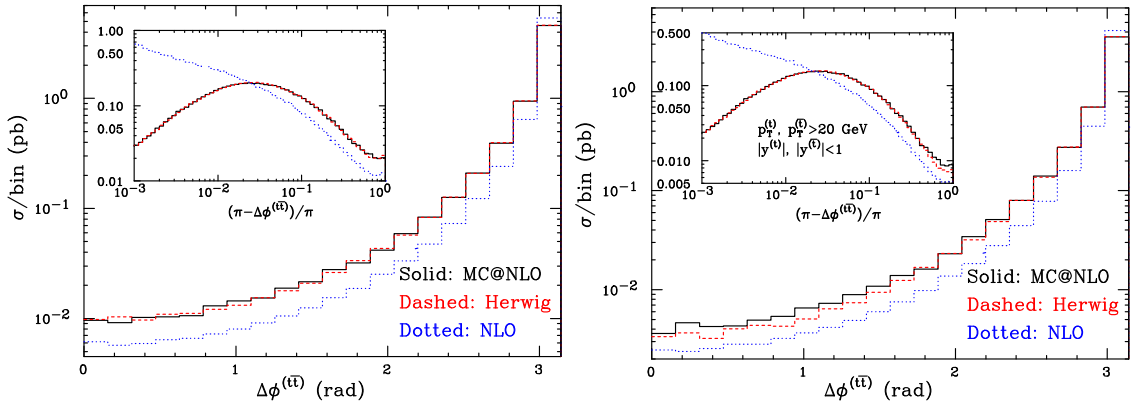

  \begin{center}
    \epsfig{figure=top_tev_dphi.eps,width=0.49\textwidth}
    \epsfig{figure=top_tev_dphi_cut.eps,width=0.49\textwidth}
\caption{\label{fig:top_tev_dphi} 
  As in fig.~\ref{fig:top_tev_ptpair},
  for the difference in the azimuthal angles
  of the $t$ and $\bar t$.
}
  \end{center}
\end{figure}

Corresponding results for Tevatron Run II ($p\bar p$ at $\sqrt{S}=2$~TeV)
are shown in figs.~\ref{fig:top_tev_yandpt}-\ref{fig:top_tev_dphi}.
An interesting new feature of the top quark rapidity distribution
is its forward-backward asymmetry, which cannot appear in $pp$ collisions 
and is also absent in $p\bar p$ at the Born level, and hence also in the MC 
result. To document this, we plot in the left panel of 
fig.~\ref{fig:top_tev_yandpt} the rapidity asymmetry rather than the
rapidity itself; cuts $\ptt,\pttbar>20$~GeV have been imposed.
We see a fair agreement between MC@NLO and NLO, whereas
\HW\ predicts an asymmetry compatible with zero.
In the right panel of the same figure, the top quark transverse momentum
(with a $\abs{\yt}<1$ cut) is shown. As in the case of LHC,
we see no substantial differences between NLO, MC@NLO and \HW.

The picture for $\pttt$ (fig.~\ref{fig:top_tev_ptpair})
is broadly similar to that at the LHC,
except for the reduced tail at high $\pttt$: the MC@NLO
prediction makes a smooth transition from the NLO to the resummed
MC form as $\pttt$ decreases.

The situation for $\dphitt$, on the other hand, is slightly different.
At Tevatron energies the influence of hard emissions is not so strong
as at the LHC. Consequently (see fig.~\ref{fig:top_tev_dphi})
the MC and MC@NLO predictions coincide quite closely over the whole range
of $\dphitt$, whereas the NLO remains lower, regardless of whether or
not cuts are applied.

\section{Results on bottom quark production}\label{sec:bot}
In the case of bottom production, many complications arise that are not 
present in top production. In a standard Monte Carlo, all of the three 
mechanisms discussed in sect.~\ref{sec:processes}, namely flavour creation 
(FCR), flavour excitation (FEX), and gluon splitting (GSP), need to be 
considered; they have rather different kinematic signatures, and
they are dominant in different regions of the phase space.
From the point of view of perturbative computations, bottom
cross sections are characterized by a fairly large value of
the coupling constant, which implies sizable K factors;
also, the importance of large logarithmic terms arising at
all orders is manifest in many observables, and suitable
resummations are often necessary for sensible comparisons 
with data.

As in the case of top production, we shall not attempt here
to discuss the phenomenological implications of our findings.
We shall rather emphasize the kind of problems encountered
in $b$-physics simulations with standard MC's, and the way
in which MC@NLO solves some of them. Comparisons between
MC@NLO and NLO results will also be presented here. We shall
not discuss the well-known limitations of fixed-order computations,
but refer the reader to ref.~\cite{Frixione:1997ma} for further 
details. Unlike the case of top production, in $b$ physics we
can compare MC@NLO results with MC and NLO ones either for 
$b$ quarks or for $b$-flavoured hadrons. The former option is
clearly preferable, if we aim at understanding the extent of the
validity of the MC@NLO approach, and the improvements with respect
to traditional formalisms. Technically, in MC@NLO and in \HW\ we
define the $b$-quark level as the stage which comes immediately
before the gluon-splitting phase,\footnote{This gluon splitting
is non-perturbative, and it has nothing to do with the perturbative
branching of a gluon which takes place during a shower; see
app.~\ref{sec:Cluform} for more details.} corresponding to \HW\ 
status code {\tt ISTHEP}=2.

The results presented in this section have been obtained at
the Tevatron Run II ($p\bar p$ at $\sqrt{S}=2$~TeV), for 
a bottom mass of 5~GeV. The other parameters have been chosen as in 
sect.~\ref{sec:top}. In order to simplify the analysis procedure, pair 
observables are defined by considering all possible pairs in the event,
regardless of their charge. Thus, in \HW\ and MC@NLO $b\bar{b}$, $bb$, and
$\bar{b}\bar{b}$ pairs are treated on an equal footing (at the NLO, one 
has just one $b\bar{b}$ pair), and will collectively be denoted as 
$b\bar{b}$ pairs. From the practical point of view this detail is almost 
irrelevant, since we find that the probability of having more than 
two $b$'s in an event with at least two $b$'s is of the order of 0.1\%.

\subsection{$b$-production issues in HERWIG}\label{sec:binHerwig}
We start by discussing the problems arising in the simulation of $b$ 
production with \HW. We stress that similar problems are present in 
any standard parton shower MC. As we discussed in sect.~\ref{sec:processes}, 
FEX and GSP contributions are considered in heavy flavour production
simply because FCR alone is not capable of describing the kinematics 
of observed events. It should be noted that FEX and GSP are somewhat
anomalous from the point of view of MC's, since usually the simulation
relevant to a given hard system involves the production of such a system
at the level of hard process generation. This 
fact has profound consequences: MC's cannot simulate
small-$\pt$ production of heavy quarks, since FEX and GSP matrix
elements are diverging for $\pt\to 0$. This poses a practical problem,
which is easily circumvented by cutting off the matrix element divergencies.
In \HW, this is achieved by requiring the transverse momenta of the
primary partons to be larger than a given quantity, called
$\ptmin$. In addition to this, \HW\ has an effective cutoff at the
level of hard matrix elements for FEX processes, which prevents
the generation of primary partons with $\pt=0$ even if $\ptmin=0$ (see
app.~\ref{sec:dyn}). In $b$ production, GSP processes also have an 
effective cutoff, but of a different nature. The probability of getting 
a showering scale large enough to produce a $b\bar{b}$ pair vanishes 
as $\pt\to 0$ in the hard process. Still, this doesn't allow one
to set $\ptmin=0$ in GSP, since the hard process is generated 
independently of the shower.

Although $t$-channel singularities are cut off by $\ptmin$, a problem of 
principle remains: if one interprets the output of a given showering event 
as a Feynman diagram, one obtains a contribution which in perturbation theory 
is only relevant in the large-momentum regime $\pt\gg m$. Thus, strictly 
speaking one should run FEX and GSP, and keep only those events with a 
large-$\pt$ heavy quark. It is customary to ignore this problem, and to keep 
all the events generated. The results are in general biased by $\ptmin$,
and it is therefore necessary to insure that this bias does not affect
the predictions in the kinematical regions of interest.

\begin{figure}[htb]
  \begin{center}
    \epsfig{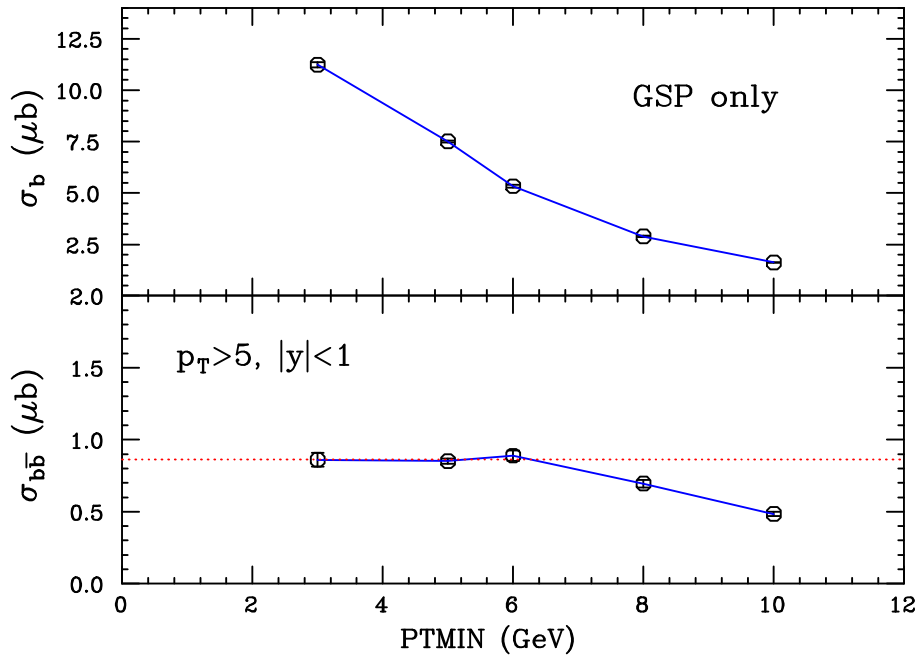}
\caption{\label{fig:ptmindep} 
  $b$ (upper panel) and $b\bar{b}$ (lower panel) GSP cross sections 
  within the cuts given in eqs.~(\ref{eq:SIcuts}) and~(\ref{eq:DDcuts})
  respectively, as predicted by \HW\ for various choices of $\ptmin$.
}
  \end{center}
\end{figure}

We have studied the bias due to the choice of $\ptmin$ by considering 
the GSP contribution. An analogous study can be done for the FEX
contribution, but in this case the presence of an effective cutoff at 
the level of hard matrix elements complicates the discussion unnecessarily.
We have considered the inclusive $b$ cross section, requiring
\beq
\ptb\,>\,5~{\rm GeV}\,,\;\;\;\;\;\;
\abs{\yb}<1\,,
\label{eq:SIcuts}
\eeq
and the pair cross section, requiring
\beq
\ptb,\ptbbar\,>\,5~{\rm GeV}\,,\;\;\;\;\;\;
\abs{\yb},\abs{\ybbar}<1\,.
\label{eq:DDcuts}
\eeq
The results are presented in fig.~\ref{fig:ptmindep}; open points 
are \HW\ predictions for the corresponding $\ptmin$ value, and the
solid lines are there just to guide the eye. The dotted line is
the weighted average of the results for the pair cross sections
obtained at $\ptmin=3,5$, and 6~GeV.
The lower panel of fig.~\ref{fig:ptmindep} has an easy interpretation:
for $\ptmin$ values smaller than 6 GeV, \HW\ predictions are independent
of $\ptmin$, i.e., they are not biased. Clearly, these values of $\ptmin$
are correlated to the choices of cuts made in eq.~(\ref{eq:DDcuts}) and,
to a smaller extent, to the fact that the observable chosen is fully
inclusive. It seems however safe to choose $\ptmin=5$~GeV for studying
any type of pair correlations with the cuts of eq.~(\ref{eq:DDcuts}).

The upper panel of fig.~\ref{fig:ptmindep} displays a less pleasant
behaviour; no range in $\ptmin$ can be found where the inclusive-$b$ 
cross section is independent of $\ptmin$. This happens because, at
small but finite $\pt$ of the primary partons, \HW\ can choose a
colour flow according to which the evolution scale is almost equal
to $\hat{s}$. Thus, a gluon acquires a large enough virtuality to
split into a $b\bar{b}$ pair. When boosted to the lab frame, one of 
the $b$'s can have a transverse momentum exceeding the $\pt$ cut (the 
probability for this to happen is small, and thus the probability 
of getting both $b$'s above the cut is negligible, which explains
the difference between the two panels of fig.~\ref{fig:ptmindep}). As 
already discussed before, the selection of such a colour flow is less and
less probable with decreasing $\pt$. Thus, there must exist a $\ptmin$
which returns unbiased single-inclusive cross sections. We didn't try 
to find such a $\ptmin$ value here, since for single-inclusive distributions
it is more sensible to compare MC@NLO results to NLO ones. However, our
exercise proves that it is very time-consuming to get unbiased predictions 
with \HW: the efficiency for generating events passing the cuts of
eqs.~(\ref{eq:SIcuts}) and~(\ref{eq:DDcuts}) is very rapidly decreasing with
decreasing $\ptmin$, because of the divergence of the matrix elements at
$\ptmin=0$.

\begin{figure}[htb]
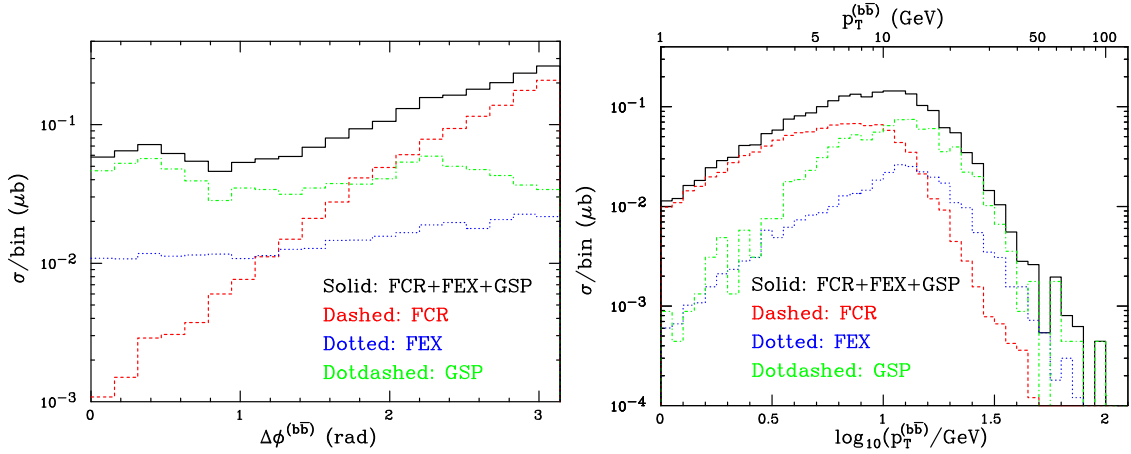

  \begin{center}
    \epsfig{figure=herwdphi.eps,width=0.49\textwidth}
    \epsfig{figure=herwptbb.eps,width=0.49\textwidth}
\caption{\label{fig:bbcorrherw} 
  Azimuthal $b\bar{b}$ distance (left panel) and transverse momentum
  of the $b\bar{b}$ pair (right panel) as predicted by \HW\ (solid); the
  contributions of FCR (dashed), FEX (dotted), and GSP (dot-dashed)
  are also separately presented.
}
  \end{center}
\end{figure}

To document the relative importance of the mechanisms contributing to 
\HW\ predictions, we show in fig.~\ref{fig:bbcorrherw} the results for the 
azimuthal distance $\dphibb$ between the $b$ and $\bar{b}$, and the transverse
momentum $\ptbb$ of the $b\bar{b}$ pair. 
The dashed, dotted, and dot-dashed histograms are the FCR, 
FEX, and GSP contributions respectively, whereas the solid histogram is 
the sum of the three. It is apparent that FCR is important only for those 
kinematic configurations which are almost $\twototwo$, namely 
$\dphibb\simeq\pi$ and $\ptbb\simeq 0$; elsewhere, FEX and GSP
contributions cannot be neglected. This implies that $b$-physics
simulations in standard MC's are always computing intensive.
We performed our GSP runs setting $\ptmin=5$~GeV; we found that
the fraction of events with at least a $b\bar{b}$ pair is
about 3.7$\cdot 10^{-3}$; when the cuts of eq.~(\ref{eq:DDcuts}) are
applied, this fraction is reduced to 1.6$\cdot 10^{-4}$. The dot-dashed
histograms in fig.~\ref{fig:bbcorrherw} contain only 1900 events,
obtained by generating 1.2$\cdot 10^{7}$ GSP events with \HW.
Although techniques are known to increase the efficiency, and
the figures quoted above depend on the particular MC used, it
is at present unknown how to generate GSP $b\bar{b}$ events
with high efficiency.

\subsection{$b$-production issues in MC@NLO}\label{sec:binMCatNLO}
In standard shower Monte Carlo programs, the bottom flavour
is included in the evolution of the initial-state and final-%
state showers, consistently with the fact that FEX and GSP 
processes need also to be considered. On the other hand, as
discussed in sect.~\ref{sec:processes} the implementation of
MC@NLO is based on FCR processes only. Thus, if we implemented 
bottom production in exactly the same way as we did with top, we would 
have to switch off the bottom flavour in all places where it appears in 
the shower Monte Carlo. This would be particularly problematic for the 
initial-state shower. Since we normally use five-flavour parton densities,
switching off the bottom flavour in backward evolution may lead
to inconsistencies. 

This problem is not peculiar to MC@NLO implementation.
The original NLO calculation of heavy flavour production was carried
out in the decoupling scheme of ref.~\cite{Collins:wz}, and thus
should be used in conjunction with a 4-flavour coupling constant and
4-flavour parton densities. In practice, five-flavour parton densities
were always used, since, as pointed out in eq. (3.11) of
ref.~\cite{Nason:1989zy}, this is correct (up to a numerically small
effect in the gluon parton density) as long as one neglects
parton densities with incoming heavy quarks.

In ref.~\cite{Cacciari:1998it}, the exact prescription for a change
of scheme in the heavy flavour production formulae (in order to go
from the decoupling scheme of ref.~\cite{Collins:wz}
to the full $\MSbar$ scheme with 5 flavours)
was given. It is summarized as follows:
\def \mur  {\mu_{\rm \scriptscriptstyle{R}}}
\def \muf  {\mu_{\rm \scriptscriptstyle{F}}}
\def \Tf  {T_{\rm \scriptscriptstyle{F}}}
\begin{itemize}
\item
Use five-flavour parton densities and strong coupling constant.
\item
Ignore the $b$ and $\bar{b}$ flavours in the parton densities.
\item
Add a term $-\as \frac{2\Tf}{3\pi}\log\frac{\mur^2}{m^2}\;
\sigma^{(0)}_{q\bar{q}}$ (where $\mur$ is the renormalization scale)
 to the $q\bar{q}$ channel cross section.
\item
Add a term $-\as \frac{2\Tf}{3\pi}\log\frac{\mur^2}{\muf^2}\;
\sigma^{(0)}_{gg}$ (where $\muf$ is the factorization scale)
to the $gg$ channel cross section.
\end{itemize}
This change of scheme was implemented in the MC@NLO code. The
corrections in the last two items above, although necessary to maintain 
formal correctness at the NLO level, have very small numerical impact.  
The implementation of the MC@NLO for bottom is then identical to that 
for top, except for the corrections listed above.

Unlike the case of top production, typical studies of $b$ production at hadron
colliders are carried out in a relatively large transverse momentum regime,
where the bottom flavour behaves in part as a light flavour, and the
resummation to all orders of $L=\log\pt/m$ terms may therefore be necessary 
in order to obtain sensible predictions. In perturbation theory, and for
single-inclusive observables, such a resummation is carried out to NLL
accuracy by considering all $\twototwo$ and $\twotothree$ partonic processes
(i.e., including those with one or no $b$ in the final state), and convoluting
them with NLL perturbative fragmentation functions~\cite{Mele:1990cw}, which
describe the evolution of a light parton or a $b$ quark into a $b$ quark. The
resummed formulae can be consistently combined~\cite{Cacciari:1998it} with the
exact ${\cal O}(\as^3)$ formulae, in order to obtain reliable predictions for
any $\pt$ value.

This procedure has not been extended yet to more exclusive observables.
MC@NLO would seem a natural way to achieve this goal, at least at the
leading logarithmic accuracy; this would imply implementing, besides 
the exact ${\cal O}(\as^3)$ FCR formulae, also the ${\cal O}(\as^2)$ ones
for light-parton scatterings. The role of the perturbative fragmentation
functions would then be played by the showers. The presence of light-parton
processes would be taken into proper account in the definition of the MC 
subtraction terms, in order to avoid double counting. Although this task
is not totally out of reach, it is very difficult to achieve it in practice,
since it would lead to a great increase in complexity in comparison to the 
top case.

\begin{figure}[htb]
  \begin{center}
    \epsfig{figure=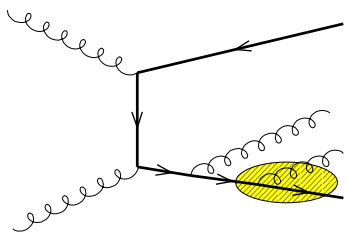,width=0.3\textwidth}
\caption{\label{fig:LLterms} 
  Enhanced terms present in the MC@NLO implementation, accurate
  to the leading-logarithmic level in the MC@NLO approach. The
  shaded area represents an almost collinear branching due to parton 
  showers.
}
  \end{center}
\end{figure}

In the present work we shall restrict ourselves to implementing
bottom production with the same accuracy as standard NLO calculations,
using the five-flavour strong coupling constant and parton densities in
the scheme described above. This allows us to include some leading-log 
effects (i.e., terms like $\as^2 (\as L)^k$), but not all of them.
More precisely, single-log terms (of order $\as^3 L$) are included 
exactly in our calculation. The terms of order $\as^2 (\as L)^k$ 
that are included for any $k$ are those that result from multiple 
final-state radiation from the heavy quark lines, 
one example of which is illustrated in fig.~\ref{fig:LLterms}.
Observe that, although the hardest emission is exactly described
at the NLO level, subsequent gluon emissions have only a leading-log
accuracy, since going to next-to-leading log accuracy would
require the implementation of emission kernels (i.e., of showers) 
at comparable accuracy. 

There are also other effects of order $\as^2 (\as L)^k$ that are automatically
included in our MC@NLO implementation, which are due to branchings that
involve $b$ quarks. One such contribution arises when, by backward evolution,
a gluon fusion FCR process is connected to a $b$ quark line via the 
initial-state shower. A second type of contribution arises when a 
final-state gluon, emerging from a FCR hard process, splits into a $b\bar{b}$
pair. This splitting process, together with the associated virtual process,
gives rise to the correction to the running of $\as$ due to the bottom
flavour. The consistency between the shower evolution including $b$ quarks,
and the NLO FCR cross sections, is in fact what dictates the use of the
five-flavour scheme for $\as$ and for the parton densities.

However, it is clearly impossible to include all effects of order 
$\as^2 (\as L)^k$ starting from FCR processes, even with a five-flavour
scheme. Most noticeably, the $\as^2 (\as L)^2$ GSP and FEX contributions 
depicted in fig.~\ref{fig:as4logs} are not included in MC@NLO.

\begin{figure}[htb]
  \begin{center}
    \epsfig{figure=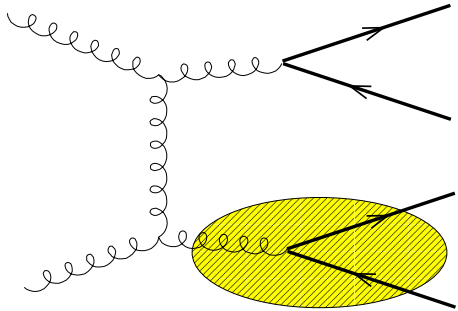,width=0.4\textwidth}
\caption{\label{fig:fourq} 
Heavy flavour production by gluon splitting, followed by a further 
splitting process (by showering) of the recoiling gluon, represented
by a shaded area.
}
  \end{center}
\end{figure}

In the case of final states containing two or more heavy flavour pairs,
the level of precision of the MC@NLO treatment is unclear because one of
the pairs (the one produced at the primary NLO vertex) is described
differently from those produced by parton showering, as illustrated
in  fig.~\ref{fig:fourq}.  As a consequence, the weight factors in the
prediction of the bottom pair multiplicity, for example, may not be
correct.  We postpone discussion of this point to a later date,
since from a practical viewpoint multiple pair production is a rare
phenomenon (0.1\% of single-pair production) which has a negligible 
effect on the plots shown here.

\subsection{Pair correlations}\label{sec:bbcorr}
\begin{figure}[htb]
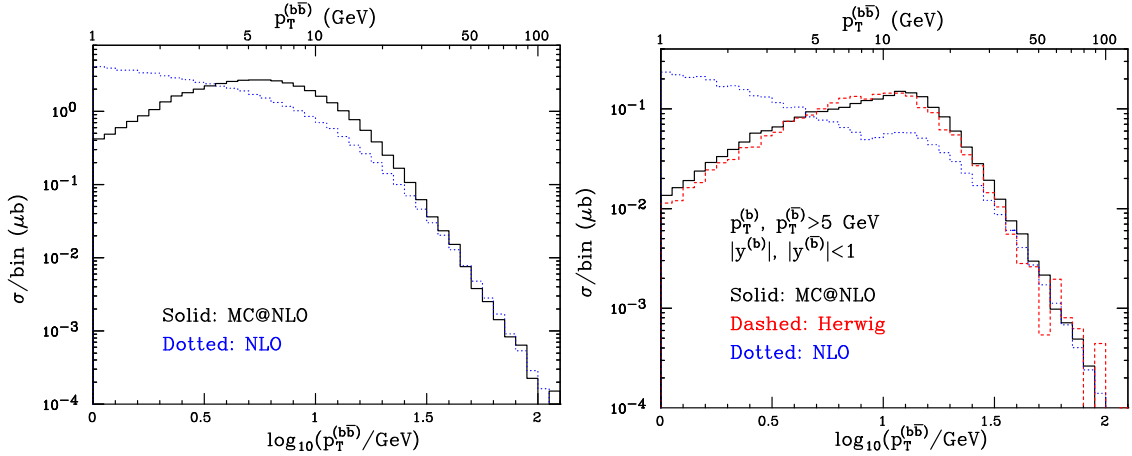

  \begin{center}
    \epsfig{figure=bbpt.eps,width=0.49\textwidth}
    \epsfig{figure=bbpt_cut.eps,width=0.49\textwidth}
\caption{\label{fig:bbpt} 
  MC@NLO (solid), \HW\ (dashed), and NLO (dotted) results for the pair 
  transverse momentum, without (left panel) and with (right panel)
  the cuts of eq.~(\ref{eq:DDcuts}).
}
  \end{center}
\end{figure}

In this section, we present predictions for $b\bar{b}$ pair observables.
We start by considering the transverse momentum of the pair,
$\ptbb$. The MC@NLO (solid) and NLO (dotted) results are shown
in fig.~\ref{fig:bbpt}; in the left panel no cuts have been applied,
whereas the right panel includes the effect of the cuts given
in eq.~(\ref{eq:DDcuts}). Regardless of the presence of the cuts,
the two plots display the same pattern as the analogous plots
for top production (figs.~\ref{fig:top_lhc_ptpair} 
and~\ref{fig:top_tev_ptpair}): the NLO distributions diverge
for $\ptbb\to 0$, whereas MC@NLO results have a regular behaviour.
On the other hand, in the large-$\ptbb$ region, dominated by 
hard-parton emission, MC@NLO and NLO coincide in shape and 
normalization.

When the cuts of eq.~(\ref{eq:DDcuts}) are applied, we can also
consider the \HW\ result (dashed histogram), which has been already
shown in fig.~\ref{fig:bbcorrherw}. Unlike the case of top production,
here we take \HW\ result with its normalization; a rescaling by
the K factor would be more appropriate only if FCR process alone
were included. The agreement between MC@NLO and \HW\ is remarkable.
In the low-$\ptbb$ region, we can see that the shape of the MC@NLO
result is basically the same as that of \HW, analogously to what
we have seen in the case of top production.
The comparison at large $\ptbb$ is hampered by the poor statistics
of the \HW\ result, since this region receives its main contribution
from the GSP process (see fig.~\ref{fig:bbcorrherw}); however, it
appears that MC@NLO and \HW\ agree well. This is not surprising, 
since a $b\bar{b}$ pair produced through the GSP mechanism mainly recoils
against a hard gluon, exactly as in the NLO matrix elements which are
implemented in MC@NLO. The inability of the MC to produce hard emissions
is only evident in the FCR contribution (dashed histogram in
fig.~\ref{fig:bbcorrherw}), which in fact has a much softer behaviour
than GSP or MC@NLO.

\begin{figure}[htb]
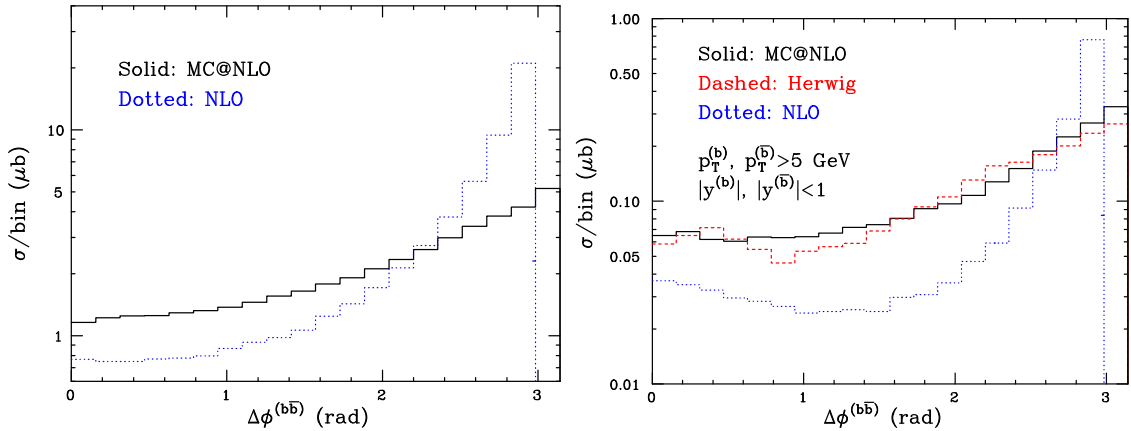

  \begin{center}
    \epsfig{figure=bbdphi.eps,width=0.49\textwidth}
    \epsfig{figure=bbdphi_cut.eps,width=0.49\textwidth}
\caption{\label{fig:bbdphi} 
  As in fig.~\ref{fig:bbpt}, for the azimuthal distance.
}
  \end{center}
\end{figure}
We now turn to the case of the azimuthal distance, $\dphibb$.
As discussed in sect.~\ref{sec:top}, this observable receives 
non-negligible contributions to the tail $\dphibb\simeq 0$ from
both hard emissions and multiple soft or collinear emissions.
The presence of large logarithms in the perturbative expansion
is manifest at the NLO, since the prediction is strongly peaked
towards $\dphibb\to\pi$; the correct result for the total rate
implies that the cross section at $\dphibb=\pi$ is negative.
This behaviour is apparent in the dotted histograms in fig.~\ref{fig:bbdphi}.
The shapes of the MC@NLO results are similar to the NLO ones
in the small-$\dphibb$ region, but do strongly differ elsewhere.
One could perhaps expect that, upon applying the cuts of 
eq.~(\ref{eq:DDcuts}), MC@NLO and NLO would be closer in normalization
for $\dphibb\to 0$; this doesn't happen because MC@NLO and NLO have
non-negligible differences in the intermediate-$\pt$ region, which dominates
the cross section when the cuts of eq.~(\ref{eq:DDcuts}) are applied.
This fact is not accidental, and will be discussed in 
sect.~\ref{sec:bSingInc}.

As in the case of the transverse momentum of the pair, when the
cuts are applied we can also consider the \HW\ prediction (dashed
histogram in the right panel of fig.~\ref{fig:bbdphi}). In this
case, too, the agreement between MC@NLO and \HW\ is remarkable,
and emphasises again the importance of the GSP contribution
(see fig.~\ref{fig:bbcorrherw}). MC@NLO appears to be somewhat 
more peaked than \HW\ towards $\dphibb\to\pi$.

\subsubsection{Impact of initial-state radiation}\label{sec:bbISR}
The results presented above imply that multiple radiation is a
crucial effect in $b$ production. Lacking an N$^k$LO computation,
a typical way of estimating the effects of multiple 
{\em initial-state} emissions on heavy-quark distributions
is to supplement an NLO calculation with an intrinsic transverse 
momentum for the incoming partons (denoted as NLO+$\kt$-kick hereafter). 
We stress that this procedure is ill-defined, since there is no way of 
avoiding double counting. There are however cases in which it is 
justified from the phenomenological point of view, allowing a much better
description of the data than NLO predictions alone (see
ref.~\cite{Frixione:1997ma} for a discussion of the implementation of
$\kt$-kick in heavy flavour production, and its implications).

\begin{figure}[htb]
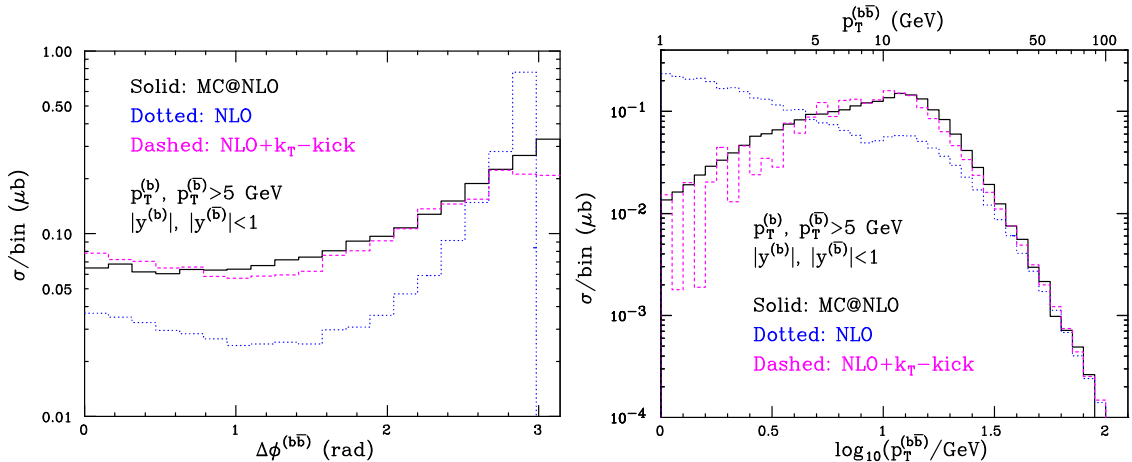

  \begin{center}
    \epsfig{figure=bbdphi_kick.eps,width=0.49\textwidth}
    \epsfig{figure=bbpt_kick.eps,width=0.49\textwidth,
                                 height=0.405\textwidth}
\caption{\label{fig:bbktkick} 
  $\dphibb$ (left panel) and $\ptbb$ (right panel) predicted by MC@NLO 
  (solid), NLO+$\kt$-kick (dashed), and NLO (dotted). The cuts of 
  eq.~(\ref{eq:DDcuts}) have been applied.
}
  \end{center}
\end{figure}
We define our $\kt$-kick by assuming a gaussian transverse momentum 
distribution for the incoming partons; we fix the free parameter of
this distribution by requiring that the average $\ptbb$ predicted by
NLO+$\kt$-kick be equal to that obtained from MC@NLO. In this way,
the average transverse momentum of the incoming partons turns out to
be $\langle\kt\rangle\simeq 4$~GeV. With this choice, 
the NLO+$\kt$-kick results for the $\ptbb$ and $\dphibb$ distributions
are seen to agree reasonably with the corresponding MC@NLO results
(see fig.~\ref{fig:bbktkick}). This confirms the importance
of multiple parton radiation, on top of the possible hard emission
present at the NLO level, and shows that the vast majority of the 
effect is due to emissions from initial-state partons.

The reader is urged not to take NLO+$\kt$-kick results too seriously,
since they are based on a model with little theoretical justification.
In fact, an average intrinsic transverse momentum of 4~GeV is too
large to be considered a typical non-perturbative effect. On the
contrary, in the MC@NLO implementation this effect has a purely 
perturbative origin, being due to multiple initial-state emissions. 

\subsection{Single-inclusive observables}\label{sec:bSingInc}
We finally turn to the case of single-inclusive $b$-quark distributions.
In this section, we shall not consider \HW\ results, because 
of the findings of sect.~\ref{sec:binHerwig}; on the other hand,
we do consider the effect of the NLL resummation of large logs
$L=\log\pt/m$, as discussed in sect.~\ref{sec:binMCatNLO}.

\begin{figure}[htb]
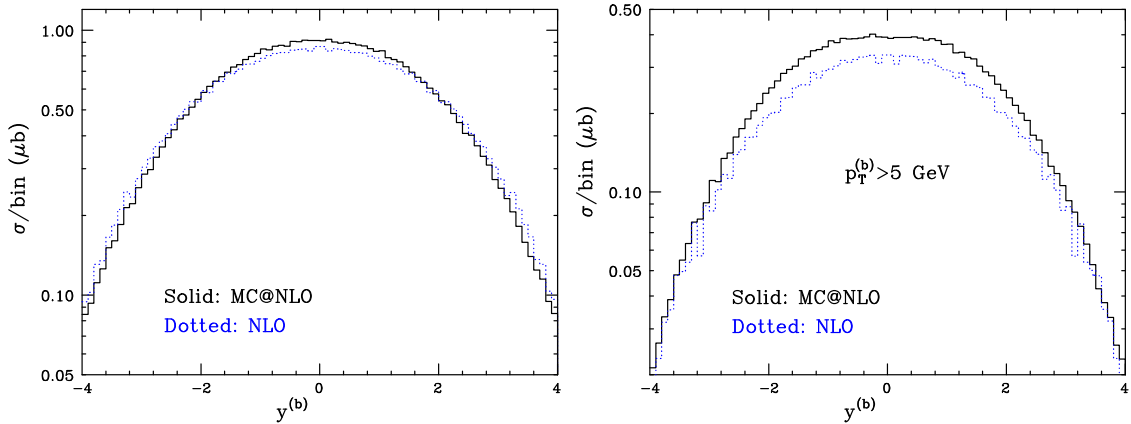

  \begin{center}
    \epsfig{figure=by.eps,width=0.49\textwidth}
    \epsfig{figure=by_cut.eps,width=0.49\textwidth}
\caption{\label{fig:bycut}
  MC@NLO (solid) and NLO (dotted) results for single-inclusive $b$
  rapidity spectrum, without (left panel) and with (right panel) a
  $\pt$ cut.
}
  \end{center}
\end{figure}
We start with the rapidity of the $b$ quark, presented
in fig.~\ref{fig:bycut} without and with a transverse momentum cut
$\ptb>5$~GeV. Regardless of the presence of this cut, MC@NLO (solid)
and NLO (dotted) results agree well. This is to be expected, since
for such an observable NLO results are in general reliable. 

\begin{figure}[htb]
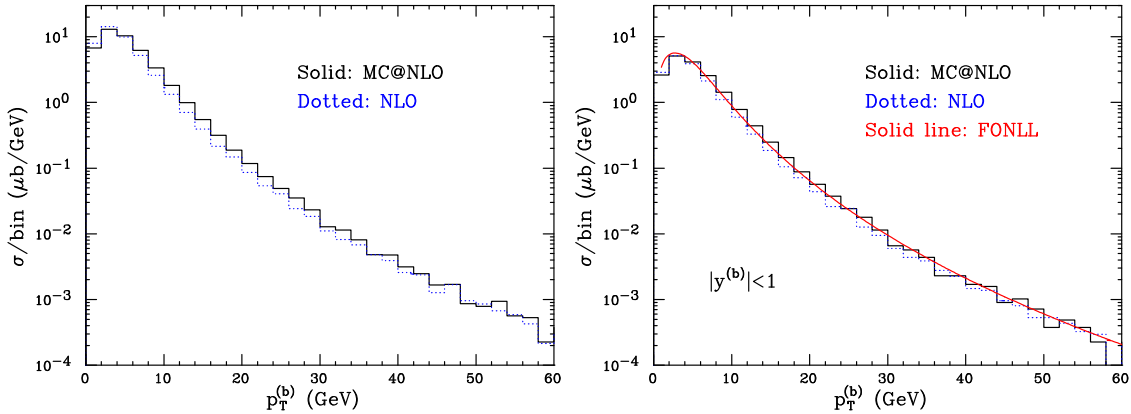

  \begin{center}
    \epsfig{figure=bpt.eps,width=0.49\textwidth}
    \epsfig{figure=bpt_cut.eps,width=0.49\textwidth}
\caption{\label{fig:bptcut}
  MC@NLO (solid) and NLO (dotted) results for single-inclusive $b$
  $\pt$ spectrum, without (left panel) and with (right panel) a
  rapidity cut. In the latter case, the FONLL~\cite{Cacciari:1998it} 
  result (solid line) is also shown.
}
  \end{center}
\end{figure}
The differences between MC@NLO and NLO results are larger in the
case in which a $\pt$ cut is applied. The reason becomes clear if one plots
the $\pt$ spectrum, shown in fig.~\ref{fig:bptcut}. Regardless
of the presence of the rapidity cut, the MC@NLO predictions (solid) 
are above the NLO ones (dotted) in the intermediate-$\pt$ range. 
When we apply a rapidity cut, we can also consider FONLL
predictions~\cite{Cacciari:1998it}, based on a formalism that
includes not only NLO, ${\cal O}(\as^3)$ terms, but also log-enhanced 
terms of order $\as^2 (\as\log \pt/m)^k$ (LL) and $\as^3 (\as\log \pt/m)^k$ 
(NLL). The FONLL result is shown as a solid line in the right panel
of fig.~\ref{fig:bptcut}.

\begin{figure}[htb]
  \begin{center}
    \epsfig{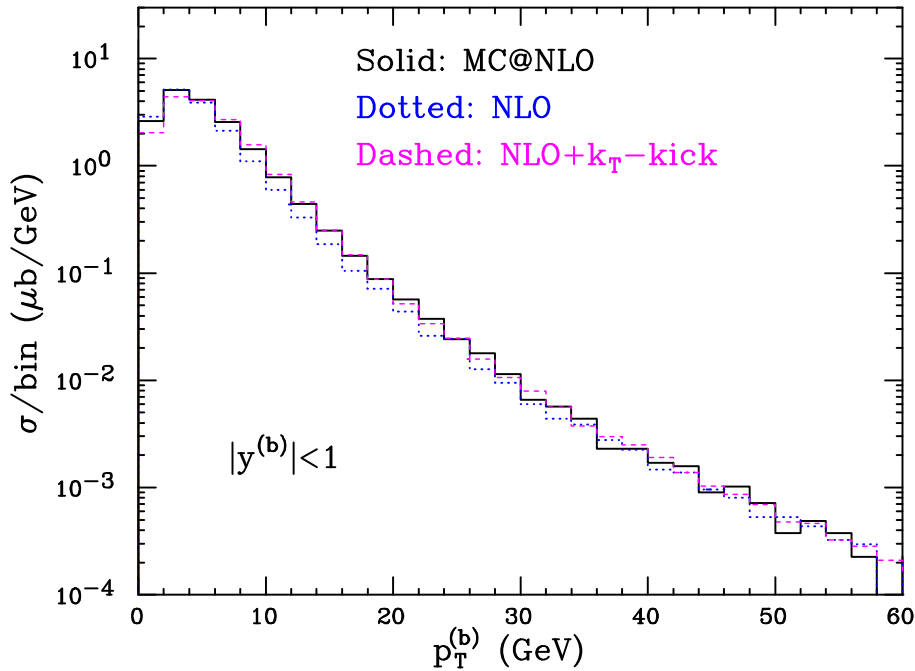}
\caption{\label{fig:bptkick}
Comparison of the $b$-quark $\pt$ spectrum obtained with MC@NLO (solid),
NLO+$\kt$-kick (dashed), and NLO (dotted).
}
  \end{center}
\end{figure}

Similarly to MC@NLO, the FONLL result is slightly above the NLO calculation
in the $\pt$ region from about 5 to 50 GeV. In the FONLL calculation,
this small excess in the intermediate region was attributed
to the inclusion in the calculation of higher-order corrections
to light parton production (in particular $gg\to gg$) followed
by gluon fragmentation into a heavy quark (see
ref.~\cite{Cacciari:1998it} for details).
Neither the NLO, nor the MC@NLO calculations include these corrections.
On the other hand, in the MC@NLO implementation the enhancement
is very likely due to the transverse momentum boost given to
the hard process by the initial state showers. In order to
verify this possibility, we use again the NLO+$\kt$-kick approach
used in sect.~\ref{sec:bbISR}, with the same parameter setting
adopted there. The result is presented in fig.~\ref{fig:bptkick}.
The MC@NLO and NLO+$\kt$-kick results are again in fair agreement.
As in the previous section, we interpret this fact as a hint of the
relevance of initial-state multiple emissions in $b$ production at 
the Tevatron. Such effects are not included in FONLL. 
We thus tentatively speculate that a formalism able to take into account all
these effects (i.e. FONLL effects plus multiple soft radiation from the
initial-state partons) may result in further enhancement of the 
intermediate-$\pt$ region with respect to both MC@NLO and FONLL results.

\section{Conclusions and future prospects}\label{sec:conc}

In this paper we have applied the MC@NLO method to the process of
heavy quark production at hadron colliders. The method was developed in
ref.~\cite{Frixione:2002ik} and applied there to the process of gauge
boson pair production.  For heavy quark production, the basic formalism
remains the same; it was only necessary to compute the relevant MC
subtraction terms and associated colour flows.  However, this proved
a good deal more complicated than for gauge boson pair production,
owing to the presence of coloured objects in the final state even at
the Born level, giving rise to different patterns of momentum reshuffling
in the MC and more alternative colour flows.  In fact, the implementation
for heavy quark production demonstrates that MC@NLO can be applied to
arbitrarily complicated processes, in terms of kinematics, colour flows
and large K factors.

Results were presented on single-particle distributions and correlations
at the heavy quark level, with parton showering and hadronization
but showing the distributions  of the heavy quarks before decay
(in the case of top) or hadronization (in the case of bottom),
to facilitate comparisons with purely perturbative predictions.

In the case of top quark production, results appear to be fully under
perturbative control at Tevatron and LHC energies. The MC@NLO results
on single-particle distributions are generally close to those at NLO.
A forward-backward asymmetry is seen in the top-quark rapidity distribution
at the Tevatron, as expected in NLO but not in LO or in standard MCs
which lack exact NLO corrections.  In $t\bar t$ correlations, MC@NLO
combines the more correct treatment of hard emissions in NLO with
the MC resummation of soft and collinear contributions, giving
a smooth distribution of $\pttt$ and an enhancement in the
$\dphitt$ distribution when the $t$ and $\bar t$ are not
back-to-back. The effects of hard emissions are naturally more visible
at the LHC.

Results on $b$ production (presented for the Tevatron only) show more markedly
the advantages of the MC@NLO approach compared with conventional MCs. The fact
that NLO contributions are included eliminates the need for separate flavour
creation (FCR), flavour excitation (FEX) and gluon splitting (GSP)
contributions, which are required in standard MCs but are plagued with
inefficiencies, ambiguities and cutoff dependences. In fact, we did not
attempt to generate single-$b$ distributions with the standard \HW\ MC for
comparison with MC@NLO and NLO, due to its low efficiency for the GSP
contribution.  The MC@NLO results, on the other hand, tend to be even more
stable than pure NLO, since cancellations between large numbers occur at the
matrix-element level, rather than in histograms. They do not depend on
cutoffs or on $b$ PDFs close to threshold (a source of uncontrolled errors),
and thus one can sensibly predict cross sections all the way down to $\pt=0$.

MC@NLO predictions for single $b$-quark distributions at the Tevatron are
quite similar to those at NLO, with some enhancement at intermediate
transverse momenta, which can be interpreted as resulting from 
Sudakov effects due to multiple initial-state gluon emission.  

In the case of $b\bar b$ correlations at the Tevatron, we could
obtain stable \HW\ MC predictions for reasonable values of the
cutoff parameter {\tt PTMIN}, provided the cuts of eq.~(\ref{eq:DDcuts})
were imposed. With the resulting mixture of FCR, FEX and GSP contributions,
the  \HW\ results were then in quite good overall agreement with those of
MC@NLO.  However, as stressed above, the arbitrariness and inefficiency
of the standard MC prescription make the MC@NLO approach vastly
preferable from all points of view.

One feature of MC@NLO that might seem unattractive compared with standard
MCs is the presence of a fraction (10--20\%) of negative weights.
As may be seen from the histograms presented here (all generated with
unweighted events) this does not cause problems in practice.
Most of the negative weights arise from the small-$\pt$ region where
the NLO real emission contribution is not resolvable.
It may be possible to reduce their contribution further
by tuning the subtraction procedure.
In spite of the presence of negative weights, MC@NLO is more efficient
that standard MCs for $b$ production, since it  generates the GSP
contribution with efficiency 1.

We have not attempted any detailed phenomenological study or comparisons
with experimental data in this paper, since our primary objectives
were to present the MC@NLO method and compare it with others in
the context of heavy quark production.  An obvious next step
would be to compare with Tevatron data on final-state properties
in top and bottom production. In the case of top production,
this will require the inclusion of decay correlations due to
polarization of the $t$ and $\bar t$, which is straightforward
in principle but not yet included in the NLO calculation that
we have used.  In the case of bottom production, careful attention
must be paid to the hadronization model used to connect $b$-quark
and B-hadron distributions (see app.~\ref{sec:bhad} and
ref.~\cite{Cacciari:2002pa}).

Other obvious future objectives are the application of MC@NLO
to Higgs boson, single vector boson, and jet production.
The only extra complication, compared with heavy quark production,
arises in the latter case from collinear singularities due to
emission from massless final-state partons. The necessary MC subtraction
terms can be computed using the kinematics already presented
in the present paper.  The resulting terms will cancel the
additional singularities and should provide more reliable
predictions of jet production and fragmentation when combined
with a subtraction-method NLO calculation of this process.

\section*{Acknowledgements}
We thank Rick Field for helpful correspondence. S.F.\ is grateful for
the hospitality of CERN Theory Division during parts of this project.

\newpage
\appendix

\noindent{\large\bf Appendices}

\section{Heavy flavour production in HERWIG}
We describe here the main features of the \HW\ event generator that
are relevant to heavy flavour production. We should distinguish between
the standard \HW\ Monte Carlo program and the \HW\ implementation of
parton showering and hadronization that forms part of MC@NLO.
In the former case the program includes the generation of a hard
subprocess as outlined in app.~\ref{sec:hardsub}. In the case of MC@NLO,
only app.~\ref{sec:showers} and \ref{sec:hadrize} are relevant,
since the hard process configurations are read from the input
file as explained in sect.~\ref{sec:interface}.

\subsection{Hard subprocess}\label{sec:hardsub}
The primary heavy quark processes included in the standard
\HW\ program are the $\twototwo$ processes of flavour creation 
(FCR, eq.~(\ref{eq:FCRtwo})), and flavour excitation 
(FEX, eq.~(\ref{eq:FEXtwo}) and charge-conjugate processes).  
However, the hard process configuration is
always chosen {\em as if it were FCR}, i.e.\ the kinematics are always chosen
for $\bp_1\bp_2\to\bk_1,\bk_2$ with $\bp_{1,2}$
massless and $\bk_{1,2}$ having the heavy quark mass $m$.

\subsubsection{Kinematics}
The algorithm for selecting the $\twototwo$ subprocess kinematics
is as follows:
\begin{enumerate}
\item Choose the heavy quark transverse mass $\mt=\sqrt{\bkT^2+m^2}$
from a power distribution $\dsb/d\mt\propto \mt^{-p}$ where
$p=${\tt PTPOW}[4]\footnote{Default values (in GeV units) are given in
square brackets.} with {\tt PTMIN}[10]$<\bkT<${\tt PTMAX}[$\sqrt S/2$].
\item Choose the outgoing parton rapidities uniformly in
{\tt YJMIN}[--8]$<y_{3,4}<${\tt YJMAX}[8] (or in the kinematically allowed 
region if smaller).
\item Compute the incoming parton light-cone momentum fractions according
to
\beq
\bx_1=(e^{y_3}+e^{y_4})\mt/\sqrt S\;,\;\;\;\;
\bx_2=(e^{-y_3}+e^{-y_4})\mt/\sqrt S\;.
\eeq
The program also computes at this stage the c.m.\ scattering angle
\beq
\cos\bh = (e^{y_3}-e^{y_4})\mt/\sqrt{e^{y_3+y_4}(\bx_1\bx_2 S-4m^2)}\,.
\eeq
\item Compute the kinematic invariants according to
\beq
\bs=\bx_1\bx_2 S\;,\;\;\;\;\bt=-(1+e^{y_4-y_3})\mt^2
\;,\;\;\;\;\bu=-\bs-\bt\;.
\eeq
\end{enumerate}

\subsubsection{Dynamics}\label{sec:dyn}
Having chosen a kinematic configuration, \HW\ calls for PDF evaluations
at momentum fractions $\bx_1, \bx_2$ and hard process scale {\tt EMSCA}
given by
\beq\label{eq:emsca}
{\tt EMSCA} = \sqrt{\frac{2\bs\bt\bu}{\bs^2+\bt^2+\bu^2}}\;,
\eeq
which is also used as the argument of $\as$ in the calculation of
hard process matrix elements.

The program then cycles through all possible FCR and FEX subprocesses,
computing the product of relevant PDFs times the on-shell differential
cross section evaluated at $(\bs,\bt,\bu)$, times a Jacobean
factor. The event weight {\tt EVWGT} is the sum of all these contributions.
In unweighted event generation, if this is greater than
the maximum weight times a random number, the configuration is accepted.
The program then multiplies {\tt EVWGT} by another random number, and
cycles through the FCR and FEX subprocesses again until the partial sum of
contributions exceeds this value.  The last subprocess selected is then
loaded into the event record as an on-shell $\twototwo$ process, with the
parton momenta reconstructed from the $\bx_1,\bx_2,\cos\bh$ values
computed previously.

Note that:
\begin{enumerate}
\item Values of $(\bs,\bt,\bu)$ are generated only inside the
FCR phase space,
and so values accessible only in FEX are not explored.  In particular
the singular FEX point $\bt=0$ is {\em never generated}, even if one sets
{\tt PTMIN}=0.
\item If the process selected is FEX instead of FCR, the reconstructed
$\pt=\pt^{\mbox{\tiny FEX}}$ of the heavy quark is not equal to the
$\pt=\pt^{\mbox{\tiny FCR}}$ originally selected assuming FCR kinematics.
An elementary calculation shows that
\beq
\pt^{\mbox{\tiny FEX}} =  \pt^{\mbox{\tiny FCR}}\frac{\bs}
{\sqrt{(\bs+m^2)(\bs-4m^2)}}
\eeq
Thus one always has $\pt^{\mbox{\tiny FEX}} > \pt^{\mbox{\tiny FCR}}$.
\item In all the above, the masses of light quarks and the effective
mass of the gluon are neglected, whereas \HW\ uses these masses (only)
in the final step, when it reconstructs the parton momenta from the
values of $\bx_1,\bx_2,\cos\bh$.
\end{enumerate}

\subsubsection{Colour structure}\label{app:col}
\HW\ treats all colour flows as distinct subprocesses.
Where the colour flow is ambiguous, it is assigned according to the
$N\to\infty$ limit.  For example, in the FCR process $gg\to Q\bQ$
the colour-averaged matrix element squared is given by eq.~(\ref{eq:MEgg}).
In the $N\to\infty$ limit the two colour flows are proportional to
$\bu/\bt$ and $\bt/\bu$ factors in the first bracket, and therefore 
we assign~\cite{Odagiri:1998ep}
\beq\label{eq:dsbtu}
\dsb_{gg}^{(t)} = \dsb_{gg}\frac{\bu/\bt}{\bu/\bt+\bt/\bu}
 =  \frac{\dsb_{gg}}{1+\bt^2/\bu^2}\;,\;\;\;
\dsb_{gg}^{(u)} = \frac{\dsb_{gg}}{1+\bu^2/\bt^2}\;,
\eeq
with $\dsb_{gg}$ given in eq.~(\ref{eq:Bornxsec}).
In $\dsb_{gg}^{(t)}$, which we shall call the $t$-flow contribution,
gluon 1 is colour-connected to gluon 2 and the heavy quark $Q$,
while gluon 2 is connected to 1 and $\bQ$.
In $\dsb_{gg}^{(u)}$, the $u$-flow contribution,
gluon 1 is colour-connected to 2 and $\bQ$,
while 2 is connected to 1 and $Q$.

In the other FCR process $q\bq\to Q\bQ$, the colour connection is
uniquely $q-Q$ and $\bar q-\bQ$.

\subsection{Parton showers}\label{sec:showers}
For a general introduction to the parton shower approximation, see Chapter 5 
of ref.~\cite{Ellis:1996qj}. The parton shower in \HW\ \cite{Corcella:2000bw}
is performed using an angular variable  $\xi$ \cite{Marchesini:1983bm} and
energy fraction $z$.  In the parton branching $i\to jk$, we have
\beq
\xi_{jk}= \frac{p_j\cdot p_k}{E_j E_k}\;,\;\;\;
z_j=\frac{E_j}{E_i}\;,\;\;\;z_k=\frac{E_k}{E_i}=1-z_j\;.
\eeq
Thus $\xi_{jk}=1-\cos\theta_{jk}$ for massless partons. The evolution scale
variable is $Q=E\sqrt\xi$; thus the scale set by the above branching is
$Q_i=E_i\sqrt\xi_{jk}$. Colour coherence is simulated by angular ordering,
which implies that the initial scales for showering on partons $j$ and
$k$ are $Q_j=z_jQ_i$ and $Q_k=z_kQ_i$ respectively. 

The extension of the angular-ordered shower approximation to heavy quark
processes is described in ref.~\cite{Marchesini:1989yk}.

\subsubsection{Initial conditions}\label{app:init}
The initial conditions for parton showering on a given line $i$ are
determined by the invariant quantity $E_{ij}^2=p_i\cdot p_j$, where $j$
is the colour partner of $i$.  If $i$ is a gluon line, it has two colour
partners; one of these is selected at random, with equal probability.
The showering on any line $i$ is performed in a standard frame,
in which the initial energy of $i$ is $E_0=E_{ij}$, its direction
is along the $z$-axis while that of the colour partner $j$ is in the
$(xz)$-plane, and the  upper limit on the angular evolution variable
is $\xi_0=1$. Thus the initial scale for the shower is $Q=E_0$.

In the FCR processes $gg\to Q\bQ$, when the colour flow is the
$t$-flow defined in app.~\ref{app:col}
we have $E_0^2=-\bt/2$ for the outgoing heavy quarks, while
for each incoming gluon we choose (separately) between $E_0^2=-\bt/2$
and $E_0^2=\bs/2$ with equal probability. When the colour flow is the
$u$-flow, $\bt$ is replaced by $\bu$. In the process
$q\bq\to Q\bQ$, we have $E_0^2=-\bt/2$ for all the parton showers.

\subsubsection{Shower algorithm}
In all parton branchings, the leading-order massless splitting
functions are used.  In timelike (final-state) showering,
the allowed region of $z$ for the branching $q\to qg$ is
$Q_q/Q < z < 1-Q_g/Q$ where $Q_q=m_q+${\tt VQCUT} and
$Q_g=m_g+${\tt VGCUT}, $m_q$ ($m_{u,d,s}$[0.32,0.32,0.5]) and $m_g$[0.75] 
being the quark and gluon effective masses, and {\tt VQCUT}[0.48],
{\tt VGCUT}[0.10] minimum virtuality parameters. The argument of
$\as$ is $z(1-z)Q$, which is an approximation to the $\kt$ of the
emitted gluon\footnote{The missing factor of $\sqrt 2$ is absorbed 
into the definition of the QCD scale $\Lambda$.} -- see eq.~(\ref{eq:ktsq}). 
The corresponding Sudakov form factor is thus
\beq\label{eq:sud}
\Delta(Q) = \exp\left(-\frac{C_F}{2\pi}\int_{Q_q+Q_g}^Q\frac{dQ'}{Q'}
\int_{Q_q/Q'}^{1-Q_g/Q'}dz\,\as(z(1-z)Q')\frac{1+z^2}{1-z}\right)\;.
\eeq
The next value $Q'$ for the evolution
variable is selected by solving $\Delta(Q')=\Delta(Q)/R$, where $Q$ is
the current value and $R\in [0,1]$ is a random number. Branching
stops whenever $R<\Delta(Q)$ is selected.  The minimal value of
$Q$ is $Q_q+Q_g$.

Spacelike (initial-state) showering is performed backwards,
i.e.\ starting from the hard process.  The evolution equation becomes
\beq
\Delta(Q')/f(x,Q')=[\Delta(Q)/f(x,Q)]/R\,,
\eeq
where $f$ is the relevant PDF and $x$ is the current value of the energy
fraction of the spacelike parton. A compensating factor of $f(x/z,Q)$
multiples the probability distribution in $z$.  This ``guides'' the
parton distribution to follow the input PDF.  For example, since
$f(x/z,Q)=0$ for $z<x$, branching to $x$ values above 1 is prohibited.

Perturbative branching of the spacelike parton in an initial-state shower
stops when a value of $Q <$ {\tt QSPAC}[2.5]\footnote{The default value
is relatively high because the input PDF parametrizations
may be unreliable (even negative) at lower scales.} is selected.
However, if the parton is not a valence constituent
of the corresponding beam particle, further non-perturbative branching
is forced until a valence parton is generated.  This is done in order
to have a simple model of the beam remnant, composed of the other valence
constituents carrying the remainder of the beam momentum.

\subsubsection{Azimuthal correlations}\label{sec:azcorr}
After the shower, all the kinematics can be reconstructed from the values
of $z$ and $\xi$, except for the azimuthal angles. The distribution of
these is isotropic unless dictated otherwise by the logical parameters
{\tt AZSOFT}[{\tt .TRUE.}] and {\tt AZSPIN}[{\tt.TRUE.}].
If {\tt AZSOFT} is {\tt .TRUE.}, the azimuth of each branching is
distributed according to the eikonal formula within the cone defined by the
previous branching.  In addition, if {\tt AZSPIN} is also {\tt .TRUE.},
the azimuthal correlation in gluon branching due to gluon polarization is
included. Polarization correlations for quarks are neglected, since they
vanish due to helicity conservation in the massless limit.

In the processes $gi\to Q\bQ i$, where $i=q,\bq$ or $g$,
there is an azimuthal correlation between the scattering plane
and the plane of virtual gluon ($g^*$) emission in initial-state
branching, of the form~\cite{Frixione:1995ms} (in the collinear limit)
\beq
{\cal M}\propto \ME{gg}\,P^{(0)}_{gi}(z)
+ C_{gg}\,Q_{g^*i}(z)\cos 2\phi\,,
\eeq
where $\phi$ is the azimuthal angle between the planes,
$P^{(0)}_{gi}$ is the leading-order $i\to g$ splitting function,
\beq\label{eq:Cgg}
C_{gg}=-\frac{g^4 N}{N^2-1}\,\frac{1}{2\bs}
\left(\frac{\bu}{\bt} +\frac{\bt}{\bu}
-\frac 1{N^2}\frac{\bs^2}{\bt\bu}\right)\left(
\frac{m^2}{\bs}-\frac{m^4}{\bt\bu}\right),
\eeq
and
\beq
Q_{g^*i}(z) = -4 C_i\left(\frac{1-z}{z}\right),
\eeq
with $C_g = C_A = N$ and $C_q = C_{\bq} = C_F = (N^2-1)/2N$.

Since this correlation vanishes in the limit $m\to 0$, it is also neglected
in \HW. However, this means that there is matching between the parton
showers and the matrix elements in the collinear limit only after azimuthal
averaging of the latter. This poses a problem in MC@NLO implementation,
whose solution is discussed in app.~\ref{sec:MCsubt}.

\subsubsection{Momentum reshuffling}\label{app:reshuf}
After showering, each external parton line in the hard process has become
a jet.  The parton momenta have to be replaced by the jet momenta in
such a way that energy-momentum is conserved, without seriously affecting
the dynamics.  This is achieved by the following momentum reshuffling
procedure.

Each final-state jet is first rotated from the standard showering frame 
(see sect.~\ref{app:init}) to the direction of the corresponding parton in 
the hard process c.m.\ frame, with a rotation about the jet axis to give the 
correct correlation with the direction of the colour partner.
The magnitudes of the jet three-momenta in the hard process c.m.\ frame
are computed as follows.  Suppose initially that each jet were given
a three-momentum equal to that of the parton it replaces. Since the
jet masses are not equal to the parton masses,  this would violate
energy-momentum conservation.   However, energy-momentum conservation
can be restored by rescaling the outgoing parton three-momenta in the
hard process c.m.\ frame by a common overall factor, to obtain the jet
three-momenta in that frame.  Once this factor has been computed, each
jet is boosted along its axis to the required three-momentum.
Note that the overall four-momentum of the hard process is not
changed by this procedure.

In the case of initial-state jets, the partons connected to the incoming
beam particles are aligned with the beam directions (smeared by some
intrinsic $\pt$ if that is requested through the input parameter
{\tt PTRMS}[0]).  Each of the two initial-state jets is then boosted
longitudinally (i.e. along the beam directions)
until the (now off-shell) momenta of the partons
entering the hard process are consistent with the original kinematics.
This is not a unique procedure, since not all kinematic variables can be
restored to the values they had before showering.  There are at present
two options, controlled by the logical parameter
{\tt PRESPL}[{\tt .TRUE.}]:
\begin{itemize}
\item If {\tt PRESPL} is {\tt .TRUE.}, then the c.m.\ energy and longitudinal 
momentum of the hard process are preserved (``$p$-scheme'');
\item If {\tt PRESPL} is {\tt .FALSE.}, then the c.m.\ energy and rapidity of 
the hard process are preserved (``$y$-scheme'').
\end{itemize}

Finally the transverse momenta of the initial-state partons entering the
hard process are combined to give the transverse momentum of the hard
process, and all the final-state jets are boosted transversely, by the
amount required for transverse momentum conservation.

In the previous \HW\ version (6.4) the longitudinal and transverse
boosts were combined into a single boost in the direction of the
new hard process momentum. However, this makes the matching of
MC and NLO very complicated.  Therefore the two boosts are performed
separately in version 6.5. The difference between the
two procedures corresponds to a rotation in the hard process
c.m.\ frame (Thomas precession).

One can argue as follows that the effects of momentum reshuffling
are beyond the next-to-leading order: to give at least one jet a mass
requires one power of $\as$, and the average mass-squared is of order
$\as$. Thus the fractional change in observables due to momentum
reshuffling is expected to be of relative order $\as^2$.

\subsection{Hadronization}\label{sec:hadrize}
\subsubsection{Cluster formation}\label{sec:Cluform}

After the perturbative parton showering and momentum reshuffling,
all outgoing gluons
are split non-pertur\-batively, into light quark-antiquark or
diquark-antidiquark pairs (the default option is to disallow diquark
splitting). At this point, each jet consists of a set of outgoing quarks
and antiquarks (also possibly some diquarks and antidiquarks) and,
in the case of spacelike jets, a single incoming valence quark or antiquark.
The latter is replaced by an outgoing spectator carrying the opposite 
colour and the residual flavour and momentum of the corresponding beam hadron.

In the limit of a large number of colours, each final-state colour line
can now be followed from a quark/anti-diquark to an antiquark/diquark
with which it can form a colour-singlet cluster. By virtue of the
preconfinement property of the shower~\cite{Amati:1979fg},
these clusters have a distribution
of mass and spatial size that peaks at low values, falls rapidly for large
cluster masses and sizes, and is asymptotically independent of the hard
subprocess type and scale.

The clusters thus formed are fragmented into hadrons.
If a cluster is too light to decay into two hadrons, it is taken to
represent the lightest single hadron of its flavour. Its mass is shifted
to the appropriate value by an exchange of 4-momentum with a neighbouring
cluster in the jet.  Similarly, any diquark-antidiquark clusters with
    masses below threshold for decay into
  a baryon-antibaryon pair are shifted to the  threshold via a transfer of
    4-momentum to a neighbouring cluster.

Those clusters massive enough to decay into two hadrons, but below
a fission threshold to be specified below, decay isotropically%
\footnote{Except for those containing a `perturbative' quark when
{\tt CLDIR}=1 -- see below.}
into pairs of hadrons selected in the following way. A flavour $f$ is
chosen at random from among $u$, $d$, $s$, the six corresponding diquark
flavour combinations, and $c$. For a cluster of flavour $f_1 \bar{f_2}$, this
specifies the flavours $f_1 \bar{f}$ and $f \bar{f_2}$ of the decay
products, which are then selected at random from tables of hadrons of
those flavours.
The selected choice of decay products is accepted in proportion
to the density of states (phase space times spin degeneracy) for that
channel. Otherwise, $f$ is rejected and the procedure is repeated.

A fraction of clusters have masses too
high for isotropic two-body decay to be a reasonable ansatz, even
though the cluster mass spectrum falls rapidly (faster than any power) at
high masses. These clusters are fragmented using an iterative fission 
model until the masses of the fission products fall below the fission
threshold. In the fission model the
produced flavour $f$ is limited to $u$, $d$ or $s$ and the product clusters
$f_1 \bar{f}$ and $f \bar{f_2}$ move in the directions of the original
constituents $f_1$ and $\bar{f_2}$ in their c.m.\ frame. Thus the fission
mechanism is not unlike string fragmentation~\cite{Andersson:ia}.

In \HW\ there are three main fission parameters,
{\tt CLMAX}[3.35], {\tt CLPOW}[2] and {\tt PSPLT}[1].
The maximum cluster mass parameter {\tt CLMAX} and {\tt CLPOW} specify the
fission threshold $M_f$ according to the formula
\beq
M_f^{{\tt CLPOW}} = {\tt CLMAX}^{{\tt CLPOW}}+ (m_1 + m_2)^{{\tt CLPOW}}\,,
\eeq
where $m_1$ and $m_2$ are the quark mass ({\tt RMASS})
parameters for flavours $f_1$ and $f_2$.
The parameter {\tt PSPLT} specifies the mass spectrum of the
produced clusters, which is taken to be $M^{{\tt PSPLT}}$ within the allowed
phase space. Provided the parameter {\tt CLMAX} is not chosen too small,
the gross features of events are insensitive to the details of the
fission model, since only a small fraction of clusters undergo fission.
However, the production rates of high-$\pt$ or heavy particles (especially
baryons) are affected, because they are sensitive to the tail of the cluster
mass distribution.  Reducing {\tt CLPOW} increases the
yield of heavier clusters (and hence of baryons) for heavy quarks, without
affecting light quarks much.  For example, the default value gives no
$b$-baryons (for the default value of {\tt CLMAX}) whereas {\tt CLPOW}=1.0
makes the ratio of $b$-baryons to $b$-hadrons about 1/4.

There is also a switch {\tt CLDIR}[1] for cluster decays. For the default
value, a cluster that contains a `perturbative' quark, i.e.\ one coming from
the perturbative stage of the event (the hard process or perturbative gluon
splitting) `remembers' its direction. Thus when the cluster decays, the hadron
carrying its flavour continues in the same direction (in the cluster c.m.\ 
frame) as the quark. This considerably hardens the spectrum of heavy hadrons,
particularly of $c$- and $b$-flavoured hadrons. {\tt CLDIR}=0 turns off this
option, treating clusters containing quarks of perturbative and
non-perturbative origin equivalently.
  
In the {\tt CLDIR}=1 option, the parameter {\tt CLSMR}[0] allows for a
Gaussian smearing of the direction of the perturbative quark momentum.  The
smearing is actually exponential in $(1-\cos\theta)$ with mean value 
{\tt CLSMR}.  Thus increasing {\tt CLSMR} decorrelates the cluster decay 
from the initial quark direction.

\subsubsection{$b$-quark hadronization}\label{sec:bhad}

The process of $b$-quark hadronization requires special treatment and
the results obtained using \HW\ are still not fully satisfactory.
Generally speaking, it is difficult to obtain a sufficiently hard B-hadron
spectrum and the observed $b$-meson/$b$-baryon ratio. These depend not only
on the perturbative subprocess and parton shower but also on
non-perturbative issues such as
the fraction of $b$-flavoured clusters that become a single B
meson, the fractions that decay into a B meson and another meson, or into
a $b$-baryon and an antibaryon, and the fraction that are split into
more clusters. Thus the properties of $b$-jets depend on the parameters
{\tt RMASS(5)}, {\tt CLMAX}, {\tt CLPOW} and {\tt PSPLT} in a rather
complicated way. In practice these parameters are tuned to global
final-state properties and one needs extra parameters to describe $b$-jets.

A parameter {\tt B1LIM}[0] has therefore been introduced to allow clusters
somewhat above the B$\pi$ threshold mass $M_{th}$ to form a single B meson
if
\beq
M < M_{lim} = (1+{\tt B1LIM}) M_{th}\;.
\eeq
The probability of such single-meson clustering is assumed to decrease
linearly for $M_{th} < M < M_{lim}$.
This has the effect of hardening the B spectrum if {\tt B1LIM} is increased
from the default value.

In addition, in \HW\ version~6, the parameters {\tt PSPLT}, {\tt CLDIR} and
{\tt CLSMR} have been converted into two-dimensional arrays, with the first
element controlling clusters that do not contain a $b$-quark and
the second those that do. Thus tuning of $b$-fragmentation can now
be performed separately from other flavours, by setting
{\tt CLDIR(2)}=1 and varying {\tt PSPLT(2)} and {\tt CLSMR(2)}.
By reducing the value of {\tt PSPLT(2)}, further hardening of the
B-hadron spectrum can be achieved.

\begin{figure}[htb]
\begin{center}
\epsfig{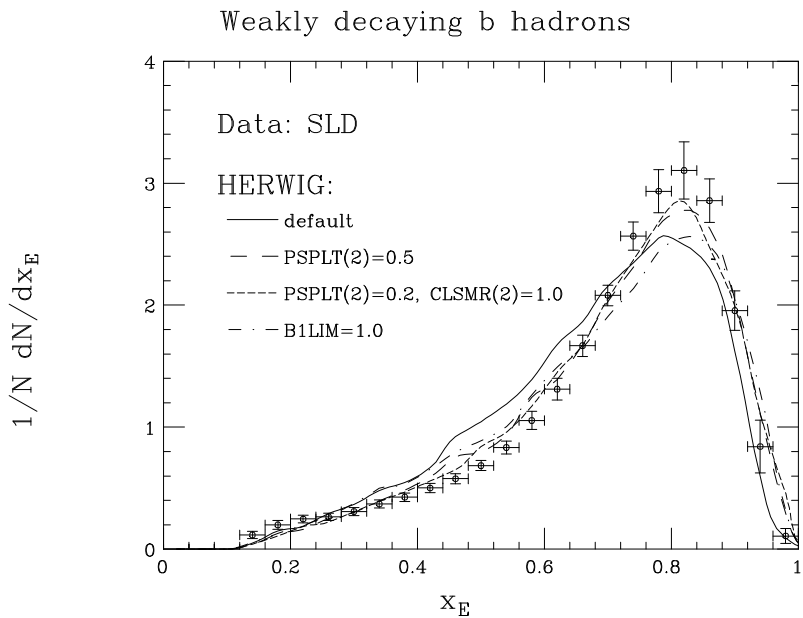}
\end{center}
\caption{\label{fig:bfrag}
Effect of \HW\ parameters on the $b$ fragmentation function.}
\end{figure}

Figure~\ref{fig:bfrag} shows the $b$ fragmentation function in $Z^0$ 
decay, i.e.\ the distribution of the energy fraction $x_E=2E_B/M_Z$ 
in $e^+e^-$ collisions, where $E_B$ is
the energy of a weakly-decaying $b$-flavoured hadron in the $Z^0$
rest frame.  The \HW\ predictions for different values of some of the
parameters discussed above are compared with the data of the
SLD Collaboration~\cite{Abe:2002iq}. For the MC@NLO predictions
in sect.~\ref{sec:bot}, parameters were left at their default values
except for {\tt PSPLT(2)}=0.5, corresponding to the long-dashed curve.
However, as explained in sect.~\ref{sec:bot}, all the results presented
there are at the $b$-quark level, and are therefore insensitive to the
$b$-hadronization parameters.

\section{MC subtraction terms}\label{sec:MCsubt}
In this section, we derive the MC subtraction terms $d\bSigma_{ab}\xMC$
needed in order to define MC@NLO, eq.~(\ref{eq:rMCatNLO}). We follow 
$\FW$, sect.~A.5. In the present case, the ${\cal O}(\as^3)$ 
term in the expansion of the result of \HW\ is given in
eq.~(\ref{MCatas}), which we report again here
\beq
d\sigma\xMCB=\sum_{ab}\sum_{L}\sum_{l}d\sigma_{ab}^{(L,l)}\xMCB .
\label{appMCatas}
\eeq
This equation is identical to \FWeq{A.58}, except for the fact that
more emitting legs and colour structures have been considered.

We start from the case of initial-state radiation, considering emission
from leg 1 for definiteness. We can rewrite the relevant part of \FWeq{A.58}) 
as follows:
\beqn
d\sigma_{ab}^{(+,l)}\xMCB&=&d\bx_{1i} d\bx_{2i} 
f_a^{\Hone}(\bx_{1i}/z^{(l)}_+)f_b^{\Htwo}(\bx_{2i})
\MEc{cb}{l}\left(\bx_{1i},\bx_{2i},\bt_+\right) 
d\phi_2\left(\bx_{1i}\bx_{2i} S\right)
\nonumber \\*&\times&
\frac{d\xi^{(l)}_+}{\xi^{(l)}_+}\frac{dz^{(l)}_+}{z^{(l)}_+}
\frac{d\varphi_+}{2\pi}
\frac{\as}{2\pi} P^{(0)}_{ca}(z^{(l)}_+)
\stepf\left((z_+^{(l)})^2-\xi_+^{(l)}\right),
\label{eq:sMCplus}
\eeqn
where we have explicitly indicated the integration over the azimuthal
angle $\varphi_+$ of the branching parton (which is trivial here but
will not be so in the following). 
We have used the findings of sects.~\ref{sec:ISR} 
and~\ref{sec:HW_ini}. The momentum fractions of the partons entering
the $\twototwo$ hard process are $\bx_{1i}$, and $\bx_{2i}$, given in
eqs.~(\ref{eq:ini_x12_p}) and~(\ref{eq:ini_x12_y}) in the $p$- and
$y$-scheme respectively (see sect.~\ref{app:reshuf}). The $t$-channel
$\twototwo$ invariant is $\bt_+$, given in eq.~(\ref{eq:ini_inv}).
The set \mbox{$(\bx_{1i},\bx_{2i},\bt_+)$} fully specifies the 
momenta of the $\twototwo$ hard process in the lab frame. In fact,
the $s$-channel invariant is
\beq
\bs=\bx_{1i} \bx_{2i} S\equiv \bs_+ ,
\eeq
where $\bs_+$ is given in eq.~(\ref{eq:ini_inv}), and the last equality
holds thanks to eq.~(\ref{eq:ini_x1x2}). With $\bs_+$, $\bt_+$, and
$\bu_+=-\bs_+ -\bt_+$ one gets immediately the $\twototwo$ momenta in
the hard c.m.\ frame:\footnote{In order to simplify the notation, the momenta
have been rotated in the transverse plane along the direction 
corresponding to $\varphi_+=0$.}
\beq
\bp_{1,2} = \bE(1,0,0,\pm 1)\;,\;\;\;
\bk_{1,2} = (\bE,\pm\bkT,0,\pm\bkL)\;,
\label{eq:app2to2cm}
\eeq
where from eq.~(\ref{eq:stutwo})
\beq
\bE=\half\sqrt{\bs_+}\;,\;\;
\bkT=\sqrt{\frac{\bt_+\bu_+}{\bs_+}-m^2}\;,\;\;
\bkL=\frac{\bt_+-\bu_+}{2\sqrt{\bs_+}}\;.
\label{eq:Ektkl}
\eeq
Finally, the momenta in eq.~(\ref{eq:app2to2cm}) are boosted to the lab
frame with
\beq
y_{\sss CM}^{(i)}=\half\log\frac{\bx_{1i}}{\bx_{2i}}\,.
\label{eq:ybstini}
\eeq
In eq.~(\ref{eq:sMCplus}), $\MEc{cb}{l}$ is the Born matrix element squared, 
summed over spin and colour degrees of freedom (the average factor for 
initial-state ones is understood), times the flux factor, times a factor
which depends on the partonic process and the colour flow. When ${cb}={q\bq}$
this factor is one; when ${cb}={gg}$, it is one for $l=s$, whereas for
$l=t,u$ we have $\MEc{cb}{l}=\dsb_{gg}^{(l)}/d\phi_2$, with $\dsb_{gg}^{(l)}$
given in eq.~(\ref{eq:dsbtu}). $\MEc{cb}{l}$ depends on $\bx_{1i}$ and 
$\bx_{2i}$ only through their product; however, we use the notation of
eq.~(\ref{eq:sMCplus}) order to remember the boost of eq.~(\ref{eq:ybstini}).

We now perform some changes of integration variables in eq.~(\ref{eq:sMCplus}).
We replace $(\bx_{1i},\bx_{2i})$ with $(x_1,x_2)$, the fractional parton
momenta of the $\twotothree$ process. We also replace 
$(z^{(l)}_+,\xi^{(l)}_+,\varphi_+)$ with $(\xin,\yin,\vpin)$. Here, we
do not need to define these variables precisely; they are used in order
to eliminate the dependence upon the index $l$ in the integration
measure, and will be related to the three-body phase-space variables in
the following. Finally, we use the fact that, in the c.m.\ frame of the
$\twototwo$ hard process, the two-body phase space can be written as follows
\beq
d\phi_2(s)=\frac{\bb(s)}{16\pi}d\cos\thin\,,
\label{eq:twobps}
\eeq
where the function $\bb(\bs)$ is given in eq.~(\ref{eq:bbdef}). We thus get
\beq
d\sigma_{ab}^{(+,l)}\xMCB=
{\cal L}_{ab}^{(+,l)}\,{\cal B}_{ca}^{(+,l)}\,{\cal H}_{cb}^{(+,l)}\,
dx_1\,dx_2\,d\xin\,d\yin\,d\vpin\,d\cos\thin\,.
\label{eq:sMCplusch}
\eeq
The factor ${\cal L}$ contains the dependence upon parton densities
\beq
{\cal L}_{ab}^{(+,l)}=\frac{\partial(\bx_{1i},\bx_{2i})}{\partial(x_1,x_2)}
\frac{1}{z^{(l)}_+} f_a^{\Hone}(\bx_{1i}/z^{(l)}_+)f_b^{\Htwo}(\bx_{2i}).
\eeq
The actual form of the Jacobean that appears in this equation depends on
the momentum reshuffling scheme (see app.~\ref{app:reshuf}). By explicit
computation, using eqs.~(\ref{eq:ini_x12_p}) and (\ref{eq:xydef}) we
find, in the $p$-scheme, 
\beq
\frac{\partial(\bx_{1i},\bx_{2i})}{\partial(x_1,x_2)}
= \frac{xx_+}{\sqrt{x_+^2 - x_1x_2v_1v_2/s^2}}\,.
\label{eq:xzjacpsch}
\eeq
In the $y$-scheme, we use eqs.~(\ref{eq:ini_x12_y}) and (\ref{eq:xydef})
to obtain
\beq
\frac{\partial(\bx_{1i},\bx_{2i})}{\partial(x_1,x_2)} = x\,.
\label{eq:xzjacysch}
\eeq
The factor ${\cal B}$ collects the terms related to parton branching
\beq
{\cal B}_{ca}^{(+,l)}=\frac{1}{2\pi}
\frac{\partial(z^{(l)}_+,\xi^{(l)}_+,\varphi_+)}
{\partial(\xin,\yin,\vpin)}
\frac{\as}{2\pi} 
\frac{P^{(0)}_{ca}(z^{(l)}_+)}{\xi^{(l)}_+}
\stepf\left((z_+^{(l)})^2-\xi_+^{(l)}\right).
\eeq
In contrast to the case of $\FW$ (see \FWeq{A.72}) the Jacobean has 
in general a non-trivial dependence upon the azimuthal angles; this is 
so in the cases $l=t,u$, and it is due to the complicated dependence
of the \HW\ scale $E_0$ upon the phase-space variables.
Finally, the factor ${\cal H}$ contains the full information on
the hard $\twototwo$ process, and phase-space factors
\beq
{\cal H}_{cb}^{(+,l)}=\frac{\bb(\bx_{1i}\bx_{2i}S)}{16\pi}
\MEc{cb}{l}\left(\bx_{1i},\bx_{2i},\bt_+\right).
\eeq

We now turn to the case of final-state branching, considering emission
from the heavy quark. The analogue of eq.~(\ref{eq:sMCplus}) reads
\beqn
d\sigma_{ab}^{(Q,l)}\xMCB&=&d\bx_{1f} d\bx_{2f} 
f_a^{\Hone}(\bx_{1f})f_b^{\Htwo}(\bx_{2f})
\MEc{ab}{l}\left(\bx_{1f},\bx_{2f},\bt_Q\right) 
d\phi_2\left(\bx_{1f}\bx_{2f} S\right)
\nonumber \\*&\times&
\frac{d\xi^{(l)}_Q}{\xi^{(l)}_Q} dz^{(l)}_Q
\frac{d\varphi_Q}{2\pi}
\frac{\as}{2\pi} P^{(0)}_{qq}(z^{(l)}_Q)
\stepf\left((z_Q^{(l)})^2-\frac{2m^2}{|\bl_Q|\xi_Q^{(l)}}\right).
\label{eq:sMCQ}
\eeqn
Here, the results of sects.~\ref{sec:FSR} and~\ref{sec:HW_fin} have to be
used. The momentum fractions of the incoming partons are not affected by
final-state branching, and eq.~(\ref{eq:fin_x12}) holds; therefore
\beq
\bs=\bx_{1f} \bx_{2f} S = s\equiv \bs_Q .
\label{eq:bsfin}
\eeq
The $t$-channel invariant is given in eq.~(\ref{eq:fin_inv}). The momenta
of the hard $\twototwo$ process are given in eq.~(\ref{eq:app2to2cm}) in the 
parton c.m.\ frame, with $\bE$, $\bkT$, and $\bkL$ computed as in 
eqs.~(\ref{eq:Ektkl}) after the formal replacements 
$(\bs_+,\bt_+,\bu_+)\to (\bs_Q,\bt_Q,\bu_Q)$. These momenta are subsequently
boosted to the lab frame by means of
\beq
y_{\sss CM}^{(f)}=\half\log\frac{\bx_{1f}}{\bx_{2f}}\,.
\label{eq:ybstfin}
\eeq
Performing changes of variables analogue to those in eq.~(\ref{eq:sMCplus}),
we get the analogue of eq.~(\ref{eq:sMCplusch})
\beq
d\sigma_{ab}^{(Q,l)}\xMCB=
{\cal L}_{ab}^{(Q,l)}\,{\cal B}^{(Q,l)}\,{\cal H}_{ab}^{(Q,l)}\,
dx_1\,dx_2\,d\xout\,d\yout\,d\vpout\,d\cos\thout\,,
\label{eq:sMCQch}
\eeq
where now\footnote{For final-state emission the Jacobean factor
for the Bjorken $x$'s is trivial, see eq.~(\ref{eq:fin_x12}).} 
\beqn
{\cal L}_{ab}^{(Q,l)}&=&f_a^{\Hone}(\bx_{1f})f_b^{\Htwo}(\bx_{2f}),
\label{eq:lumifin}
\\
{\cal B}^{(Q,l)}&=&\frac{1}{2\pi}
\frac{\partial(z^{(l)}_Q,\xi^{(l)}_Q,\varphi_Q)}
{\partial(\xout,\yout,\vpout)}
\frac{\as}{2\pi} 
\frac{P^{(0)}_{qq}(z^{(l)}_Q)}{\xi^{(l)}_Q}
\stepf\left((z_Q^{(l)})^2-\frac{2m^2}{|\bl_Q|\xi_Q^{(l)}}\right),
\\
{\cal H}_{ab}^{(Q,l)}&=&\frac{\bb(\bx_{1f}\bx_{2f}S)}{16\pi}
\MEc{ab}{l}\left(\bx_{1f},\bx_{2f},\bt_Q\right).
\label{eq:hardfin}
\eeqn

Equation~(\ref{eq:sMCplusch}) and~(\ref{eq:sMCQch}) are closely related
to the MC subtraction terms $d\bSigma_{ab}\xMC/d\phi_3$ needed for the
definition of MC@NLO. We use \FWeq{A.61} 
\beq
d\sigma_{ab}^{(+,l)}\xMC=dx_1 dx_2 
\frac{d\bSigma_{ab}^{(+,l)}}{d\phi_3}\xMCBB
d\phi_3\,.
\eeq
In order to proceed, we introduce an explicit parametrization for the
three-body phase space. We use (see \FWeq{A.10})
\beq
d\phi_3(s)=\frac{s\bb(xs)}{1024\pi^4}(1-x)dx\,dy\,d\vpq\,d\cos\thq\,,
\label{eq:thrbps}
\eeq
where $\thq$ is the scattering angle of $Q$ in the $Q\bQ$ c.m.\ frame. 
The variables $x$, $y$ and $\vpq$ refer to the emitted parton in the parton 
c.m.\ frame of the $\twotothree$ process: $(1-x)$ is the energy in units 
$\sqrt{s}/2$, and $y$ and $\vpq$ are the cosine of the polar angle and the
azimuthal angle respectively; $x$ and $y$ can be expressed in
terms of invariants as shown in eq.~(\ref{eq:xydef}).
The ranges of the angular variables are $0\leq\thq\leq\pi$, 
$0\leq\vpq\leq 2\pi$. The parametrization of eq.~(\ref{eq:thrbps})
is identical to that of eq.~(2.12) of ref.~\cite{Mangano:jk}, provided 
that a (physically irrelevant) rotation of $\vpq-\theta_2$ is performed
there, in the transverse plane in the $Q\bQ$ c.m.\ frame. The momenta 
parametrizations of eqs.~(2.8) of that paper can thus be used here.
We use eq.~(\ref{eq:thrbps}) to rewrite eqs.~(\ref{eq:sMCplusch}) 
and~(\ref{eq:sMCQch}) as follows
\beqn
\frac{d\bSigma_{ab}^{(+,l)}}{d\phi_3}\xMCBB&=&
{\cal L}_{ab}^{(+,l)}\,{\cal B}_{ca}^{(+,l)}\,{\cal H}_{cb}^{(+,l)}\,,
\label{eq:sMCplustwo}
\\
\frac{d\bSigma_{ab}^{(Q,l)}}{d\phi_3}\xMCBB&=&
{\cal L}_{ab}^{(Q,l)}\,{\cal B}^{(Q,l)}\,{\cal H}_{ab}^{(Q,l)}\,,
\label{eq:sMCQtwo}
\eeqn
where we have chosen $(\xin,\yin,\vpin)$ and $(\xout,\yout,\vpout)$ 
to coincide with $(x,y,\vpq)$, and\footnote{Here and in what follows
we have some abuse of notation, since we denote by the same symbols
${\cal L}$, ${\cal B}$, and ${\cal H}$ quantities which possibly differ
by multiplicative factors from those previously introduced.}
\beqn
{\cal L}_{ab}^{(+,l)}&=&\frac{\partial(\bx_{1i},\bx_{2i})}{\partial(x_1,x_2)}
\frac{1}{z^{(l)}_+} f_a^{\Hone}(\bx_{1i}/z^{(l)}_+)f_b^{\Htwo}(\bx_{2i}),
\label{eq:Lin}
\\
{\cal B}_{ca}^{(+,l)}&=&\as
\frac{\partial(z^{(l)}_+,\xi^{(l)}_+,\varphi_+)}
{\partial(x,y,\vpq)}
\frac{P^{(0)}_{ca}(z^{(l)}_+)}{(1-x)\xi^{(l)}_+}
\stepf\left((z_+^{(l)})^2-\xi_+^{(l)}\right),
\label{eq:Bin}
\\
{\cal H}_{cb}^{(+,l)}&=&\frac{16\pi}{x_1 x_2 S}
\frac{d\cos\thin}{d\cos\thq}
\MEc{cb}{l}\left(\bx_{1i},\bx_{2i},\bt_+\right),
\label{eq:Hin}
\\
{\cal L}_{ab}^{(Q,l)}&=&f_a^{\Hone}(x_1)f_b^{\Htwo}(x_2),
\label{eq:Lout}
\\
{\cal B}^{(Q,l)}&=&\as
\frac{\partial(z^{(l)}_Q,\xi^{(l)}_Q,\varphi_Q)}
{\partial(x,y,\vpq)}
\frac{P^{(0)}_{qq}(z^{(l)}_Q)}{(1-x)\xi^{(l)}_Q}
\stepf\left((z_Q^{(l)})^2-\frac{2m^2}{|\bl_Q|\xi_Q^{(l)}}\right),
\label{eq:Bout}
\\
{\cal H}_{ab}^{(Q,l)}&=&\frac{16\pi}{x_1 x_2 S}
\frac{\bb(x_1 x_2 S)}{\bb(\bx_{1i}\bx_{2i}S)}
\frac{d\cos\thout}{d\cos\thq}
\MEc{ab}{l}\left(x_1,x_2,\bt_Q\right).
\label{eq:Hout}
\eeqn
Here, we have used $xs=\bx_{1i}\bx_{2i}S$ and $s=\bx_{1f}\bx_{2f}S$.
The quantities $\bx_{1i}$ and $\bx_{2i}$ depend on $x_1$, $x_2$, and
the invariants; thus, the Jacobean in eq.~(\ref{eq:Lin}) must be
computed at fixed $(x,y,\vpq,\thq)\equiv\phi_3$. Furthermore, the 
Jacobeans in eq.~(\ref{eq:Bin}) and~(\ref{eq:Bout}) have to be 
computed at fixed $\thin$ and $\thout$ respectively.

The MC subtraction terms appear twice in the definition of 
MC@NLO, eq.~(\ref{eq:rMCatNLO}), since they contribute to the
weights of $\clH$ and $\clS$ events. These weights are treated
as ordinary MC weights in the MC evolutions, whose generating
functionals are ${\cal F}_{\mbox{\tiny MC}}^{(3)}$ and
${\cal F}_{\mbox{\tiny MC}}^{(2)}$ respectively. Implicit
in these generating functionals is also the dependence upon
the initial conditions for the showers, i.e. the momenta of
the $\twotothree$ and $\twototwo$ processes. As the integral in
eq.~(\ref{eq:rMCatNLO}) requires, these momenta need to be
specified for each point in the integration range $(x_1,x_2,\phi_3)$.
This is straightforward in the case of $\clH$ events, after a definite
parametrization is chosen for the angles $\thin$ and $\thout$, which 
allows the computation of the Jacobeans that appear in eqs.~(\ref{eq:Hin})
and~(\ref{eq:Hout}).

The situation is more involved in the case of $\clS$ events. Since here
the initial condition for the shower is a $\twototwo$ process, a mapping
of the three-body phase space onto the two-body one is necessary.
This map is provided by the MC, which relates the momenta of the
partons entering the $\twototwo$ hard process to those of the partons 
emerging from the shower. In the case of a single branching, this map has
been denoted by $\EVprjmap$ in sect.~\ref{sec:rev}. In this section,
we have given explicit implementations of $\EVprjmap$: in the case
of initial-state emission, one uses 
eqs.~(\ref{eq:app2to2cm})--(\ref{eq:ybstini}), where the $\twototwo$
invariants are expressed in terms of the $\twotothree$ ones as explained
in sect.~\ref{sec:ISR}. For final-state emission, analogous relations
hold (see the discussion after eq.~(\ref{eq:bsfin})). These maps
also provide us with the definitions of the scattering angles $\thin$
and $\thout$ in terms of the three-body phase-space variables. In
fact, the $t$-channel invariant is always given by the function
(see eq.~(\ref{eq:tuinv}))
\beq
\bt(\bs,\bh) = -\half\bs (1-\bb(\bs)\cos\bh)\,,
\eeq
$\bs$ and $\bh$ being the $s$-channel invariant and scattering angle
in the hard c.m.\ frame respectively. Thus, taking into account 
eq.~(\ref{eq:ini_bs}), and the fact that $\bs_Q=s$, we have
\beqn
&&\bt(s+v_1+v_2,\thin)=\bt_+\,,
\label{eq:solthin}
\\
&&\bt(s,\thout)=\bt_Q\,,
\label{eq:solthout}
\eeqn
which can be solved for $\thin$ and $\thout$ using 
eqs.~(\ref{eq:ini_inv}) and~(\ref{eq:fin_inv}) respectively.

However, eqs.~(\ref{eq:solthin}) and~(\ref{eq:solthout}) manifestly 
define two {\em different} maps $\EVprjmapi$ and $\EVprjmapo$, whereas
eq.~(\ref{eq:rMCatNLO}) requires $\EVprjmapo=\EVprjmapi\equiv\EVprjmap$.
In fact, if $\EVprjmapo\ne\EVprjmapi$, for a given three-body configuration
$(x_1,x_2,\phi_3)$ one would get two two-body configurations 
generated by $\EVprjmapi$ and $\EVprjmapo$, and the choice of which one 
to use in  ${\cal F}_{\mbox{\tiny MC}}^{(2)}$ as initial condition for 
the shower would result in an ambiguity in MC@NLO. Besides, the 
complicated functional relations implicit in eq.~(\ref{eq:solthin}) 
and~(\ref{eq:solthout}) would make the task of the implementation
of event projection in the NLO cross section a difficult one.

A solution for the first problem mentioned above is to treat initial-
and final-state emissions independently in MC@NLO. The NLO cross
section must also be written accordingly, since MC subtraction terms
act as local counterterms. This can be achieved for example using the 
technique of ref.~\cite{Frixione:1995ms}, which basically amounts to
partitioning the phase space into regions dominated by initial- or
final-state emissions.\footnote{Ref.~\cite{Frixione:1995ms} is based
on the subtraction method, so no approximation is involved in this
phase-space partition.} In terms of numerical accuracy and unweighting 
efficiency this may be the best strategy in those cases in which final-state
collinear emissions are singular at the level of short-distance cross
sections, such as in jet physics. However, it would still leave us with
the problem of an easy implementation of event projection. 

We argue that, regardless of the presence of final-state singularities 
in real matrix elements, the simplest possible definitions of the
maps $\EVprjmapi$ and $\EVprjmapo$ should be adopted. We now show
how this can be achieved; as a by product, we also show that, in
the present case, initial- and final-state emissions can indeed
be treated simultaneously.

We start by observing that momentum reshuffling is an effect beyond
NLO (see sect.~\ref{app:reshuf}). It follows that two $\twotothree$ 
kinematic configurations are equivalent at the NLO if they coincide 
in the limit in which the off-shell parton (i.e., the parton which
has branched) has its on-shellness restored. This is a consequence of
the fact that the branching procedure is exact only for zero-angle
emission. Thus, for a given $\twotothree$ configuration we can take its 
zero-angle-emission limit, and extract from there the $\twototwo$ momenta 
we seek. This is most easily done in the partonic c.m.\ frame, where
eq.~(\ref{eq:app2to2cm}) holds; eqs.~(\ref{eq:Ektkl}) are then used to
obtain $\bE$, $\bkT$, and $\bkL$, with the $\twototwo$ invariants expressed
in terms of the zero-angle limits of the $\twotothree$ invariants. The limits
can be computed by choosing any parametrization for the three-body
phase space; in the following, we shall use eq.~(\ref{eq:thrbps}),
exploiting the explicit formulae of ref.~\cite{Mangano:jk}.

For initial-state emission (say, from parton 1), from eq.~(\ref{eq:ini_inv})
we get, in the zero-angle-emission limit
\beqn
\bs_+^{(0)} &=& xs\,,
\nonumber \\
\bt_+^{(0)} &=& -\half xs \left[1-\bb(xs)\cos\thq\right]\,.
\label{eq:ini_invza}
\eeqn
Replacing $\bt_+$ with $\bt_+^{(0)}$ in eq.~(\ref{eq:solthin}), 
and solving for $\thin$, we get 
\beq
\cos\thin=\cos\thq\,.
\label{eq:solthinex}
\eeq 
This is what we expect, since for strictly collinear initial-state emission,
the parton $\twototwo$ c.m.\ frame (where $\thin$ is defined) and the $Q\bQ$
c.m.\ frame (where $\thq$ is defined) coincide.

For final-state emission from the quark leg, from eq.~(\ref{eq:fin_inv})
we get
\beqn
\bs_Q^{(0)} &=& s\,,
\nonumber \\
\bt_Q^{(0)} &=& -\half s \left[1-\bb(s)\cos\thout(x,\cos\thq)\right]\,,
\label{eq:fin_invza}
\eeqn
where
\beq
\cos\thout(x,\cos\thq)=
-\frac{1-x-(1+x)\cos\thq}{1+x-(1-x)\cos\thq}\,.
\label{eq:solthoutex}
\eeq
Note that
\beq
\cos\thout(1,\cos\thq)=\cos\thq\,,
\label{eq:bhvsthq}
\eeq
which is what one expects on physical grounds, since $x\to 1$ is the
soft limit in the parametrization of eq.~(\ref{eq:thrbps}), and it is
only in this limit that the parton $\twototwo$ c.m.\ frame and the $Q\bQ$ 
c.m.\ frame coincide in the case of final-state emission.

We can now use eqs.~(\ref{eq:solthinex}) and~(\ref{eq:solthoutex}) in
eqs.~(\ref{eq:Hin}) and~(\ref{eq:Hout}) respectively. This fully
defines the MC subtraction terms, at least in the case of $\clH$
events, where the initial condition for the MC shower is given
by the $\twotothree$ configuration, as specified by the integration
variables $(x_1,x_2,\phi_3)$. However, we still have the problem
that $\EVprjmapi\ne\EVprjmapo$. To see this explicitly, we write
the $\twototwo$ momenta that we obtain with the zero-angle-emission 
prescription:

\noindent
$\bullet$ for initial-state emission
\beq
\bp_{1,2} = \half\sqrt{xs}\,(1,0,0,\pm 1)\;,\;\;\;
\bk_{1,2} = \half\sqrt{xs}\,(1,\pm\bb(xs)\sin\thq,0,\pm\bb(xs)\sin\thq)\;;
\label{eq:2to2iniza}
\eeq

\noindent
$\bullet$ for final-state emission
\beq
\bp_{1,2} = \half\sqrt{s}\,(1,0,0,\pm 1)\;,\;\;\;
\bk_{1,2} = \half\sqrt{s}\,(1,\pm\bb(s)\sin\thout,0,\pm\bb(s)\sin\thout)\;,
\label{eq:2to2finza}
\eeq
with $\thout$ given in eq.~(\ref{eq:solthoutex}).
The momenta in eq.~(\ref{eq:2to2iniza}) are then boosted to the lab frame
using eq.~(\ref{eq:ybstini}), those in eq.~(\ref{eq:2to2finza}) using
eq.~(\ref{eq:ybstfin}). We now observe that the momenta in 
eq.~(\ref{eq:2to2iniza}) can be obtained with the parametrization
of eq.~(\ref{eq:thrbps}) by imposing the soft limit $x\to 1$, but 
freezing the c.m.\ energy to the value $xs$. Furthermore, as described 
in sect.~\ref{sec:FSR}, for final-state emission the reshuffling does 
not change the direction of the outgoing heavy quark in the partonic 
c.m.\ frame; thanks to eq.~(\ref{eq:bhvsthq}), a practical way of
computing it is to consider the soft-emission limit in the
parametrization of eq.~(\ref{eq:thrbps}). Thus, in the soft limit
the scattering angles in the parton $\twototwo$ c.m.\ frames coincide
for initial- and final-state emission. Unfortunately, the $\twototwo$ 
momenta do not coincide, because of the different $s$-channel invariants
and boosts to the lab frame. However, we can force them to coincide with
a simple formal manipulation, similar to what is done for event projection.
We can use the identity $xs=\bx_{1i}\bx_{2i}S$ in eq.~(\ref{eq:2to2iniza}), 
and $s=x_1 x_2 S$ in eq.~(\ref{eq:2to2finza}). Next, we formally
replace $x_1$ with $\bx_{1i}$ and $x_2$ with $\bx_{2i}$ in
eq.~(\ref{eq:sMCQtwo}) (which is allowed, these variables being just
integration variables there), and then we change integration variables
back to $x_1$ and $x_2$. Eqs.~(\ref{eq:Lout}) and~(\ref{eq:Hout}) become
\beqn
{\cal L}_{ab}^{(Q,l)}&=&\frac{\partial(\bx_{1i},\bx_{2i})}{\partial(x_1,x_2)}
f_a^{\Hone}(\bx_{1i})f_b^{\Htwo}(\bx_{2i}),
\label{eq:LoutS}
\\
{\cal H}_{ab}^{(Q,l)}&=&\frac{16\pi}{\bx_{1i}\bx_{2i} S}
\frac{d\cos\thout}{d\cos\thq}
\MEc{ab}{l}\left(\bx_{1i},\bx_{2i},\bt_Q\right).
\label{eq:HoutS}
\eeqn
In may appear counterintuitive to have a dependence on $\bx_{1i}$ and
$\bx_{2i}$ in these equations. However, it must be clear that this has
nothing to do with initial-state emission. Simply, the freedom of changing
integration variables allowed us to scale the partonic c.m.\ energy in
eq.~(\ref{eq:sMCQtwo}) by a factor of $x$. Thus, eq.~(\ref{eq:2to2iniza})
now holds for final-state emission as well. Furthermore, the boost to
the lab frame is now performed with eq.~(\ref{eq:ybstini}) rather than
with eq.~(\ref{eq:ybstfin}). This is precisely what we wanted to achieve,
since a unique $\EVprjmap$ has now been defined. For a given choice
of $(x_1,x_2,\phi_3)$, one reconstructs the $\twototwo$ momenta in the hard
c.m.\ frame by taking the soft limit with the $s$-channel invariant fixed 
to $xx_1x_2S$, and then boosts to the lab frame using eq.~(\ref{eq:ybstini}).

We stress that, in the case of final-state emissions, 
we define the MC subtraction terms using
eqs.~(\ref{eq:Lout})--(\ref{eq:Hout}) for $\clH$ events, and
eqs.~(\ref{eq:LoutS})--(\ref{eq:HoutS}) for $\clS$ events.
In principle, we should therefore introduce the notations
$d\bSigma_{ab}^{(\clH)}\xMC$ and $d\bSigma_{ab}^{(\clS)}\xMC$,
but we prefer to avoid it, since it has been shown that the difference
between the two amounts to a change in the integration variables.

The zero-angle procedure implies the possibility of several 
definitions of the MC subtraction terms, which differ beyond NLO.
For example, one could use the $\twototwo$ invariants obtained in
the zero-angle-emission limit, eqs.~(\ref{eq:ini_invza}) 
and~(\ref{eq:fin_invza}), in the computation of the hard $\twototwo$ 
matrix elements, for both $\clH$ and $\clS$ event, or only for
$\clS$ events. We prefer not to consider the latter option, which
results in a total rate computed by the MC@NLO different from
the NLO one. The former choice has indeed been implemented; the
results have been found to be identical to those obtained with our 
default choice in the case of top production. When bottom production
is considered, very small differences are visible in the tails of 
those distributions which are effectively of leading-order accuracy in
the fixed-order computation of ref.~\cite{Mangano:jk}, such as the
transverse momentum of the pair or $\Delta\phi$. However, these
differences never exceed a few percent, and are much smaller than
the uncertainties of the fixed-order results due to scale variation.
This confirms that reshuffling effects are beyond the accuracy 
of MC@NLO.

As pointed out in $\FW$, sect.~A.5, the MC subtraction terms given
in eqs.~(\ref{eq:sMCplustwo}) and~(\ref{eq:sMCQtwo}) cannot act as
{\em local} counterterms for real emission matrix elements, since the
angular distribution of a soft gluon emitted by the MC does not agree
with the corresponding perturbative result. However, it has been argued in 
$\FW$ that this effect must be irrelevant in the definition of observables,
at least for infrared-safe ones. Thus, in $\FW$ the MC subtraction terms
were defined by smoothly matching what is obtained from the
perturbative expansion of the MC result with the leading singular
behaviour of the real matrix elements in the soft limit. The matching 
has been defined through a parameter-dependent damping function, \FWeq{A.86}.
The physical observables were found to be independent of the parameters
used to define the damping function, thus practically confirming that
infrared-safe observables are insensitive to the angular distribution
of soft-gluon emission.

Here, we shall follow the same strategy. Since the soft singularity
structure is more complicated than in the case of gauge-boson pair
production, and in particular cannot be associated with the soft divergence
of an Altarelli-Parisi splitting function, we generalize \FWeq{A.83}
and \FWeq{A.84} as follows:
\beqn
\frac{d\bSigma_{ab}}{d\phi_3}\xMCB&=&
\Gfun_{s}(x)\Gfun_{c}(y)
\sum_{L}\sum_{l}\frac{d\bSigma_{ab}^{(L,l)}}{d\phi_3}\xMCB
+(1-\Gfun_{s}(x)){\cal M}_{ab}({\bf S})
\nonumber \\*&+&
(1-\Gfun_{c}(y)){\cal M}_{ab}({\bf C})
-(1-\Gfun_{s}(x))(1-\Gfun_{c}(y)){\cal M}_{ab}({\bf SC}),
\label{eq:MCsubtfinal}
\eeqn
where ${\cal M}_{ab}({\bf S})$, ${\cal M}_{ab}({\bf C})$ and 
${\cal M}_{ab}({\bf SC})$
denote the leading singular behaviour of the real matrix element in the
soft, collinear, and soft-collinear limits respectively. The functional
form of the damping functions $\Gfun_{s}(x)$ and $\Gfun_{c}(y)$ is the
same, and is given in \FWeq{A.86} (with $x_{\sss DZ}=0$); the subscript 
is to remind us that the parameters these functions contain can be varied 
independently. The function $\Gfun_{s}(x)$ smoothly approaches zero 
in the soft limit
\beq
\lim_{x\to 1}\Gfun_s(x)=0\,,\;\;\;\;\;\;
x=1+\frac{v_1+v_2}{s}\,,
\label{eq:Gfuns}
\eeq
and the function $\Gfun_{c}(y)$ smoothly approaches zero 
in the initial-state collinear limits
\beq
\lim_{y\to\pm 1}\Gfun_c(y)=0\,,\;\;\;\;\;\;
y=\frac{v_2-v_1}{v_1+v_2}\,.
\label{eq:Gfunc}
\eeq
Note that the function $\Gfun_c(y)$ would not be necessary away from the
soft limit, if azimuthal correlations (see app.~\ref{sec:azcorr}) were
either absent or properly described by the MC.
As in $\FW$, no evidence has been found of a dependence of the physical
results upon the parameters entering the damping functions.

In summary, eq.~(\ref{eq:MCsubtfinal}) is our final formula for 
MC subtraction terms. The quantities $d\bSigma_{ab}^{(L,l)}\xMC$
are defined in eqs.~(\ref{eq:sMCplustwo}) and~(\ref{eq:sMCQtwo})
for initial- and final-state emissions respectively. ${\cal L}$, 
${\cal B}$, and ${\cal H}$ appearing there are given in 
eqs.~(\ref{eq:Lin})--(\ref{eq:Hin}) for initial-state emissions.
For final-state emissions, we use eqs.~(\ref{eq:Lout})--(\ref{eq:Hout}) 
for $\clH$ events, and eqs.~(\ref{eq:Bout}), (\ref{eq:LoutS}), 
(\ref{eq:HoutS}) for $\clS$ events.

\section{Colour flow codes}\label{sec:codes}
We list in tables~\ref{tab:Sflows} and~\ref{tab:Hflows} the codes {\tt IC}
used in MC@NLO to transmit colour flow information to the \HW\ event generator.
The convention is that $c_i$ and $\bar c_i$
represent the colour and anticolour of parton $i$. If $i$ is a quark
(antiquark) then its anticolour (colour) is zero.  Thus for example when
${\tt IC}=1$, corresponding to $q \bq\to gQ\bQ$ (see table~\ref{tab:Hflows}),
we have $c_q=c_g=1$,  $\bar c_{\bq}=\bar c_{\bQ}=2$, $\bar c_g=c_Q=3$, 
meaning that $q$ and $g$ are colour-connected,
$\bq$ and $\bQ$ are anticolour-connected, and the anticolour of $g$ is
connected to the colour of $Q$.  In accordance with the Les Houches convention,
the non-zero colour labels entered into the event common block {\tt HEPEUP}
are actually ${\tt ICOLUP}(1,i)=500+c_i$ and ${\tt ICOLUP}(2,i)=500+\bar c_i$. 

\begin{table}[htb]
\begin{center}
\begin{tabular}{|r|c|cccc|}\hline
{\tt IC} & $12\to 34$ & $c_1\bar c_1$ & $c_2\bar c_2$ & $c_3\bar c_3$ &
$c_4\bar c_4$\\
\hline\hline
1 & $q \bq\to Q\bQ$ & 10 & 02 & 10 & 02 \\
2 & $\bq q\to Q\bQ$ & 01 & 20 & 20 & 01 \\
3 & $g g  \to Q\bQ$ & 12 & 23 & 10 & 03 \\
4 & $g g  \to Q\bQ$ & 12 & 31 & 30 & 02 \\
\hline
\end{tabular}
\end{center}
\caption{\label{tab:Sflows}
Colour flow codes for $\twototwo$ configurations}
\end{table}

\begin{table}[htb]
\begin{center}
\begin{tabular}{|r|c|ccccc|}\hline
{\tt IC} & $12\to 345$ & $c_1\bar c_1$ & $c_2\bar c_2$ & $c_3\bar c_3$ &
$c_4\bar c_4$ & $c_5\bar c_5$\\
\hline\hline
1 & $q \bq\to gQ\bQ$ & 10 & 02 & 13 & 30 & 02 \\
2 & $q \bq\to gQ\bQ$ & 10 & 02 & 32 & 10 & 03 \\
3 & $q  g \to qQ\bQ$ & 10 & 21 & 30 & 20 & 03 \\
4 & $q  g \to qQ\bQ$ & 10 & 23 & 20 & 10 & 03 \\
5 & $\bq q\to gQ\bQ$ & 01 & 20 & 23 & 30 & 01 \\
6 & $\bq q\to gQ\bQ$ & 01 & 20 & 31 & 20 & 03 \\
7 & $\bq g\to \bq Q\bQ$ & 01 & 23 & 03 & 20 & 01 \\
8 & $\bq g\to \bq Q\bQ$ & 01 & 12 & 03 & 30 & 02 \\
9 & $g q  \to qQ\bQ$ & 12 & 20 & 30 & 10 & 03 \\
10 & $g q \to qQ\bQ$ & 12 & 30 & 10 & 30 & 02 \\
11 & $g\bq\to \bq Q\bQ$ & 12 & 03 & 02 & 10 & 03 \\
12 & $g\bq\to \bq Q\bQ$ & 12 & 01 & 03 & 30 & 02 \\
13 & $g g  \to gQ\bQ$ & 12 & 23 & 14 & 40 & 03 \\
14 & $g g  \to gQ\bQ$ & 12 & 34 & 32 & 10 & 04 \\
15 & $g g  \to gQ\bQ$ & 12 & 23 & 43 & 10 & 04 \\
16 & $g g  \to gQ\bQ$ & 12 & 31 & 34 & 40 & 02 \\
17 & $g g  \to gQ\bQ$ & 12 & 34 & 14 & 30 & 02 \\
18 & $g g  \to gQ\bQ$ & 12 & 31 & 42 & 30 & 04 \\
\hline
\end{tabular}
\end{center}
\caption{\label{tab:Hflows}
Colour flow codes for $\twotothree$ configurations}
\end{table}

\section{$\zeta$ subtraction}
In \FW, an NLO subtraction scheme was introduced, called $\zeta$ 
subtraction, which allows one to reduce the number of negative-weight
events occurring in MC@NLO, compared to that resulting from the
implementation of the ``standard'' subtraction formulae of 
ref.~\cite{Frixione:1993yp}, in which the computation of 
W$^+$W$^-$ cross sections to NLO accuracy was originally performed.
The $\zeta$ subtraction is essentially defined by the requirement that
the subtraction terms be non-zero only in the region
\beq
\frac{4\,k_{\sss {\rm T}}^2}{s}\,<\,\zeta\,,
\label{eq:zetasubdef}
\eeq
$k_{\sss {\rm T}}$ being the transverse momentum of the real parton 
emitted at the NLO. In terms of the variables $x$ and $y$ (see 
eqs.~(\ref{eq:Gfuns}) and~(\ref{eq:Gfunc}) for their boost-invariant
definitions, and eq.~(\ref{eq:thrbps}) for an explicit phase-space
parametrization which uses them), eq.~(\ref{eq:zetasubdef}) becomes
\beq
\pfun(x,y)\,\equiv\,(1-x)^2\,(1-y^2)\,<\,\zeta\,.
\label{eq:zetasubt}
\eeq
As pointed out in \FW, eq.~(\ref{eq:zetasubt}) is not specific to vector
boson pair production, and can be applied to any NLO cross section.
However, the formulae for the partonic cross sections derived in 
app.~A.3 of \FW\, need to be generalized. That is the aim of this
section. Although we shall closely follow the notation of 
ref.~\cite{Mangano:jk}, it will be clear that our results are
relevant to any hard production process of a heavy (or hard) system,
for which the matrix elements have all possible soft and collinear
divergencies, except those due to final-state collinear emissions.

We start by observing that eq.~(\ref{eq:zetasubt}) implies that there 
is a minimum value of $x$, for fixed $y$, equal to
\beq
\tilde\rho \equiv \min(x) = {\rm max}\,
\left(1-\sqrt{\frac{\zeta}{1-y^2}}\,,\,\rho\right),
\label{eq:xmin}
\eeq
where $\rho$ is the minimum value of $x$ allowed by kinematics (for heavy 
flavour production, $\rho=4m^2/s$). We then write the analogue of eq.~(3.5) 
of ref.~\cite{Mangano:jk} in a more general form\footnote{Consistently
with app.~\ref{sec:MCsubt}, and unlike the convention of 
ref.~\cite{Mangano:jk}, we have here $0\leq \varphi \leq 2\pi$.}
\beq \label{eq:sigggr}
d\siggg^{(r)}=
d\Phi_r^{(x)}\frac{s^{1-\ep}}{64\pi^3}
dx\,dy\,d\thd\;
(1-x)^{-1-2\ep}\,(1-y^2)^{-1-\ep}\,|\sin\thd|^{-2\ep}\,\fgg(x,y,\thd)\;,
\eeq
where $d\Phi_r^{(x)}$ is the $d$-dimensional ($d=4-2\ep$) reduced phase 
space for the production of the heavy system with a squared invariant mass 
equal to $sx$. In $d\Phi_r^{(x)}$ we also incorporate further 
normalization factors, that reduce to 1 in 4 dimensions.
Thus, the phase space of the heavy system plus the light parton is
\beq
d\Phi=d\Phi_r^{(x)}\frac{s^{1-\ep}}{64\pi^3}
\,dx\,dy\,d\thd\;
(1-x)^{1-2\ep}\,(1-y^2)^{-\ep}\,|\sin\thd|^{-2\ep}.
\label{eq:Fullphsp}
\eeq
In the present context,
the precise  $d$-dimensional form of $d\Phi_r^{(x)}$ is irrelevant.
The only thing we need to know is that in 4 dimensions it is equal to
the phase space of the reduced system. Thus, in the case of heavy flavour
production
\beq
d\Phi_r^{(x)}\vert_{d=4}=\frac{\bb(xs)}{16\pi} d\cos\thu\,,
\eeq
which coincides with eq.~(\ref{eq:twobps}) for $s\to xs$. Using this 
equation in eq.~(\ref{eq:Fullphsp}), we get $d\Phi=d\phi_3(s)$ in 4 
dimensions, with $d\phi_3(s)$ given in eq.~(\ref{eq:thrbps}).
In eq.~(\ref{eq:sigggr}), we have also defined $\fgg$ as
\beq
\fgg = (1-x)^2\,(1-y^2)\, {\cal M}_{ab}(x,y,\thd)\;,
\eeq
where ${\cal M}_{ab}(x,y,\thd)$ is the invariant spin-averaged squared
amplitude divided by the flux factor. Its
dependence upon the kinematic variables of the reduced system
($\thu$ in the case of heavy flavour production) is not explicitly
shown.

Following the reasoning of ref.~\cite{Mangano:jk}, we now use the expansion
\beq
(1-x)^{-1-2\ep}=-\frac{\tilde\beta^{-4\ep}}{2\ep}\delta(1-x)
+\omxr - 2\ep \lomxr
+{\cal O}(\ep^2)\,,
\eeq
with $\tilde\beta=\sqrt{1-\tilde\rho}$ (notice that here $\tilde\rho$ 
depends on $y$, which is not the case in ref.~\cite{Mangano:jk}).
This yields 
\beqn
d\siggg^{(r)}&=&d\siggg^{(s)}+
d\Phi_r^{(x)}\frac{s^{1-\ep}}{64\pi^3}
\,dx\,dy\,d\thd\,
(1-y^2)^{-1-\ep}\,|\sin\thd|^{-2\ep}
\nonumber \\ &\times&
\left[\omxr-2\ep\lomxr\right]
\,\fgg(x,y,\thd),
\label{eq:splusns}
\eeqn
where
\beq
d\siggg^{(s)}=
d\Phi_r \frac{s^{1-\ep}}{64\pi^3}
\,dy\, d\thd\,
(1-y^2)^{-1-\ep}\,|\sin\thd\,|^{-2\ep} 
\left[-\frac{\tilde\beta^{-4\ep}}{2\ep} \right]
\,\fgg(x,y,\thd)\vert_{x=1}\;,
\eeq
with $d\Phi_r=d\Phi_r^{(x)}\vert_{x=1}$.
We can now expand
\beq
(1-y^2)^{-1-\ep} =
-[\delta(1+y)+\delta(1-y)]\frac{(2\tilde\omega)^{-\ep}}{2\ep}
+\frac{1}{2}\left[\omyo+\opyo\right]+{\cal O}(\ep),
\label{eq:yiden}
\eeq
where we define
\beq
\tilde{\omega}=1-\sqrt{{\rm max}\left(1-\frac{\zeta}{(1-x)^2}, 0\right)}\;.
\eeq
This quantity has the same formal meaning as $\omega$ in 
ref.~\cite{Mangano:jk}; unlike $\omega$, however, it does depend
on $x$. Using eq.~(\ref{eq:yiden}) in eq.~(\ref{eq:splusns}) we get
\beq
d\siggg^{(r)}=d\siggg^{(s)}+d\siggg^{(c+)}+d\siggg^{(c-)}+d\siggg^{(f)},
\eeq
where
\beqn
d\siggg^{(c\pm)}&=&
d\Phi_r^{(x)}\frac{s^{1-\ep}}{64\pi^3}
\,dx\,d\thd\,|\sin\thd|^{-2\ep}\, \Bigg\{
\left[\omxrr-2\ep\lomxrr\right]
\nonumber \\ &&
\times \left[-\frac{2^{-\ep}}{2\ep}\right]
+ \frac{1}{2} \frac{\log\tilde\omega}{1-x} \Bigg\}
\,\fgg(x,y,\thd)\vert_{y=\pm 1}\,,
\label{sigmacpm}
\eeqn
and
\beq
d\siggg^{(f)}= \frac{1}{2}\,
\omxr\left[\omyo+\opyo\right]
\frac{\fgg(x,y,\thd)}{1-x}\,d\Phi\,.
\label{sigmaf}
\eeq
Notice that $\rho$ prescriptions, rather than $\tilde\rho$ prescriptions,
appear in eq.~(\ref{sigmacpm}), since $\tilde\rho=\rho$ for
$y\to\pm 1$. Furthermore, the expression $\log\tilde\omega/(1-x)$ 
in eq.~(\ref{sigmacpm}) does not need a regularization prescription, 
since as $x\to 1$ also $\tilde\omega\to 1$.

To see precisely the structure of the generated
counterterms in $d\siggg^{(f)}$, let us expand the expression
\beq
dx\, dy \omxr\omyo V(x,y)
\eeq
according to the definition of the distributions. Here $V(x,y)$ represents 
symbolically some function of the final-state kinematics, regular for 
$x\to 1$ and $y\to 1$. We get
\beqn
&& \int dx\, dy \omxr\omyo V(x,y) \nonumber 
\\&&\;\;
=\int\frac{ dx\, dy }{1-x}\left[\omyo V(x,y) - 
\stepf(x-\tilde\rho)\omyp V(1,y)\right]
\\&&\;\;
=\int\frac{ dx\, dy}{1-x}\left\{\frac{ V(x,y)
 - \stepf(\tilde\omega-(1-y)) V(x,1)}{1-y}
-\frac{\stepf(x-\tilde\rho) V(1,y)-\stepf(y)V(1,1)}{1-y}\right\}\nonumber 
\eeqn
where in the middle (last) expression we have used the fact that
$\tilde\omega \to 1$ when $x\to 1$ ($\tilde\rho\to\rho$ when $y\to 1$). 
Using the relation
\beq
\stepf(\tilde\omega-(1-y))=\stepf(x-\tilde\rho) \stepf(y)
\eeq
we can rewrite the above expression as
\beqn
&& \int dx\, dy \omxr\omyo V(x,y) \nonumber 
\\&&\phantom{aaaa}
=\int \frac{ dx\, dy}{1-x}\Bigg\{\frac{ V(x,y)}{1-y}
-\frac{\stepf(x-\tilde\rho) (V(1,y)+\stepf(y)(V(x,1)-V(1,1)))}{1-y}
\nonumber \\ &&\phantom{\int \frac{ dx\, dy}{1-x}aaaaaa}
+\frac{\stepf(\tilde\rho-x)\stepf(y) V(1,1)}{1-y}\Bigg\}.
\eeqn
The last term is finite by itself, since
\beq
\stepf(\tilde\rho-x)=\stepf\left((1-x)^2(1-y^2)-\zeta\right).
\eeq
It is a soft term, which we can rewrite by explicitly computing the
integral over $y$
\beq
\int \frac{ dx\, dy}{1-x}\frac{\stepf(\tilde\rho-x)\stepf(y) V(1,1)}{1-y}
=- V(1,1) \int_\rho^1 dx \frac{\log\tilde\omega}{1-x}\,.
\eeq
For consistency with \FW , we include this term in $d\siggg^{(c+)}$,
which is then modified with the replacement
\beq
\frac{\log\tilde\omega}{1-x}
\Longrightarrow \frac{\log\tilde\omega}{1-x} - \delta(1-x) \int_\rho^1 dx \frac{\log\tilde\omega}{1-x}
\equiv \left(\frac{\log\tilde\omega}{1-x}\right)_\rho\;.
\eeq

We now present the corrections to be {\em added} to the soft and collinear
terms in order to go from the standard subtraction scheme of
ref.~\cite{Mangano:jk} (computed with $\min(x)=\rho$ and 
$\omega=1$) to the $\zeta$-subtraction scheme.
In the soft term, the correction is given by
\beqn
&&\Delta d\siggg^{(s)}=
-d\Phi_r\frac{s^{1-\ep}}{128\pi^3\ep}
\nonumber \\&&\;\;\;\;\times
\int dy \, d\thd \,(1-y^2)^{-1-\ep}\,|\sin\thd|^{-2\ep}
\,(\tilde\beta^{-4\ep}-\beta^{-4\ep})\; \fgg(x,y,\thd)\vert_{x=1}\,.
\eeqn
We find
\beqn
\tilde\beta^{-4\ep}-\beta^{-4\ep}&=&
\stepf\left(\beta-\sqrt[4]{\frac{\zeta}{1-y^2}}\right)
\left[ \left(\frac{\zeta}{1-y^2}\right)^{-\ep}-\beta^{-4\ep}\right]
\nonumber \\
&=& \stepf\left(1-y^2 - \frac{\zeta}{\beta^4}\right) \beta^{-4\ep}
\left[ \left(\frac{\zeta/\beta^4}{1-y^2}\right)^{-\ep}-1\right].
\eeqn
Since the $\stepf$ function implies $|y|<1$, we only need to expand
the square bracket to order $\ep$, thus getting
\beq
\tilde\beta^{-4\ep}-\beta^{-4\ep}
= \stepf\left(1- \frac{\zeta}{\beta^4}-y^2 \right)
\left[ \log \frac{\beta^4}{\zeta} + \log(1-y^2)\right]\,\ep\,.
\eeq
So, our final formula for the correction is
\beqn
&&\Delta d\siggg^{(s)}=
-d\Phi_r\frac{s}{128\pi^3}
\nonumber \\&&\;\;\;\;\times
\int_{-\bar{y}}^{\bar{y}} dy \int d\thd \,
\left[ \frac{\log \frac{\beta^4}{\zeta}}{1-y^2} + 
\frac{\log(1-y^2)}{1-y^2}\right]
 \fgg(x,y,\thd)|_{x=1}\,,
\label{eq:softcorrzeta}
\eeqn
with (see \FWeq{A.27})
\beq
\bar{y}=\stepf\left(1- \frac{\zeta}{\beta^4}\right)
\sqrt{1- \frac{\zeta}{\beta^4}}\;.
\eeq
The soft limit $\fgg(x,y,\thd)|_{x=1}$ is non-zero only if the radiated 
parton is a gluon. It is given in general by a sum over eikonal factors
\beq
\fgg(x,y,\thd)|_{x=1} =
\sum_{lm} (q_l,q_m)\;{\cal C}_{lm}\,,\;\quad\quad
(q_l,q_m)\equiv\frac{q_l\cdot q_m}{q_l\cdot k\; q_m\cdot k}\,,
\label{eq:fggsoft}
\eeq
where $q_{l/m}$ are the external momenta of the reduced process,
$k$ is the momentum of the soft gluon, and ${\cal C}_{lm}$ are
functions of the momenta of the reduced process.
Expressions for $\fgg(x,y,\thd)|_{x=1}$ in the case of heavy
quark production can be easily obtained (after correcting a couple
of misprints\footnote{$1/2 C_{\sss A}$ in the last line of eq.~(A.12)
should read $C_{\sss A}$, and the $-2$ in the last line of eq.~(A.21) 
should read $+2$}) from Appendix A of ref.~\cite{Mangano:jk}. Terms 
arising from the eikonal factor $(p_1,p_2)$ associated with initial-state 
soft emissions can be integrated analytically over $y$ and $\thd$ in
eq.~(\ref{eq:softcorrzeta}). For the other terms, only the azimuthal
integration can be easily computed analytically. The $y$ integration is
performed numerically.

\noindent
The collinear correction is
\beq
\Delta d\siggg^{(c\pm)} = d\Phi_r^{(x)}\frac{s}{128\pi^3}
dx\, d\thd
 \left(\frac{\log\tilde\omega}{1-x}\right)_\rho
\fgg(x,y,\thd)\vert_{y=\pm 1}\;.
\eeq
Finally, for the finite part of the real cross section an equation
identical to \FWeq{A.26} holds:\footnote{In \FWeq{A.26} a different notation
is used, which leads to an error if one forgets to freeze ${\cal P}(x,y)$
when applying the ${\cal P}$ prescriptions, as required in 
eqs.~({\bf I}.A.21)--({\bf I}.A.23).}
\beq\label{eq:zetasub}
d\siggg^{(f)}=\frac{1}{2}\,
\left[ \left(\frac{1}{(1-x)(1-y)}\right)_{\cal P}
+\left(\frac{1}{(1-x)(1+y)}\right)_{\cal P}\right]
\,\frac{\fgg(x,y,\thd)}{1-x}\,d\Phi\;,
\eeq
where the ${\cal P}$-distribution prescription is defined by
\beqn\label{eq:pprescription}
&& \left(\frac{1}{(1-x)(1\pm y)}\right)_{\cal P} V(x,y) \equiv
\\ &&\quad\quad
\frac{1}{1-x}\left\{\frac{ V(x,y)}{1\pm y}
-\frac{\stepf(x-\tilde\rho) (V(1,y)-\stepf(\mp y)(V(x,\mp 1)-V(1,\mp 1)))}
{1\pm y}\right\}.\nonumber 
\eeqn

We now discuss the relation of the above results with the results of \FW. 
The collinear correction is equivalent to the term
\begin{displaymath}
\left(\frac{\log(1-{\cal F}_c(x))}{1-x}\right)_{\cal P}
\end{displaymath}
of \FWeq{A.25}. In the case of the soft correction, only the terms
corresponding to \mbox{$\{l,m\}=\{1,2\}$} are non-zero in 
eq.~(\ref{eq:fggsoft}); these can be integrated analytically,
yielding the ${\cal F}_s$ term in \FWeq{A.24}.

In the case of heavy flavour production, we find that $\zeta$ subtraction
reduces only marginally (1\%--2\%) the number of negative weights with 
respect to standard subtraction. In fact, due to the simultaneous
presence of several colour flows which induce different dead zones
(see figs.~\ref{fig:fin} and~\ref{fig:ini}), the subtraction region
of eq.~(\ref{eq:zetasubt}) never matches closely the region in which
\HW\ radiation is allowed. Nevertheless, it is still advantageous to
use $\zeta$ subtraction, since the parameter tuning, necessary to reduce as
much as possible the number of negative weights, is easier than in the case of
standard subtraction (simply because the latter depends on two parameters
rather than one).

\end{document}